\documentclass{aa}
\usepackage[varg]{txfonts}
\usepackage{graphicx}
\usepackage{hyperref}
\hypersetup{colorlinks,allcolors=blue}
\usepackage{color, colortbl}
\usepackage{natbib}
\bibpunct{(}{)}{;}{a}{}{,} 

\newcommand{\lya}{Ly$\alpha$}
\newcommand{\ha}{H$\alpha$}
\newcommand{\hi}{H\textsc{i}}
\definecolor{Gray}{gray}{0.9}

\begin{document} 

\title{The Lyman Alpha Reference Sample}
\subtitle{XVI. Global 21cm \hi\ properties of Lyman-$\alpha$ emitting galaxies}

\titlerunning{Global 21cm \hi\ properties of Lyman-$\alpha$ emitting galaxies}
\authorrunning{Le Reste et al.}

\author{Le Reste A. \inst{1,2}
\and Hayes M. J.\inst{1}
\and Cannon J. M.\inst{3}
\and Melinder J.\inst{1}
\and Runnholm A.\inst{1}
\and Rivera-Thorsen T. E.\inst{1}
\and \"{O}stlin G.\inst{1}
\and Adamo A.\inst{1}
\and Herenz E. C.\inst{4}
\and Schaerer D.\inst{5}
\and Scarlata C.\inst{2}
\and Kunth D. \inst{6}
}

\institute{The Oskar Klein Centre, Department of Astronomy, Stockholm University, AlbaNova, SE-10691 Stockholm, Sweden.
\and Minnesota Institute for Astrophysics, University of Minnesota, 116 Church Street SE, Minneapolis, MN 55455, USA. 
\and Department of Physics and Astronomy, Macalester College, 1600 Grand Avenue, Saint Paul, MN 55105, USA.  
\and Inter-University Centre for Astronomy and Astrophysics (IUCAA), Pune University Campus, Pune 411 007, India. 
\and Observatoire de Genève, Université de Genève, Chemin Pegasi 51, 1290, Versoix, Switzerland.
\and Institut d’Astrophysique de Paris, 98 bis Boulevard Arago, F-75014 Paris, France. } 

\date{Received date /
Accepted date }

\abstract
{The Lyman-$\alpha$ (\lya) line of hydrogen is a well-known tracer of galaxies at high-redshift,observed at $z\gtrsim0.002$, all the way into the Epoch of Reionization. However, the connection between \lya\ observables and galaxy properties has not fully been established, limiting the use of the line to probe the physics of galaxies.}
{Here, we derive global neutral hydrogen gas (\hi) properties of nearby \lya-emitting galaxies to assess the impact of neutral gas on the \lya\ output of galaxies.}
{We observed 21cm line emission using the Karl G. Jansky Very Large Array in D-array configuration ($\sim$55" resolution, $\sim38\,$kpc) for 37 star-forming galaxies with available \lya\ imaging from the Lyman Alpha Reference Samples (LARS and eLARS). We compare these observations to \lya\ properties from Hubble Space Telescope imaging. }
{We detect 21cm emission for 33 out of the 37 galaxies observed. We find no significant correlation of global \hi\ properties (including \hi\ mass, column density, gas fraction, depletion time, line width or velocity shift between \hi\ and \lya), with \lya\ luminosity, escape fraction or equivalent width derived with HST photometry. Additionally, both \lya-emitters and weak or non-emitters are distributed evenly along the \hi\ parameter space of optically-selected $z=0$ galaxies.
 Around 74\% of the sample is undergoing galaxy interaction, this fraction is higher for \lya-emitters (83\% for galaxies with $EW\geq20\AA$) than for non or weak emitters (70\%). Nevertheless, galaxies identified as interacting have \lya\ and \hi\ properties statistically consistent with those of non-interacting galaxies. } 
{Our results show that global \hi\ properties (on scales $>30$kpc) have little direct impact on the \lya\ output from galaxies. Instead, neutral gas likely regulates \lya\ emission on small scales: statistical comparisons of \lya\ and high angular resolution 21cm observations are required to fully assess the role of \hi\ in \lya\ radiative transfer. While our study indicates that major and minor galaxy mergers could play a role in the emission of \lya\ photons in the local universe, especially for galaxies with high \hi\ fractions, the line-of-sight through which a system is observed ultimately determines \lya\ observables.}

\keywords{Galaxies: starburst -- Galaxies: ISM -- Radio lines: galaxies -- ISM: lines and bands}

\maketitle

\section{Introduction}
The Lyman-$\alpha$ (\lya, 2p$\rightarrow$1s, $\lambda=1215.67\,$\textrm{\AA}) line of hydrogen is a fundamental tracer of the physical processes occurring in galaxies. In star-forming galaxies, it is primarily produced during recombination of interstellar hydrogen after ionization by Lyman Continuum radiation from O and B stars. During hydrogen recombination, \lya\ photons are emitted with a probability of 68\% assuming case B recombination \citep{Dijkstra2014}, making \lya\ the intrinsically strongest line emitted by galaxies. While the line is emitted in the rest-frame ultraviolet (UV), \lya\ is redshifted to the optical range at $z>2$ and the infrared at $z\gtrsim5$, making it an excellent observational probe of the high-redshift universe with current ground-based and space telescopes. Consequently, this emission line has enabled the discovery of several tens of thousands of galaxies through narrow-band imaging and spectroscopy \citep[][]{Ouchi2020,MentuchCooper2023}. Furthermore, \lya\ is an excellent tool to study the Epoch of Reionization, an important cosmological epoch that saw the transition of the Universe from mostly neutral to predominantly ionized \citep[e.g.][]{Haiman1999,Stark2010,Konno2018}.

However, the use of the \lya\ line to extract information about galaxies is limited by the complex radiative transfer that \lya\ photons undergo within the neutral interstellar medium (ISM). The line is resonant, undergoing spatial and spectral scattering in neutral hydrogen gas (\hi) that modifies the line shape \citep{Verhamme2006}. Furthermore, dust within the interstellar medium absorbs \lya\ photons, and the probability of absorption increases with the number of scattering events, and thus, with the column density of neutral gas. As a consequence, the \lya\ output of galaxies is thought to strongly depend on the neutral gas content, geometry, and kinematics of galaxies \citep[e.g.][]{Giavalisco1996,Kunth1998,MasHesse2003,Atek2009,Wofford2013,Ostlin2021,Hayes2023}. This also means that \lya\ profiles could potentially be used to recover information about the neutral gas properties of individual galaxies, and vice versa. Assessing the impact of the neutral interstellar medium on \lya\ emission and whether ISM properties can be recovered from \lya\ profiles requires detailed comparisons between \hi\ properties and the \lya\ output of galaxies.

Most of the available information on the neutral interstellar medium of \lya-emitting galaxies has been obtained through observations of low ionisation state (LIS) lines from metals \citep{Savage1996,Shapley2003,Quider2009,Henry2015,Rivera-Thorsen2015,Jaskot2019,Reddy2022,Hayes2023a,Parker2024}. These absorption lines can be observed in the same spectra as \lya, providing a very useful proxy for neutral gas column density, kinematics and covering fraction, at a limited observational cost. Studies using metal absorption lines have shown that outflows are necessary for \lya\ escape from high neutral gas column density environments, and \lya\ emission is generally associated with low neutral gas covering fractions. These results indicate that \hi\ column density, regulated by early time radiative and wind feedback, is likely a driver in the emission of \lya\ photons.
However, the use of LIS absorption lines presents a few drawbacks. Among those are the fact that LIS absorption lines trace neutral gas on small scales within the spectroscopic aperture. While these scales are the same as those where \lya\ is measured, absorption lines only yield information on the gas directly in front of the source, when \lya\ is expected to scatter in surrounding gas, and on the back-side of continuum sources \citep{Verhamme2006}. Furthermore, many studies have pointed to the fact that metal absorption lines generally underestimate the neutral gas column density  \citep{Reddy2016,Gazagnes2018,Huberty2024}, and the most commonly used lines such as C\textsc{ii} and Si\textsc{ii} are also found in  Hydrogen-ionized gas. For these reasons, LIS absorption lines from metals do not directly relate \lya\ emission to the neutral gas content and distribution within galaxies. Lyman series absorption lines such as Lyman-$\beta$ or Lyman-$\gamma$, originate directly from neutral hydrogen. They have been used as a complement to LIS absorption lines in a handful of studies \citep{Henry2015,Steidel2018,Gazagnes2020}, mostly confirming the results obtained with metal absorption lines, but also demonstrating their limitations. Nevertheless, whether they originate from metals or hydrogen, absorption lines remain limited as means to characterize the \hi\ content of \lya-emitting galaxies. Similarly to metal lines, Lyman series absorption lines do not probe the full physical scale of neutral gas impacting \lya\ radiative transfer, as they do not probe the medium behind and around UV continuum sources, that also affects the \lya\ output of galaxies due to scattering of \lya\ photons. They have left a central question relating to \lya\ radiative transfer without answer: how does the large-scale neutral gas reservoir of galaxies impact \lya\ emission?

The 21cm line of hydrogen is one of the best tracers of neutral gas as it originates directly from neutral hydrogen, and allows for studies of the global \hi\ content and kinematics of galaxies. However, it is faint and thus only observable in the very nearby universe. The furthest direct detection of 21cm emission from an individual galaxy has been done at $z=0.376$ \cite{Fernandez2016}, and 21cm stacking measurements have included galaxies up to a higher redshift bound of $z=1.45$ \citep{Chowdhury2020}. Efforts have been made to detect lensed 21cm signal at higher redshifts, yielding one marginal detection at $z=0.407$ \citep{Blecher2019,Deane2024}. In comparison, most of \lya\ observations have been made above $z=2$, where the \lya\ line is redshifted enough to be observed by ground-based telescopes. Comparing 21cm and \lya\ emission in galaxies therefore requires UV observations in the nearby universe, which can only be obtained from space. 

The Lyman Alpha Reference Samples \citep[LARS and eLARS, also referred to as (e)LARS in the rest of the manuscript][]{Ostlin2014,Hayes2014,Melinder2023} have been assembled to provide detailed, resolved observations of \lya\ emission from 42 local galaxies. Owing to a large coverage across wavelength, these samples aim to provide a complete census of the galaxy and interstellar medium properties that drive \lya\ emission. Notably, they are the only samples of \lya-emitting galaxies with uniform 21cm \hi\ imaging. This makes the sample ideal to assess the link between the 21cm and \lya\ emission. A few 21cm studies of \lya-emitting galaxies have been conducted \citep[e.g.][]{Cannon2004,Kanekar2021,LeReste2022,Purkayastha2022} and some even found tentative correlations between the properties of the neutral interstellar medium traced by the 21cm line and \lya\ properties \citep{Pardy2014}. However until now, none had sufficiently large samples to assess the role of neutral gas on \lya-emission in a statistically robust manner. \citet{Pardy2014} evaluated the impact of global neutral gas content through single dish observations of 14 galaxies in the original LARS sample, and low angular resolution interferometric observations for 5 galaxies using the Karl G. Jansky Very Large Array (VLA) in D-configuration. This study found tentative anti-correlations between the HI line width and the \lya\ extension parameter (the ratio between \lya\ and \ha\ Petrosian radii), and between the HI mass and both the \lya\ escape fraction and equivalent width. Furthermore, a qualitative multi-wavelength study of two galaxies with unprecedented high resolution 21cm observations (3.5" synthesized beam size) in the (e)LARS samples reported morphological similarities between \ha, \lya\ and 21cm emission \citep{LeReste2022}. Both of these studies point towards the possible existence of scaling relations between 21cm and \lya\ properties. However, the samples these studies were conducted on contained few galaxies, limiting efforts to establish the general neutral gas properties of \lya-emitters and investigate correlations between 21cm \hi\ and \lya\ emission properties. 
Here, we present 21cm VLA D-configuration observations of 37 galaxies in the LARS and eLARS samples. We aim to determine if \lya-emitting galaxies belong to a different galaxy population than optically-selected galaxies per their \hi\ gas properties, and quantify the impact of the neutral gas content and kinematics of galaxies on their \lya\ emission output.

The paper is structured as follows. The methods, including the description of 21cm observations, 21cm data reduction scheme, and derivation of \lya\ and galaxy properties are presented in Section \ref{sec:methods}. Results on the 21cm properties of \lya-emitters, on the impact of global \hi\ properties on \lya\ observables, and on the role of galaxy interactions are presented in Section \ref{sec:results}. We discuss the implications of these results for \lya\ emission from galaxies in Section \ref{sec:discussion} and provide a summary and conclusion in Section \ref{sec:conclusion}.\\
Throughout this paper, we assume a standard $\Lambda$CDM cosmology with parameters H$_0=70\,$km.s$^{-1}$.Mpc$^{-1}$, $\Omega_m=0.3$ and $\Omega_{\Lambda}=0.7$. Unless stated otherwise, redshifts and velocities follow the optical convention. We use the terminologies "neutral gas" and "neutral hydrogen gas" interchangeably to refer to neutral hydrogen gas.

\begin{table*}[!t]
   \scriptsize
    \label{tab:gal_params}
    \centering
        \caption{General properties of the LARS and eLARS galaxies. The mean error on the redshift is 3e-6, yielding a mean error on the luminosity distance of 0.02 Mpc. a - From SED fitting of UV, optical and IR data in the GSWLC catalog \citep{Salim2016}. b -  GSWLC stellar masses and SFR predicted from fit to MPA-JHU values presented in \citet{Melinder2023}. A gray background indicates the galaxies that were not observed as part of the LARS VLA D-array observation campaign. On the last row, we show the average and standard deviation for the full sample, where applicable. }
    \begin{tabular}{cccccccccccc}
    \hline
    ID & RA & DEC & z & D & 
    M$_{GSWLC}^a$ & SFR$_{GSWLC}^a$ & EW$_{Ly\alpha}$ &f$_{esc,Ly\alpha}$  &L$_{Ly\alpha}$&$\Delta$v$_{Ly\alpha}^{red}$ \\
    & J2000 & J2000 &  & Mpc& 
    $10^{10}$ M$_{\sun}$& M$_{\sun}$/yr& \AA & & $10^{41}\,\rm{erg}.\rm{s}^{-1}.\rm{cm}^{-2}$ &km$.\rm{s}^{-1}$\\
    \hline
  LARS01 &  202.183400 &  43.930530 &  0.0280 &  122.4 & 
  0.01$\pm$0.01 $^b$& 1.31$\pm$0.23$^b$& $42.2^{+0.4}_{-0.5}$& $0.134^{+0.002}_{-0.002}$ &  $8.35^{+0.09}_{-0.09}$ & 102.3 $^{+ 13.8 }_{- 1.1 }$ \\ 
  LARS02 &  136.770630 &  53.449046 &  0.0298 &  130.7 &
  0.03$\pm$0.001 & 0.56$\pm$0.01& $56.7^{+1.3}_{-1.9}$& $0.299^{+0.01}_{-0.01}$ &  $4.15^{+0.09}_{-0.12}$ &  123.9 $^{+ 14.4 }_{- 14.5 }$ \\ 
  LARS03 &  198.896420 &  62.124249 &  0.0307 &  134.7 & 
  3.56$\pm$0.19 & 5.28$\pm$0.73& $34.2^{+1.3}_{-1.8}$& $0.005^{+0.000}_{-0.000}$ &  $1.92^{+0.08}_{-0.10}$ & 348.9 $^{+ 38.8 }_{- 27.8 }$ \\ 
  LARS04 &  196.867650 &  54.447443 &  0.0325 &  143.7 &
  0.07$\pm$0.04$^b$ & 1.48$\pm$0.26$^b$ & $2.1^{+0.5}_{-0.4}$& $0.006^{+0.001}_{-0.001}$ & $0.28^{+0.06}_{-0.06}$&406.7 $^{+ 14.2 }_{- 1.5 }$ \\ 
  LARS05 &  209.962580 &  57.439725 &  0.0338 &  148.6& 
  0.02$\pm$0.01$^b$ & 1.15$\pm$0.20$^b$ &$24.9^{+0.5}_{-0.5}$& $0.126^{+0.002}_{-0.002}$ &  $6.52^{+0.11}_{-0.11}$ & 152.7 $^{+ 1.1 }_{- 0.8 }$ \\ 
  LARS06 &  236.435780 &  44.263876 &  0.0341 &  150.1 &
  0.03$\pm$0.005 & 0.51$\pm$0.01&$\leq2.4$& $\leq0.012$ & $\leq0.09$ & -  \\ 
  LARS07 &  199.016280 &  29.381746 &  0.0378 &  166.6 &
  0.08$\pm$0.04$^b$ & 4.30$\pm$0.78$^b$& $38.8^{+0.9}_{-1.0}$& $0.111^{+0.002}_{-0.002}$ &  $6.66^{+0.13}_{-0.14}$ & 160.0 $^{+ 0.8 }_{- 1.1 }$ \\ 
  LARS08 &  192.557420 &   7.578947 &  0.0382 &  169.3 &
  6.37$\pm$0.61 & 5.35$\pm$0.21&$17.3^{+1.3}_{-0.9}$& $0.006^{+0.000}_{-0.000}$ & $4.02^{+0.27}_{-0.19}$  & 28.8 $^{+ 2.5 }_{- 2.5 }$ \\ 
  LARS09 &  125.978840 &  28.106336 &  0.0472 &  209.4 &
  1.91$\pm$0.76$^b$ & 9.67$\pm$1.82$^b$ & $9.8^{+0.5}_{-0.5}$& $0.016^{+0.001}_{-0.001}$ & $5.81^{+0.28}_{-0.27}$ & - \\
  \rowcolor{Gray}
  LARS10 &  195.423210 &  29.381453 &  0.0574 &  256.5 &
  1.03$\pm$0.22 & 4.53$\pm$1.03&$\leq2.1$ &$\leq0.003$  & $\leq0.24$ & - \\
  \rowcolor{Gray}
  LARS11 &  210.946580 &   6.470854 &  0.0844 &  384.5 &
  6.55$\pm$0.80 & 20.51$\pm$1.22 & $20.5^{+1.0}_{-0.7}$& $0.065^{+0.009}_{-0.009}$ &  $17.9^{+1.3}_{-0.97}$ & 155.3 $^{+ 40.6 }_{- 14.2 }$ \\ 
  \rowcolor{Gray}
  LARS12 &  144.556340 &  54.473719 &  0.1021 &  470.5 &
  0.19$\pm$0.06 & 15.17$\pm$0.18 & $18.1^{+1.0}_{-1.0}$& $0.027^{+0.001}_{-0.001}$ &  $15.4^{+0.9}_{-0.76}$& 390.6 $^{+ 1.6 }_{- 31.1 }$ \\ 
  \rowcolor{Gray}
  LARS13 &   27.618398 &  13.149793 &  0.1467 &  696.0 &
  2.07$\pm$0.40 & 19.19$\pm$1.71& $\leq1.8$& $\leq0.004$  &  $\leq2.99$& 267.8 $^{+ 17.1 }_{- 46.1 }$ \\ 
  \rowcolor{Gray}
  LARS14 &  141.501600 &  44.460023 &  0.1807 &  875.4 &
  13.87$\pm$3.08 & 40.28$\pm$17.01& $49.0^{+2.3}_{-2.1}$& $0.263^{+0.01}_{-0.01}$ &  $55.6^{+1.91}_{-1.98}$&  229.0 $^{+ 1.2 }_{- 1.5 }$ \\ 
 eLARS01 &  242.919650 &  52.456920 &  0.0295 &  129.0 &
 5.11$\pm$0.08  & 9.59$\pm$0.13 &$21.2^{+0.3}_{-0.3}$& $0.012^{+0.000}_{-0.000}$ &  $5.17^{+0.06}_{-0.06}$ & 93.4 $^{+ 27.2 }_{- 15.7 }$ \\ 
 eLARS02 &  200.287655 &  59.101636 &  0.0429 &  189.6 &
 1.29$\pm$0.01& 3.02$\pm$0.03&$12.4^{+0.4}_{-0.3}$& $0.061^{+0.002}_{-0.002}$ &  $3.31^{+0.12}_{-0.08}$ & 337.3 $^{+ 1.7 }_{- 13.3 }$ \\ 
 eLARS03 &  176.295715 &  61.708249 &  0.0353 &  154.7 & 
 3.05$\pm$1.20$^b$ & 4.84$\pm$0.88$^b$ &$4.4^{+0.5}_{-0.4}$& $0.005^{+0.001}_{-0.001}$ & $1.23^{+0.15}_{-0.12}$ & - \\
 eLARS04 &  262.099778 &  57.545171 &  0.0286 &  125.0 &
 1.09$\pm$0.10 & 3.77$\pm$0.68& $17.1^{+0.5}_{-0.4}$& $0.07^{+0.002}_{-0.002}$ & $4.42^{+0.12}_{-0.11}$ & - \\
 eLARS05 &  166.258059 &  59.684407 &  0.0337 &  148.0 &
 4.19$\pm$0.34 & 2.93$\pm$0.08&$26.5^{+1.0}_{-0.9}$& $0.148^{+0.006}_{-0.006}$ & $5.88^{+0.22}_{-0.18}$ & 199.5 $^{+ 53.0 }_{- 4.4 }$ \\ 
 eLARS06 &  179.669216 &  64.964576 &  0.0337 &  148.2 & 
 0.74$\pm$0.02 &1.20$\pm$0.03& $12.6^{+1.4}_{-1.0}$& $0.06^{+0.005}_{-0.005}$ & $1.3^{+0.14}_{-0.1}$ & - \\
 eLARS07 &  153.179202 &  61.551092 &  0.0348 &  152.8 & 
 0.03$\pm$0.02$^b$ & 0.88$\pm$0.15$^b$& $7.1^{+1.1}_{-1.4}$& $0.032^{+0.006}_{-0.006}$ &  $0.81^{+0.13}_{-0.16}$ & 
284.3 $^{+ 0.6 }_{- 0.6 }$ \\
 eLARS08 &  157.587230 &  61.263667 &  0.0307 &  134.6 & 
 1.52$\pm$0.15 & 1.64$\pm$0.37& $16.7^{+1.4}_{-1.4}$& $0.017^{+0.002}_{-0.002}$ & $1.37^{+0.12}_{-0.12}$  & - \\
 eLARS09 &  201.894782 &  66.754643 &  0.0303 &  132.9 & 
 0.15$\pm$0.03 & 1.55$\pm$0.19&$5.6^{+1.0}_{-1.0}$& $0.054^{+0.01}_{-0.01}$ & $0.46^{+0.08}_{-0.08}$ & - \\
 eLARS10 &  166.267513 &  59.665887 &  0.0332 &  146.0 & 
 1.31$\pm$0.11 & 1.25$\pm$ 0.25& $16.8^{+2.4}_{-1.9}$& $0.016^{+0.002}_{-0.002}$ & $0.93^{+0.12}_{-0.1}$ & - \\
 eLARS11 &  141.814497 &  58.615173 &  0.0302 &  132.2 & 
 0.67$\pm$0.04 & 2.01$\pm$0.28&$8.9^{+1.0}_{-0.9}$& $0.065^{+0.007}_{-0.007}$ & $0.62^{+0.07}_{-0.06}$&  - \\
 eLARS12 &  196.529145 &  59.217530 &  0.0320 &  140.5 & 
 1.92$\pm$0.14 & 1.34$\pm$0.33& $\leq1.3$& $\leq0.002$ & $\leq0.08$ & -\\
 eLARS13 &  162.752742 &  65.994651 &  0.0325 &  142.7 & 
 0.25$\pm$0.003  & 1.02$\pm$0.01&$37.7^{+1.6}_{-1.4}$& $0.207^{+0.008}_{-0.008}$ &  $2.53^{+0.09}_{-0.09}$ & 192.7 $^{+ 0.9 }_{- 0.8 }$ \\ 
 eLARS14 &  171.841551 &  60.748431 &  0.0326 &  143.3 & 
 0.29$\pm$0.04& 0.75$\pm$0.06& $\leq1.3$& $\leq0.005$ & $\leq0.07$ & - \\
 eLARS15 &  206.157772 &  61.240107 &  0.0354 &  155.7 & 
 0.79$\pm$0.13 & 2.21$\pm$0.36 &$20.0^{+3.0}_{-2.6}$& $0.124^{+0.018}_{-0.018}$ &  $0.81^{+0.12}_{-0.1}$ & 135.3 $^{+ 13.5 }_{- 2.6 }$ \\
 eLARS16 &  208.038017 &  56.108526 &  0.0350 &  154.0 &
 0.36$\pm$0.04 & 0.64$\pm$0.10&$\leq2.4$& $\leq0.009$ &$\leq0.06$ & - \\
 eLARS17 &  156.734892 &  58.828176 &  0.0311 &  136.5 &
 0.63$\pm$0.06 & 0.54$\pm$0.05&$19.4^{+3.1}_{-2.5}$& $0.082^{+0.012}_{-0.012}$ & $0.77^{+0.12}_{-0.1}$ &  - \\
 eLARS18 &  230.223349 &  57.189353 &  0.0295 &  129.1 &
 0.12$\pm$0.02  & 0.30$\pm$ 0.04& $7.9^{+2.7}_{-3.3}$& $0.03^{+0.013}_{-0.013}$ &  $0.17^{+0.06}_{-0.07}$ & - \\
 eLARS19 &  181.276581 &  56.558696 &  0.0309 &  135.3 &
 0.04$\pm$0.01 & 0.33$\pm$0.01& $10.0^{+2.0}_{-1.6}$& $0.064^{+0.01}_{-0.01}$ & $0.27^{+0.05}_{-0.04}$ & - \\ 
 eLARS20 &  204.743405 &  61.832522 &  0.0312 &  137.1 &
 0.27$\pm$0.04 & 0.83$\pm$0.53& $7.9^{+1.8}_{-2.0}$& $0.031^{+0.008}_{-0.008}$ & $0.21^{+0.05}_{-0.05}$ &  - \\
 eLARS21 &  214.797140 &  65.829525 &  0.0328 &  144.1 &
 0.08$\pm$0.02& 0.24$\pm$0.02 & $9.0^{+3.5}_{-3.1}$& $0.047^{+0.033}_{-0.033}$ & $0.11^{+0.04}_{-0.04}$ & - \\
 eLARS22 &  257.303506 &  60.830417 &  0.0471 &  208.8 &
 0.08$\pm$0.02 & 5.43$\pm$0.14& $6.8^{+0.5}_{-0.6}$ &  $0.056^{+0.005}_{-0.005}$ & $1.85^{+0.14}_{-0.15}$ & 294.7 $^{+ 13.7 }_{- 43.1 }$ \\ 
 eLARS23 &  222.114337 &  63.036274 &  0.0511 &  227.4 &
 2.48$\pm$0.39 & 4.11$\pm$0.57 & $6.5^{+1.0}_{-1.1}$ &  $0.04^{+0.007}_{-0.007}$ & $1.49^{+0.21}_{-0.24}$ & - \\
 eLARS24 &  220.772412 &  61.310608 &  0.0479 &  212.8 &
 5.53$\pm$0.17 & 11.14$\pm$0.26 & $22.3^{+1.0}_{-0.9}$ & $0.004^{+0.000}_{-0.000}$  & $4.47^{+0.18}_{-0.18}$ & 490.1$^{+ 2.1 }_{- 2.4 }$ \\ 
 eLARS25 &  153.009203 &  60.621991 &  0.0450 &  199.2 &
 1.18$\pm$0.12 & 1.49$\pm$0.08 & $9.6^{+1.1}_{-1.2}$ &  $0.094^{+0.012}_{-0.012}$ & $1.64^{+0.18}_{-0.20}$ & 228.3 $^{+ 1.8 }_{- 1.7 }$ \\ 
 eLARS26 &  178.173450 &  66.307502 &  0.0460 &  203.9 &
 1.93$\pm$0.20 &2.09$\pm$0.74 & $21.5^{+1.7}_{-1.8}$ & $0.059^{+0.005}_{-0.005}$  & $2.23^{+0.18}_{-0.17}$ &  180.2$^{+ 13.7 }_{- 29.9 }$ \\ 
 eLARS27 &  225.698237 &  62.338188 &  0.0446 &  197.5 &
 0.57$\pm$0.09 & 1.29$\pm$0.25& $18.8^{+1.6}_{-1.5}$ & $0.115^{+0.011}_{-0.011}$ &  $2.05^{+0.16}_{-0.16}$ & 289.1 $^{+ 4.9 }_{- 83.0 }$ \\ 
 eLARS28 &  178.077903 &  58.949460 &  0.0462 &  204.9 &
 1.06$\pm$0.08 & 8.61$\pm$0.36& $3.1^{+1.3}_{-1.1}$ & $0.013^{+0.005}_{-0.005}$  &  $0.37^{+0.15}_{-0.13}$ &262.3 $^{+ 14.2 }_{- 26.2 }$ \\ 
    \hline
    Average & - & - & 0.045$\pm$0.030& 203$\pm$147 & 1.3$\pm$1.8 & 4.1$\pm$5.2  & 18.4$\pm$13.0 & 0.071$\pm$0.070 & 4.75$\pm$9.44 & 233$\pm$110\\
        \hline
    \end{tabular}
\end{table*}
\section{Data and methods}\label{sec:methods}
\subsection{VLA 21cm data description and reduction}
The LARS and eLARS galaxies were observed with the Karl G. Jansky Very Large Array (VLA)\footnote{The VLA is operated by the National Radio Astronomy Observatory.} in the D configuration as part of projects 13A-181 and 14A-077 (PI Cannon). All galaxies but LARS 10, 11, 12, 13 and 14 were observed. The observations made use of the L-band centered on the respective predicted 21 cm \hi\ frequencies of the galaxies, based on optical redshifts. Information about the observations, including the integration time, percentage of flagged visibilities, sources used for calibration, beam parameters and rms can be found in Appendix Table \ref{tab:vla_obs}. Part of the D-configuration VLA 21cm observations of the LARS and eLARS galaxies have been presented in previous studies \citep{Pardy2014,LeReste2022}. \\

Data reduction followed standard prescriptions in the CASA 5.5.0 environment \citep{Casa2022}. Only a fraction of the measurement sets was selected for data reduction using the task \texttt{split}: we kept 50\% of channels centered around the theoretical 21cm line frequency according to the optical redshift. In the frequency range considered, Radio Frequency Interferences (RFIs) pose a significant problem, especially for compact array configurations. We adopted a two-step scheme to mask the visibilities impacted by RFIs. First, strong and narrow RFIs were removed automatically using the task \texttt{tfcrop} (\texttt{maxnpieces = 3}, \texttt{timecutoff = 3.0}, \texttt{freqcutoff = 3.0}), resulting in the flagging of 0-3\% of the data for each measurement set. The remaining RFIs were removed with the \texttt{flagdata} task following visual inspection of the datasets. We report the fraction of visibilities that required flagging due to RFI contamination in Table \ref{tab:vla_obs}. The flagged measurement sets were then calibrated using the bandpass and phase calibrators listed in Table \ref{tab:vla_obs}. The measurement sets containing target galaxies and calibrators were split using the task \texttt{split} to keep only the visibilities corresponding to the target galaxies.

The calibrated measurement sets were continuum subtracted in the uv-plane by fitting a polynomial of order 1 on line-free channels for each individual dataset using the task \texttt{uvcontsub}. For each measurement set, visibilities were re-weighted according to their scatter using the task \texttt{statwt} on line-free channels. For continuum subtraction and re-weighting, we typically used 100 line-free channels on each side of the line, separated from the edges of the line by at least 50 channels. The calibrated, continuum subtracted datasets were cleaned to 0.5$\sigma$ using the CASA task \texttt{tclean}. We used both the \texttt{auto-multi-thresh} algorithm \citep{Kepley2020} and visual inspection to identify 21cm emission regions in the cubes and define the regions where to apply the clean algorithm. Cleaning was performed with task \texttt{tclean} on the 21cm emission regions identified in the cube, using a Briggs weighting robust parameter of 0.5. The clean images were set to have a common beam and the spectral channel width of the cubes was set to 5 km.s$^{-1}$ (using the radio velocity definition). We corrected for the primary beam with the task \texttt{impbcor}, and produced cubes using the optical velocity definition.

Finally, we used the 21cm Source Finding Application \citep[SoFIA 2,][]{Serra2015,Westmeier2021} on the non-primary beam corrected cubes to identify the regions containing emission for subsequent analysis. The S+C algorithm was run with a threshold of 4.0$\sigma$, spatial smoothing kernels of 0, 3, 5, 10, 15 pixels and spectral smoothing kernels of 0, 3, 5, and 11 spaxels. We used a reliability threshold of 0.95 to improve the reliability of detections. The mask produced by SoFIA 2 was applied to the primary beam corrected images in order to recover accurate flux density values. We used the SoFIA Image Pipeline \citep{Hess2022} to inspect and validate the \hi\ detections made by SoFIA 2.

\begin{figure*}[!t]
    \centering
    \includegraphics[width=\textwidth]{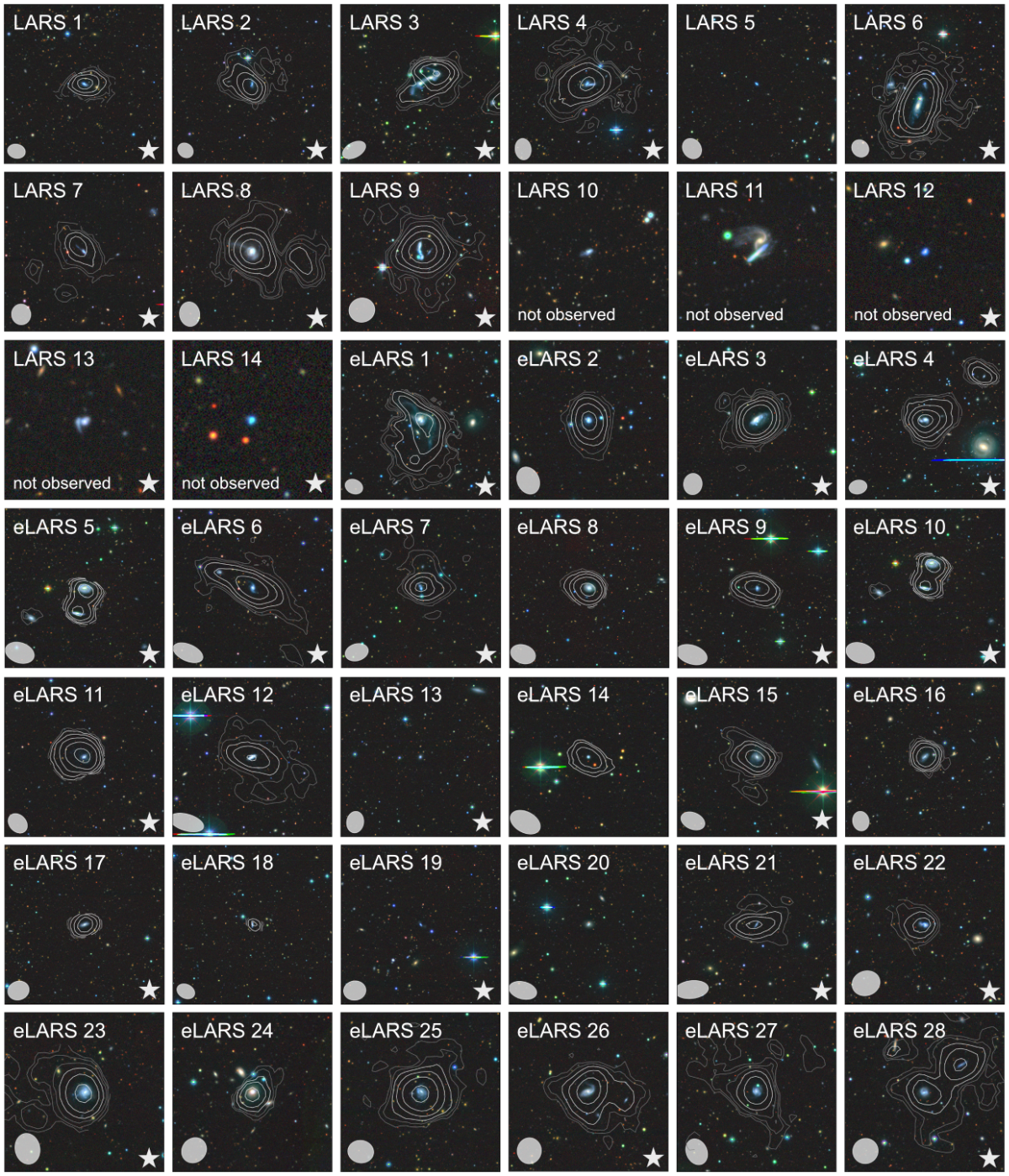}
    \caption{DECaLS optical g, r and z band colour-composite images of galaxies in the LARS and eLARS samples with \hi\ 21cm column density contours overlaid. The contours have fixed levels of $2^n \times10^{19}\,\textrm{cm}^{-2}$, with $n=0,1,2,...,5$. White contours have $\textrm{SNR}\geq3$, low signal-to-noise contours are shown in gray. All images are $300\times300\,\textrm{kpc}^2$, and are centered upon the coordinates of the galaxies. The contours show the main SoFIA 2 detection, apart from eLARS05 and eLARS10 for which an ancillary detection is shown in grey to the South East. A star on the lower right corner of a panel indicates the galaxy has been identified as undergoing interaction.}
    \label{fig:opt_nhic}
\end{figure*}

\subsection{\hi\ properties and data products}
A major limitation to comparing data observed at 21cm and in the UV is the drastic difference in both the fields of view and resolutions reached by observations. The average synthesized beam size of the VLA D 21cm cubes in our sample is 55 arcseconds, corresponding to an average 38 kpc scale covered in physical units (see Appendix Table \ref{tab:vla_obs}). Meanwhile, the HST data used to extract \lya\ properties have much higher resolution (0.04" per pixel), and the fields of view of HST images are significantly smaller than those in 21cm images. The average aperture diameter to extract the \lya\ properties used in this paper is 9.4 kpc \citep{Melinder2023}, about four times smaller than the average scale reached by the VLA synthesized beam. To alleviate the scale discrepancy between UV and radio observations, we provide measurements of the \hi\ properties of the LARS and eLARS galaxies in two different apertures. The first set of 21cm measurements was extracted from a synthesized beam-shaped aperture centered on the coordinates of the target galaxy (see Table \ref{tab:gal_params}). This is the physically smallest unit of information that can be recovered from the 21cm cubes, and is used when comparing \lya\ and 21cm \hi\ properties. Despite this effort to improve the match between \lya\ and 21cm extraction apertures, the beam-extracted 21cm emission is still extracted in more that four times the diameter of the aperture used for \lya\ photometry. The second set of measurements we provide is the total 21cm extraction obtained from the SoFIA 2 mask. We note that this total measurement encompasses all the diffuse and extended \hi\ emission, as well as the emission coming from objects interacting with the (e)LARS galaxies in some cases. Therefore, the total extracted 21cm is not as relevant for \lya\ visibility as compared to the 21cm emission extracted in the beam. We use the total measurement to assess the general 21cm properties of the galaxies in the sample. In particular, we compare the total \hi\ masses and gas fraction to those of galaxies in the local universe observed with single dish telescopes. 

We have derived several data products to characterize the properties of the 21cm emission in the LARS and eLARS galaxies. The products and methodology used to derive these are described below. Prior to deriving 21cm flux and spectra, the intensities measured in Jy/beam were converted to Jy through multiplication by $bf = (4\,\ln2\,l_{pix}^2)/(\pi\,\textrm{b$_{min}$}\,\textrm{b$_{maj}$})$, with $l_{pix}$ the pixel length, b$_{min}$ the minor beam axis and b$_{maj}$ the major beam axis in arcseconds. 

\paragraph{Integrated 21cm emission map. }
We produce moment-0 maps by integrating the 21cm data across the spectral dimension :
$$\textrm{M}_{0} = \sum_{v} I_{v} \Delta\textrm{v}$$
with $\Delta v$ the channel width in km.s$^{-1}$.

\paragraph{\hi\ column density map. }
From the integrated emission maps $\textrm{M}_{0}$ in Jy.beam$^{-1}$.km.s$^{-1}$, we derive column density $N_{HI}$ maps in atoms per cm$^{2}$, assuming the \hi\ line is optically thin, with b$_{\text{min}}$ and b$_{\text{maj}}$ in arcseconds:
$$N_{HI} = \dfrac{1.09\times10^{24}}{\textrm{b$_{\text{min}}$}\,\textrm{b$_{\text{maj}}$}}\textrm{M}_{0}$$

\paragraph{\hi\ velocity centroid map. }
We produce moment-1 maps that indicate the line centroid per pixel:
$$\textrm{M}_{1} = \frac{\sum_{v} v\, I_{v} \Delta\textrm{v}}{\textrm{M}_0}$$

\paragraph{\hi\ linewidth map. } The linewidth maps are calculated by deriving the full width half maximum of the line for each pixel in the following way:
$$\textrm{FWHM}=\sqrt{8\ln2\times\frac{\sum_{v} (v-\textrm{M}_1)^2\, I_{v} \Delta\textrm{v}}{\textrm{M}_0}}$$
For both the moment-1 and linewidth maps, we applied a signal-to-noise constraint to only show values with S/N>5. Note that we only used positive pixels to generate the linewidth maps, due to the presence of pixels with negative values away from the line center in multiple datasets. Since pixels away from the line center have larger weights, negative pixels lead to invalid values in the square root and thus, incomplete linewidth maps.

\paragraph{21cm Spectra. }
We extract two spectra for each galaxy: one from the SoFIA-masked data and one from a beam-sized aperture applied to the SoFIA-masked data. We also show values outside the mask, using either the beam size for the beam extraction or a circular mask with the average size of the SoFIA mask in the spectral direction. The noise is estimated as the rms per channel from regions outside of the SoFIA mask, taking into account the number of pixels in the mask per channel $N_v$, and a bandpass calibrator error of 3\% such that $\textrm{I}_{v,err} = rms_v\,\sqrt{N_v}+0.03\,\textrm{I}_{v}$.

\paragraph{21cm Flux. }
The \hi\ flux is calculated by summing all the pixels within the detection mask, using the optically thin regime assumption:
$$\textrm{S}_{\text{HI}} =\sum_{x,y,v} I_{x,y,v} \Delta\textrm{v}$$
The error on the flux is derived in a similar way as the error on the spectra: $\textrm{S}_{\text{HI,err}} = \sqrt{\sum_{v}(\text{rms}_v\,N_v)^2}\Delta\textrm{v}+0.03\,\textrm{S}_{\text{HI}}$. We provide upper limits for the galaxies that were not detected, using the median \hi\ linewidth within the beam and total extraction (respectively 154 and 156 km/s) and the 1$\sigma$ flux limit measured through the rms noise in the image.\\

\paragraph{Neutral gas mass. } 
The \hi\ mass in units of M$_{\sun}$ is derived from the integrated 21cm flux using the \hi\ mass equation \citep[see e.g.][]{Roberts+Haynes1994}, with the luminosity distance $D$ in Mpc, and the \hi\ flux $\textrm{S}_{HI}$ in Jy.km.s$^{-1}$ :\\
$$\textrm{M}_{HI}= 2.36 \times 10^{5}\,D^2\, \textrm{S}_{HI}$$
The error on the mass is derived taking into account the \hi\ flux error.

\paragraph{\hi\ velocity and redshift.}
The \hi\ velocities quoted for the beam and total extraction are derived by calculating the centroid within the detection mask $v_{HI}=\sum_{v}v\,I_v/(\sum_{v}I_v)$, 
the error is calculated via  $\textrm{v}_{\textrm{HI,err}}=\sqrt{(\Delta\textrm{v}/2)^2 + \sum_v(\textrm{I}_{v,err}^2\,(v-\textrm{v}_{\textrm{HI}})^2)/(\sum_v\textrm{I}_{v}^2}$) 

\paragraph{21cm line profile width.}
To derive the line width, we fit the \hi\ line profile with Gaussian profiles. For each galaxy, we visually examine the fit to the profiles with one, two and three Gaussian components, and select the fitting function yielding the best fit to the edges of the line profile. For most galaxies, we use profiles with two Gaussian components, but for a few objects these provide a poor fit. Therefore, we use a single Gaussian fit for LARS07, eLARS01 and eLARS28. For LARS06, eLARS03, eLARS05 and eLARS10, we use three Gaussian components to fit the spectra. We then measure the velocity corresponding to half the fitted peak value on each end of the profile to calculate the \hi\ line width $W_{50}$. The resulting velocity bounds used to calculate $W_{50}$ for each galaxy are indicated by dashed vertical lines in the bottom panel of Figure \ref{fig:opt_hi_el04} and Appendix Figures \ref{fig:opt_hi_l01} to\ref{fig:opt_hi_el28}.

\begin{table*}[ht]
    \centering
    \setlength{\tabcolsep}{5pt} 
    \scriptsize
        \caption{Neutral gas properties derived from 21cm VLA D configuration observations of the LARS and eLARS samples. We present both the properties extracted with a synthesized beam-shaped aperture, and with the total \hi\ detection mask made by SoFIA 2, only counting regions with SNR>3. The error on the HI redshift is on the order of 1e-5.}
    \begin{tabular}{|c|cccccc|cccccc|}
         \hline
            &\multicolumn{6}{c|}{Beam}      &\multicolumn{6}{c|}{Total}             \\
         ID & v$_{\textrm{HI,b}}$ &z$_{\textrm{HI,b}}$& S$_{\textrm{HI,b}}$ &M$_{\textrm{HI,b}}$&N$_{\textrm{HI,b}}$ & W50$_{\textrm{HI,b}}$& v$_{\textrm{HI,t}}$ & z$_{\textrm{HI,t}}$ & S$_{\textrm{HI,t}}$ &M$_{\textrm{HI,t}}$&N$_{\textrm{HI,t}}$& W50$_{\textrm{HI,t}}$\\
          & km/s & & Jy.km/s & $10^{9}\,$M$_{\sun}$ &$10^{19}$cm$^{-2}$&km/s & km/s &  &Jy.km/s & $10^{9}\,$M$_{\sun}$ &$10^{19}$cm$^{-2}$ & km/s\\
         \hline

LARS01 & 8326 $\pm$ 3 & 0.0278 & 0.32 $\pm$ 0.04 & 1.12 $\pm$ 0.13 & 6.7 $\pm$ 0.76 & 169 &
8323 $\pm$ 3 & 0.0278 & 0.65 $\pm$ 0.09 & 2.31 $\pm$ 0.33 & 3.55 $\pm$ 0.5 & 158 \\
LARS02 & 8937 $\pm$ 3 & 0.0298 & 0.27 $\pm$ 0.02 & 1.1 $\pm$ 0.1 & 7.23 $\pm$ 0.64 & 137 &
8946 $\pm$ 3 & 0.0298 & 0.84 $\pm$ 0.11 & 3.37 $\pm$ 0.44 & 2.84 $\pm$ 0.37 & 143 \\
LARS03 & 9386 $\pm$ 5.& 0.0313 & 0.17 $\pm$ 0.03 & 0.72 $\pm$ 0.12 & 3.03 $\pm$ 0.52 & 281 &
9451 $\pm$ 3 & 0.0315 & 1.91 $\pm$ 0.24 & 8.17 $\pm$ 1.02 & 3.08 $\pm$ 0.38 & 324 \\
LARS04 & 9782 $\pm$ 3 & 0.0326 & 0.51 $\pm$ 0.05 & 2.48 $\pm$ 0.23 & 12.92 $\pm$ 1.2 & 229 &
9768 $\pm$ 3 & 0.0326 & 2.17 $\pm$ 0.32 & 10.47 $\pm$ 1.56 & 5.31 $\pm$ 0.79 & 250 \\
LARS05 & --  & -- & $<$0.16 $\pm$ 0.03 & $<$0.84 $\pm$ 0.19 & --  & -- & --  & -- & 
$<$0.16 $\pm$ 0.1 & $<$0.82 $\pm$ 0.19 & --  & -- \\
LARS06 & 10459 $\pm$ 4 & 0.0349 & 0.63 $\pm$ 0.06 & 3.37 $\pm$ 0.3 & 18.06 $\pm$ 1.59 & 358
& 10372 $\pm$ 3.& 0.0346 & 5.34 $\pm$ 0.51 & 28.39 $\pm$ 2.72 & 12.75 $\pm$ 1.22 & 390 \\
LARS07 & 11332 $\pm$ 3 & 0.0378 & 0.14 $\pm$ 0.02 & 0.94 $\pm$ 0.16 & 3.18 $\pm$ 0.54 & 107
& 11329 $\pm$ 3 & 0.0378 & 0.32 $\pm$ 0.06 & 2.07 $\pm$ 0.38 & 2.09 $\pm$ 0.39 & 75 \\
LARS08 & 11464 $\pm$ 3 & 0.0382 & 0.77 $\pm$ 0.06 & 5.2 $\pm$ 0.38 & 17.83 $\pm$ 1.3 & 317
& 11456 $\pm$ 3 & 0.0382 & 2.39 $\pm$ 0.3 & 16.18 $\pm$ 2.03 & 6.06 $\pm$ 0.76 & 296 \\
LARS09 & 14029 $\pm$ 3 & 0.0468 & 0.66 $\pm$ 0.04 & 6.86 $\pm$ 0.42 & 18.69 $\pm$ 1.14 & 366
& 14020 $\pm$ 3 & 0.0468 & 1.57 $\pm$ 0.15 & 16.22 $\pm$ 1.56 & 7.44 $\pm$ 0.72 & 366 \\
eLARS01 & 8820 $\pm$ 3 & 0.0294 & 0.59 $\pm$ 0.05 & 2.33 $\pm$ 0.2 & 14.0 $\pm$ 1.17 & 122
& 8824 $\pm$ 3 & 0.0294 & 4.25 $\pm$ 0.57 & 16.7 $\pm$ 2.24 & 5.97 $\pm$ 0.8 & 153 \\
eLARS02 & 12848 $\pm$ 3 & 0.0429 & 0.29 $\pm$ 0.03 & 2.46 $\pm$ 0.24 & 6.84 $\pm$ 0.67 & 163 & 
12850 $\pm$ 3 & 0.0429 & 0.55 $\pm$ 0.07 & 4.69 $\pm$ 0.61 & 3.84 $\pm$ 0.5 & 81 \\
eLARS03 & 10594 $\pm$ 3 & 0.0353 & 1.31 $\pm$ 0.07 & 7.4 $\pm$ 0.38 & 31.55 $\pm$ 1.64 & 465 &
10598 $\pm$ 3 & 0.0353 & 3.21 $\pm$ 0.25 & 18.12 $\pm$ 1.42 & 11.47 $\pm$ 0.9 & 454 \\
eLARS04 & 8561 $\pm$ 3 & 0.0286 & 0.58 $\pm$ 0.04 & 2.15 $\pm$ 0.14 & 13.68 $\pm$ 0.92 & 190 &
8580 $\pm$ 3 & 0.0286 & 2.28 $\pm$ 0.25 & 8.39 $\pm$ 0.93 & 3.85 $\pm$ 0.43 & 174 \\
eLARS05 & 10120 $\pm$ 3 & 0.0338 & 0.36 $\pm$ 0.03 & 1.84 $\pm$ 0.15 & 5.07 $\pm$ 0.42 & 346 &
10056 $\pm$ 3 & 0.0335 & 0.83 $\pm$ 0.07 & 4.3 $\pm$ 0.34 & 3.29 $\pm$ 0.26 & 453 \\
eLARS06 & 10138 $\pm$ 3 & 0.0338 & 0.56 $\pm$ 0.04 & 2.91 $\pm$ 0.19 & 9.12 $\pm$ 0.6 & 171 &
10136 $\pm$ 3 & 0.0338 & 1.57 $\pm$ 0.17 & 8.16 $\pm$ 0.91 & 3.37 $\pm$ 0.37 & 160 \\
eLARS07 & 10481 $\pm$ 3 & 0.035 & 0.24 $\pm$ 0.03 & 1.34 $\pm$ 0.14 & 5.56 $\pm$ 0.58 & 219 &
10471 $\pm$ 3 & 0.0349 & 0.5 $\pm$ 0.07 & 2.77 $\pm$ 0.41 & 2.53 $\pm$ 0.38 & 203 \\
eLARS08 & 9200 $\pm$ 3 & 0.0307 & 0.39 $\pm$ 0.03 & 1.65 $\pm$ 0.12 & 5.99 $\pm$ 0.45 & 196 &
9194 $\pm$ 3 & 0.0307 & 0.72 $\pm$ 0.07 & 3.07 $\pm$ 0.29 & 2.81 $\pm$ 0.26 & 185 \\
eLARS09 & 9089 $\pm$ 3 & 0.0303 & 0.3 $\pm$ 0.03 & 1.26 $\pm$ 0.11 & 3.48 $\pm$ 0.3 & 154 &
9093 $\pm$ 3 & 0.0303 & 0.51 $\pm$ 0.0 & 2.11 $\pm$ 0.22 & 1.95 $\pm$ 0.2 & 127 \\
eLARS10 & 9978 $\pm$ 3 & 0.0333 & 0.3 $\pm$ 0.03 & 1.52 $\pm$ 0.13 & 4.29 $\pm$ 0.38 & 320 &
10056 $\pm$ 3 & 0.0335 & 0.83 $\pm$ 0.07 & 4.19 $\pm$ 0.33 & 3.29 $\pm$ 0.26 & 399 \\
eLARS11 & 9025 $\pm$ 3 & 0.0301 & 0.67 $\pm$ 0.04 & 2.75 $\pm$ 0.17 & 13.04 $\pm$ 0.81 & 228 &
9015 $\pm$ 3 & 0.0301 & 2.05 $\pm$ 0.16 & 8.46 $\pm$ 0.68 & 4.46 $\pm$ 0.36 & 228 \\
eLARS12 & 9636 $\pm$ 3 & 0.0321 & 0.39 $\pm$ 0.04 & 1.84 $\pm$ 0.18 & 4.96 $\pm$ 0.47 & 319 & 
9661 $\pm$ 3 & 0.0322 & 0.99 $\pm$ 0.15 & 4.59 $\pm$ 0.72 & 1.84 $\pm$ 0.29 & 303 \\
eLARS13 & --  & -- & $<$0.21 $\pm$ 0.03 & $<$1.03 $\pm$ 0.19 & -- & -- & --  & -- & 
$<$0.21 $\pm$ 0.1 & $<$1.0 $\pm$ 0.19 & --  & -- \\
eLARS14 & 9786 $\pm$ 3 & 0.0326 & 0.17 $\pm$ 0.01 & 0.81 $\pm$ 0.06 & 1.94 $\pm$ 0.14 & 175 & 
9791 $\pm$ 3 & 0.0327 & 0.26 $\pm$ 0.02 & 1.28 $\pm$ 0.1 & 0.91 $\pm$ 0.07 & 154 \\
eLARS15 & 10591 $\pm$ 3 & 0.0353 & 0.33 $\pm$ 0.03 & 1.9 $\pm$ 0.19 & 7.26 $\pm$ 0.72 & 166 & 
10587 $\pm$ 3 & 0.0353 & 0.7 $\pm$ 0.1 & 4.01 $\pm$ 0.57 & 3.46 $\pm$ 0.49 & 150 \\
eLARS16 & 10515 $\pm$ 3 & 0.0351 & 0.34 $\pm$ 0.04 & 1.89 $\pm$ 0.2 & 9.65 $\pm$ 1.02 & 203 & 
10515 $\pm$ 3 & 0.0351 & 0.64 $\pm$ 0.08 & 3.57 $\pm$ 0.44 & 5.03 $\pm$ 0.62 & 198 \\
eLARS17 & 9330 $\pm$ 3 & 0.0311 & 0.35 $\pm$ 0.03 & 1.55 $\pm$ 0.12 & 6.61 $\pm$ 0.53 & 217 & 
9332 $\pm$ 3 & 0.0311 & 0.51 $\pm$ 0.0 & 2.22 $\pm$ 0.19 & 4.07 $\pm$ 0.34 & 212 \\
eLARS18 & 8843 $\pm$ 4 & 0.0295 & 0.15 $\pm$ 0.02 & 0.58 $\pm$ 0.08 & 4.12 $\pm$ 0.54 & 222 & 
8843 $\pm$ 3 & 0.0295 & 0.16 $\pm$ 0.0 & 0.61 $\pm$ 0.08 & 3.87 $\pm$ 0.52 & 196 \\
eLARS19 & --  & -- & $<$0.16 $\pm$ 0.03 & $<$0.68 $\pm$ 0.19 & --  & -- & --  & -- 
& $<$0.15 $\pm$ 0.1 & $<$0.66 $\pm$ 0.19 & -- & -- \\
eLARS20 & --  & -- & $<$0.2 $\pm$ 0.03 & $<$0.89 $\pm$ 0.19 & --  & -- & --  & -- 
& $<$0.2 $\pm$ 0.1 & $<$0.87 $\pm$ 0.19 & -- & -- \\
eLARS21 & 9828 $\pm$ 3 & 0.0328 & 0.2 $\pm$ 0.03 & 0.99 $\pm$ 0.14 & 2.48 $\pm$ 0.35 & 159 &
9833 $\pm$ 3 & 0.0328 & 0.31 $\pm$ 0.0 & 1.53 $\pm$ 0.25 & 1.59 $\pm$ 0.26 & 154 \\
eLARS22 & 14094 $\pm$ 3 & 0.047 & 0.12 $\pm$ 0.02 & 1.23 $\pm$ 0.2 & 2.86 $\pm$ 0.47 & 268 &
14096 $\pm$ 3 & 0.047 & 0.16 $\pm$ 0.0 & 1.66 $\pm$ 0.28 & 2.15 $\pm$ 0.37 & 208 \\
eLARS23 & 15313 $\pm$ 3 & 0.0511 & 0.5 $\pm$ 0.04 & 6.09 $\pm$ 0.48 & 13.51 $\pm$ 1.06 & 127 &
15309 $\pm$ 3 & 0.0511 & 1.06 $\pm$ 0.0 & 12.89 $\pm$ 1.44 & 5.91 $\pm$ 0.66 & 116 \\
eLARS24 & 14429 $\pm$ 4 & 0.0481 & 0.2 $\pm$ 0.02 & 2.09 $\pm$ 0.26 & 6.27 $\pm$ 0.77 & 279 &
14429 $\pm$ 3 & 0.0481 & 0.29 $\pm$ 0.0 & 3.11 $\pm$ 0.42 & 3.89 $\pm$ 0.53 & 192 \\
eLARS25 & 13500 $\pm$ 3 & 0.045 & 0.53 $\pm$ 0.04 & 4.99 $\pm$ 0.36 & 13.35 $\pm$ 0.95 & 147 &
13498 $\pm$ 3 & 0.045 & 1.34 $\pm$ 0.0 & 12.55 $\pm$ 1.44 & 5.4 $\pm$ 0.62 & 153 \\
eLARS26 & 13767 $\pm$ 3 & 0.0459 & 0.38 $\pm$ 0.04 & 3.7 $\pm$ 0.36 & 8.58 $\pm$ 0.84 & 191 &
13767 $\pm$ 3 & 0.0459 & 0.86 $\pm$ 0.0 & 8.46 $\pm$ 1.33 & 4.08 $\pm$ 0.64 & 213 \\
eLARS27 & 13376 $\pm$ 3 & 0.0446 & 0.37 $\pm$ 0.04 & 3.39 $\pm$ 0.36 & 9.72 $\pm$ 1.04 & 185 &
13372 $\pm$ 3 & 0.0446 & 0.73 $\pm$ 0.0 & 6.76 $\pm$ 1.03 & 4.84 $\pm$ 0.73 & 141 \\
eLARS28 & 13850 $\pm$ 3 & 0.0462 & 0.19 $\pm$ 0.03 & 1.92 $\pm$ 0.26 & 4.42 $\pm$ 0.6 & 71 &
13853 $\pm$ 3 & 0.0462 & 0.88 $\pm$ 0.0 & 8.74 $\pm$ 1.51 & 2.87 $\pm$ 0.49 & 104 \\
    \hline     
Average & - & -& 0.38$\pm$0.23& 2.32$\pm$1.72 & 8.97$\pm$6.18& 215$\pm$101& - & - &1.14$\pm$1.15 & 6.58$\pm$6.18 &4.24$\pm$2.46 & 215$\pm$101\\
        \hline     
    \end{tabular}
    \label{tab:21cm_prop}
\end{table*}
\subsection{\texorpdfstring{\lya}{Lya} and galaxy properties}
We compare three \lya\ observables to \hi\ properties derived from 21cm measurements: the \lya\ luminosity, equivalent width, and escape fraction. These observables are extracted from continuum-subtracted HST photometry, values are presented in \citet{Melinder2023}, and are listed in Table \ref{tab:gal_params}. The offset of \lya\ peak emission on the red-side of the line as compared to systemic velocity, $\Delta v_{Ly\alpha}^{red}$, has been measured in the 2.5" COS aperture, centered on the brightest FUV pixel for all (e)LARS galaxies (see \citet{Rivera-Thorsen2015} for a description of the observations for LARS). The algorithm used to calculate $\Delta v_{Ly\alpha}^{red}$ is described in \citet[][]{Runnholm2021}, we list the values in Table \ref{tab:gal_params}. Note that \lya\ detection in HST imaging does not necessarily imply detection within the COS aperture, and vice-versa. We use the \lya\ equivalent width to separate between galaxies formally considered as \lya-emitters using the canonical $EW\geq20\AA$ value (see e.g. \citet{Ouchi2020}), and weak/non-emitters with $EW<20\AA$. We also retrieve redshifts computed through fitting of nebular emission lines from SDSS spectra \citep{Melinder2023}, and use the redshifts to calculate the luminosity distance to (e)LARS galaxies. 

In addition to comparing \hi\ and \lya\ observables, we compare the \hi\ properties of LARS galaxies to those of optically-selected galaxies at $z=0$ in the ALFALFA-SDSS Galaxy Catalog \citep{Durbala2020}. This catalog contains  measurements of the \hi\ content obtained with the Arecibo telescope ($\sim 3.6'$ beam) as part of the ALFALFA survey \citep{Haynes2018} for 30,000 SDSS galaxies. To enable comparisons between the two samples, we need stellar mass and star formation rate properties for the (e)LARS galaxies consistent with those in the ALFALFA-SDSS Galaxy Catalog. We use the stellar masses and SFRs derived through SED fitting to UV, optical and IR data from the GALEX-SDSS-WISE Legacy Catalog \citep[GSWLC][]{Salim2016}, values are listed in Table \ref{tab:gal_params}. Seven (e)LARS galaxies do not have available stellar masses and SFRs in GSWLC. For those objects, we estimate GSWLC values using a fit to the masses and SFRs presented in \citet{Melinder2023}, obtained from the SDSS DR8 MPA-JHU catalog \citep{Brinchmann2004,Salim2007}, hereafter referred to as SDSS measurements. For galaxies with available GSWLC measurements, the SDSS and GSWLC measurements are in good agreement, we thus fit the available SDSS measurements to predict the GSWLC masses and SFRs for objects with only SDSS measurements available (see Appendix D), and report the predicted values in Table \ref{tab:gal_params}. 

\subsection{Galaxy interaction identification}
The majority of LARS galaxies were classified as mergers in \citet{Guaita2015} and \citet{Micheva2018}, and \citet{LeReste2022} concluded that the neutral gas disturbance caused by the merger interactions in LARS08 and eLARS01 was likely facilitating \lya\ escape from the galaxies.
Upon inspection of the 21cm maps, we noticed the presence of morphological features indicative of galaxy interaction in a significant number of (e)LARS galaxies. Since neutral gas is less gravitationally bound than stars in galaxies, the imaging of the 21cm line in emission has long been recognised as a tool to help identify and characterize galaxy interactions \citep[see e.g.][]{Holwerda2011}. In order to assess the potential impact of galaxy interactions on the emission of \lya, we have identified galaxies that can be classified as gravitationally interacting using the following criteria:
 \begin{itemize}
     \item \textbf{Close pairs.} We search for close companions to the (e)LARS galaxies with available SDSS spectroscopy. We apply the close-pair criterion on projected proper distance $\Delta r$ and rest-frame velocity offset $\Delta v$ as defined in \citet{Ventou2019}:\\
     $\Delta r \leq  50\, \rm{kpc} $  \& $\Delta v \leq  300\, \rm{km}.\rm{s}^{-1}$\\
     or\\
     $50 \leq \Delta r \leq 100\,\rm{kpc}$ \& $\Delta v \leq 100\, \rm{km}.\rm{s}^{-1} $  \\
     
     \item \textbf{Morphology.} Not all galaxy mergers have companions, and some have companions too close for their spectra to be observed in SDSS due to fiber collision. This particularly impacts mergers close to nuclear coalescence, where the merger-driven increase in star formation and gas displacement is likely to be the the largest. For this reason, we also consider galaxies showing morphologies characteristic of merger interaction. We inspect RGB composites from HST imaging presented in \citet{Melinder2023} and DECaLS imaging \citep{Dey2019} shown in this paper for disturbed morphology indicative of galaxy interaction, such as tidal tails, or merging nuclei. We also inspect the \hi\ envelopes for signs of interactions, including \hi\ envelopes linking several objects or showing tidal components.

     \item \textbf{Kinematics.} Finally, we include galaxies that were classified as having \ha\ kinematics strongly deviating from a rotating disk in \citet{Herenz2024}. Most galaxies classified as interacting based on their optical and \hi\ morphology fall in this category, with the exception of eLARS13, that has disturbed \ha\ kinematics, but has no companion identified spectroscopically, and is too compact to be identified through morphological features. 
 \end{itemize}
We list the galaxies identified as interacting, and the methods used for identification for each galaxy in Table \ref{tab:mergers}, as well as the merger fraction for the different identification methods for all galaxies, and for the subsets classified as \lya-emitters and non or weak emitters. We use star-shaped markers in Figures \ref{fig:opt_nhic} and \ref{fig:mstarmhi_int} to indicate galaxies that were identified as interacting. 
\begin{table}[h!]
    \centering
    \scriptsize
        \caption{Galaxies identified as gravitationally interacting with a companion in the (e)LARS samples, using the methods indicated in the columns are shown by a black square. Galaxies that were not considered for a given method due to lack of data or detection are indicated by a dash. $^{a}$ Galaxies with a close companion identified spectroscopically in SDSS, using the definition in \citet{Ventou2019}. $^{b}$ Galaxies showing optical features in HST \citep{Melinder2023} or DECaLS \citep{Dey2019} imaging characteristic of a galaxy interaction. $^{c}$ Galaxies showing disturbed \hi\ envelopes characteristic of a galaxy interaction (e.g., Fig. \ref{fig:opt_hi_el01}) or envelopes with extensions encompassing a companion (e.g., Fig. \ref{fig:opt_hi_l02}). Note that for this method we display fractions considering only galaxies detected in \hi.$^{d}$ Galaxies with H$\alpha$ kinematics deviating strongly from a rotating disk case, identified in \citet{Herenz2024}. On the last lines, we present the total number and fraction of interacting galaxies for the different identification methods. The last two lines show the fraction of interacting galaxies for galaxies that are \lya-emitters (EW$_{Ly\alpha}\geq20\AA$), and non or weak-emitters EW$_{Ly\alpha}<20\AA$).}
    \begin{tabular}{c|ccccc}
    \hline
       ID  & Interacting & Close-pair$^a$ & morph$_{opt}^b$ &  morph$_{HI}^c$  & v$_{\rm{H}\alpha}^d$\\
    \hline
       LARS01 &$\blacksquare$ & &$\blacksquare$ & & \\ 
       LARS02 &$\blacksquare$ & & &$\blacksquare$  &$\blacksquare$\\ 
       LARS03 &$\blacksquare$ &$\blacksquare$ &$\blacksquare$ &$\blacksquare$& \\ 
       LARS04 &$\blacksquare$ &$\blacksquare$ &$\blacksquare$ & &$\blacksquare$ \\ 
       LARS05 & & & & - &\\ 
       LARS06 &$\blacksquare$ &$\blacksquare$ &$\blacksquare$ &$\blacksquare$ &$\blacksquare$ \\ 
       LARS07 &$\blacksquare$ & &$\blacksquare$ & & \\ 
       LARS08 &$\blacksquare$ & &$\blacksquare$ &$\blacksquare$ & \\ 
       LARS09 &$\blacksquare$ &$\blacksquare$ &$\blacksquare$ & &$\blacksquare$ \\ 
       LARS10 &$\blacksquare$ & & $\blacksquare$ & - & \\ 
       LARS11 &$\blacksquare$ & & $\blacksquare$ & - &\\ 
       LARS12 &$\blacksquare$ & &$\blacksquare$ & -&$\blacksquare$  \\ 
       LARS13 &$\blacksquare$ & &$\blacksquare$ & - &$\blacksquare$ \\ 
       LARS14 &$\blacksquare$ & &$\blacksquare$ &- &$\blacksquare$ \\ 
       eLARS01 &$\blacksquare$ &$\blacksquare$ & $\blacksquare$&$\blacksquare$ &$\blacksquare$ \\ 
       eLARS02 & & & & &\\ 
       eLARS03 &$\blacksquare$ &$\blacksquare$ & &$\blacksquare$& \\ 
       eLARS04 &$\blacksquare$ & & &$\blacksquare$ &\\ 
       eLARS05 &$\blacksquare$ &$\blacksquare$ & &$\blacksquare$ &\\ 
       eLARS06 &$\blacksquare$ & & &$\blacksquare$ &\\ 
       eLARS07 &$\blacksquare$ & &$\blacksquare$ &$\blacksquare$&$\blacksquare$  \\ 
       eLARS08 & & & & &\\ 
       eLARS09 &$\blacksquare$ &$\blacksquare$ & & &\\ 
       eLARS10 &$\blacksquare$ &$\blacksquare$ & &$\blacksquare$ &\\ 
       eLARS11 &$\blacksquare$ &$\blacksquare$ & & & \\ 
       eLARS12 & & & & &\\ 
       eLARS13 & $\blacksquare$& & & -&$\blacksquare$\\ 
       eLARS14 & & & & &\\ 
       eLARS15 &$\blacksquare$ &$\blacksquare$ & & &\\ 
       eLARS16 & & & & &\\ 
       eLARS17 &$\blacksquare$ &$\blacksquare$ & & &\\ 
       eLARS18 & & & & &\\ 
       eLARS19 &$\blacksquare$ &$\blacksquare$ &$\blacksquare$ &- &\\ 
       eLARS20 & & & & - &\\ 
       eLARS21 &$\blacksquare$ &$\blacksquare$ & & &\\ 
       eLARS22 &$\blacksquare$ &$\blacksquare$ &$\blacksquare$ & &$\blacksquare$\\ 
       eLARS23 &$\blacksquare$ &$\blacksquare$ & & & \\ 
       eLARS24 & & & & \\ 
       eLARS25 & & & & &\\ 
       eLARS26 &$\blacksquare$ &$\blacksquare$ & &$\blacksquare$ &\\ 
       eLARS27 & & & & & \\ 
       eLARS28 &$\blacksquare$ &$\blacksquare$ & &$\blacksquare$ & \\ 
    \hline
    N$_{interacting}$& 31 & 18 & 16 & 13 &11 \\
    f$_{interacting}$ & 74\% & 43\% & 38\% & 39\% & 26\%\\
    f$_{interacting}$ & 83\%  & 33\%  & 50\%  &63\% & 33\% \\
    LAE& & & & \\
    f$_{interacting}$ & 70\% & 47\%& 33\% & 32\% & 23\%\\
    non-LAE & & & & \\
    \hline
    \end{tabular}
    \label{tab:mergers}
\end{table}
 
\section{Results}\label{sec:results}
\subsection{21cm properties of LARS and eLARS galaxies}
\begin{figure*}[!t]
    \centering
    \includegraphics[width=0.8\textwidth]{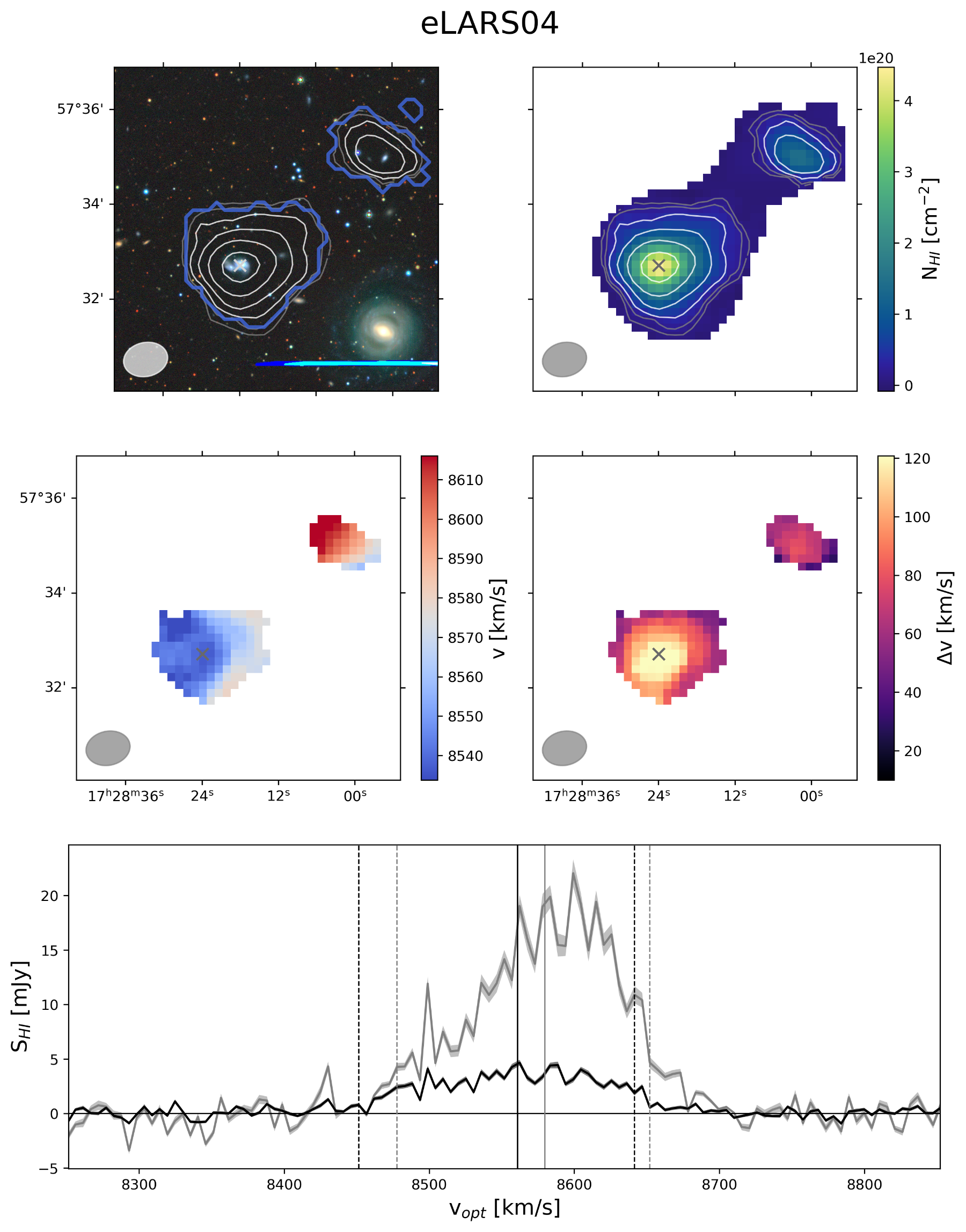}
    \caption{21cm maps and spectra of eLARS04. The field of view has been adapted to show the full extent of the main SoFIA 2 detection. \textbf{Top left:} DECaLS optical composite image with \hi\ column density contours at level $1.0\times2^n \times10^{19}\,\textrm{cm}^{-2}$, with $n=0,1,2,...,5$ overlaid. The blue solid line shows the regions with $SNR>3$. Contours fully included in the $\textrm{SNR}\geq3$ mask are shown in white, low signal-to-noise contours are shown in gray. A gray cross indicates the position of the galaxy according to optical coordinates, a synthesized beam shaped aperture centred on these coordinates was used to extract \hi\ properties in the center. \textbf{Top right:} Column density map with the same contours as on the previous panel overlaid. \textbf{Middle left:} moment-1 map. \textbf{Middle right:} linewidth map. \textbf{Bottom: } Total 21cm spectrum (gray) and beam-extracted 21cm spectrum (black). Velocity centroids are indicated by a vertical solid line, and velocity at half width on either side of the peaks by dashed lines, in either gray or black for the total or beam extraction respectively.}
    \label{fig:opt_hi_el04}
\end{figure*}
 We detect 33 of the 37 LARS and eLARS galaxies observed in 21cm with the VLA D-configuration array. Galaxies detected in 21cm in the (e)LARS samples have an average \hi\ mass $M_{HI}=6.58\times10^{9}\,M_{\odot}$, an average \hi\ column density $N_{HI}=4.24\times10^{19}\,\rm{cm}^{-2}$ and and average 21cm line width $W_{50}=215\,\rm{km}.\rm{s}^{-1}$.The galaxies that are not detected in 21cm with the VLA in D configuration are LARS05, eLARS13, eLARS19 and eLARS20, two of which are \lya-emitters ($EW\geq20\AA$). The \hi\ properties of LARS and eLARS galaxies are presented in Table \ref{tab:21cm_prop}. An overview of the \hi\ emission maps for the sample is shown on Figure \ref{fig:opt_nhic}, where fixed \hi\ column density contours are overlaid on DECaLS \citep{Dey2019} optical composite images of LARS and eLARS galaxies. To facilitate the comparison between galaxies, all panels have the same physical scale of $300 \times 300$ kpc$^2$.
We present moment maps and spectra of each galaxy on separate figures, with an example for eLARS04 shown on Figure \ref{fig:opt_hi_el04}. The moment maps and spectra for the rest of the sample are presented in the Appendix, in Figures \ref{fig:opt_hi_l01} to \ref{fig:opt_hi_el28}. In these figures, the field of view has been tailored to show the full extent of the main SoFIA 2 detection. We also indicate by a cross the coordinates of the galaxies, which are the locations around which the beam aperture was centered to extract the \hi\ properties on smaller scales.

\begin{figure*}[h]
    \centering
    \includegraphics[width=0.95\textwidth]{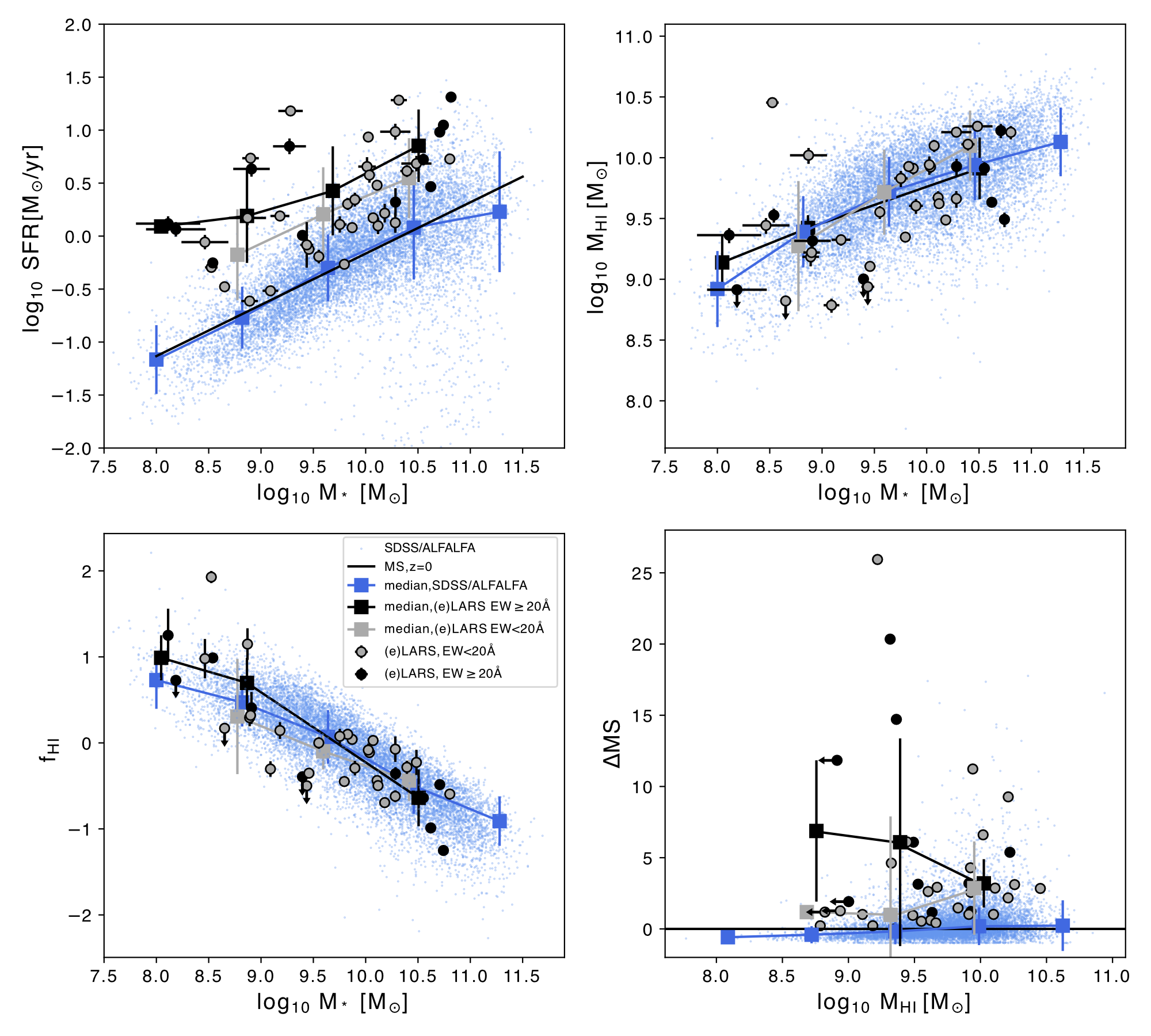}
    \caption{\hi\ properties and galaxy properties for the (e)LARS sample and optically-selected galaxies from SDSS-ALFALFA \citep{Durbala2020}. In all panels, we split the (e)LARS sample between objects considered \lya\ emitters ($EW_{Ly\alpha}\geq20\AA$, in black), and weak or non-emitters ($EW_{Ly\alpha}<20\AA$, in gray). SDSS/ALFALFA galaxies are shown in blue. Squares indicate the median in stellar mass or \hi\ mass bins, blue squares show the median for the SDSS/ALFALFA galaxies, black squares the median for the (e)LARS \lya-emitters and gray squares the median for the weak and non \lya-emitters. The error bars associated with the median points show the standard deviation within a given mass bin. In the top left panel, the solid black line shows the star forming main sequence at $z=0$ \citep{Speagle2014}. }
    \label{fig:hi_prop}
\end{figure*}

We compare the total \hi\ properties in the (e)LARS samples to those of optically-selected galaxies in the local universe from the ALFALFA-SDSS Galaxy Catalog that were observed with the Arecibo telescope ($\sim 3.6'$ beam, corresponding to physical scales between 80 and 200kpc for (e)LARS galaxies observed in \hi) in Figure \ref{fig:hi_prop}.
The position of the galaxies as compared to ALFALFA-SDSS galaxies, and to the $z=0$ star forming main sequence \citep{Speagle2014} is presented in the star formation rate to stellar mass diagram in the top left panel. (e)LARS galaxies were selected based on requirements on their \ha\ equivalent width and FUV magnitude, for this reason they have higher SFRs for a given stellar mass than the bulk of $z=0$ galaxies. \lya-emitters tend to have larger SFRs at a given stellar mass than weak and non-emitters, although there is ample overlap in the SFR-M$_*$ parameter space between \lya-emitters and non or weak emitters. The top right panel shows the \hi\ mass as a function of stellar mass: (e)LARS galaxies are distributed around the median relation for ALFALFA-SDSS galaxies. One object, LARS06 has an extremely large \hi\ mass for its stellar mass and is offset from the rest of the sample and from the ALFALFA-SDSS galaxies \citep[see also][]{Pardy2014}. As can be seen in Figure \ref{fig:opt_hi_l06}, LARS06 is a dwarf galaxy being accreted into the \hi\ envelope of the massive spiral galaxy UGC 10028, explaining the offset in \hi\ properties. Looking at the median \hi\ mass in stellar mass bins for \lya\ emitters, they have similar \hi\ masses for a given stellar mass as ALFALFA-SDSS galaxies. Weak and non \lya-emitters have a slightly higher \hi\ mass per stellar mass above $M_*\geq10^{9.5}\rm{M}_{\odot}$, but the median value in all stellar mass bins with sufficient data is in agreement with that of the ALFALFA-SDSS galaxies, and that of the \lya\ emitters. This indicates that \lya-emitters have a similar \hi\ mass for a given stellar mass as non-emitters, and as the bulk of optically-selected galaxies in the local Universe. A similar picture emerges from the lower left panel presenting the \hi\ fraction $f_{HI}=M_{HI}/M_{*}$ as a function of stellar mass.

Finally, the bottom right panel shows the offset from the star-forming main sequence $\Delta MS=\frac{\rm{SFR}_{obs}-\rm{SFR}_{MS}(M_{*,obs})}{\rm{SFR}_{MS}(M_{*,obs})}$ as a function of \hi\ mass, using the star-forming main sequence relation presented in \citep{Speagle2014} at $z=0$ to derive the offset. As can be expected from the position of the (e)LARS galaxies in the SFR-M$_*$ and M$_{HI}$-M$_*$ diagrams, (e)LARS galaxies have a larger offset from the star forming main sequence for a given \hi\ mass than the bulk of SDSS-ALFALFA galaxies. While the medians for \lya-emitters and weak or non-emitters are close, the \lya-emitters tend to have a slightly higher offset from the main sequence at a given \hi\ mass. This suggests that at a fixed stellar and \hi\ mass, \lya-emitters tend to have a larger SFR than weak or non-emitters, and than the bulk of optically selected galaxies at $z=0$. Nevertheless, the $\Delta MS-M_{HI}$ parameter space sampled by \lya-emitters and weak or non-emitters overlap.

To determine whether the high \lya\ EW sample and the weak/non-emitters are statistically different in terms of any of the properties presented in Figure \ref{fig:hi_prop}, we perform sets of Kolmogorov-Smirnov (K-S) tests. We also compare the properties of \lya-emitters to those of galaxies in the SDSS-ALFALFA sample to assess whether \lya-emitters are statistically different from the general $z=0$ population. We use the K-S test implementation \texttt{kstest} in the Python module \texttt{scipy} and require a p-value $p<0.003$ to reject the null-hypothesis that two samples are drawn from the same distribution. Results of the K-S tests can be found in Table \ref{tab:kstest}. We find that \lya-emitters and weak/non-emitters are statistically consistent in terms of all properties tested, with K-S test p-values ranging from 0.08 to 0.72. These properties include the stellar mass, star formation rate, offset to the main sequence, \hi\ mass, and \hi\ fraction. Additionally, we find that \lya-emitters have properties that are consistent with the general $z=0$ population, specifically their stellar mass, \hi\ mass, and \hi\ fraction ($p\geq0.16$). \lya-emitters in (e)LARS samples distinguish themselves from the rest of the $z=0$ galaxy population in terms of their SFR ($p=$1.4e-4) and their offset to the star-forming main sequence ($p=$8.6e-11). However, running K-S tests on these properties between  weak/non-emitters and the SDSS-ALFALFA sample leads to similar results, with $p=$1.6e-5 for the SFR, and $p=$3.5e-12 for the offset to the main sequence, indicating the difference likely stems from sample selection, rather than from star formation-related properties setting \lya-emitters apart.
\begin{table}[h]
    \centering
    \scriptsize
        \caption{Results of the K-S tests run to assess if different samples considered in this manuscript are statistically different in terms of their observables. We require a p-value $<0.003$ to reject the null-hypothesis that the two samples tested are drawn from the same distribution.}
    \begin{tabular}{llll}
    \hline
    Sample 1 & Sample 2 & Property & p-value\\
    \hline
     (e)LARS, \hi\ obs & (e)LARS, \hi\ obs  & M$_*[\rm{M}_{\odot}]$& 0.13\\
     $EW\geq20\AA$& $EW<20\AA$ & SFR$[\rm{M}_{\odot}/yr]$& 0.51\\
                   &            & $\Delta$MS & 0.08\\
                  &            &M$_{HI,t}[\rm{M}_{\odot}]$  & 0.72\\
                 &            &f$_{HI,t}$  & 0.29\\
                &            &M$_{HI,b}[\rm{M}_{\odot}]$  & 0.12\\
                 &            &f$_{HI,b}$  & 0.14\\
    \hline
     (e)LARS, \hi\ obs & SDSS-ALFALFA  & M$_*[\rm{M}_{\odot}]$& 0.42\\
     $EW\geq20\AA$&  & SFR$[\rm{M}_{\odot}/yr]$& 1.4e-4\\
                     &            & $\Delta$MS & 8.6e-11\\
                  &            &M$_{HI,t}[\rm{M}_{\odot}]$  & 0.16\\
                 &            &f$_{HI,t}$  & 0.45\\
     \hline
        Interacting&    Non-interacting   & EW [\AA] & 0.62 \\
        &            &  L$_{Ly\alpha}$[erg/s]& 0.28\\
        &            & $f_{esc}$ & 0.60\\
        &            & SFR & 0.11\\
        &            & $\Delta$MS & 0.03\\
        &            & M$_{HI,t}$ & 0.11\\
        &            & f$_{HI,t}$ & 0.11\\
        &            & M$_{HI,b}$ & 0.46\\
        &            & f$_{HI,b}$ & 0.18\\
     \hline
    Interacting&    Non-interacting & EW [\AA] & 0.25\\
     morph$_{HI}$  &  morph$_{HI}$  &   L$_{Ly\alpha}$[erg/s]& 0.23 \\
        &          & $f_{esc}$ & 0.67\\
        &            & $\Delta$MS & 0.18\\
        &            & SFR & 0.64\\
        &            & M$_{HI,t}$ & 0.06\\
        &            & f$_{HI,t}$ & 0.69\\
        &            & M$_{HI,b}$ & 0.77\\
        &            & f$_{HI,b}$ & 0.10\\
     \hline
    \end{tabular}
    \label{tab:kstest}
\end{table}

Interferometers typically underestimate the global \hi\ content of galaxies due to the lack of short-spacing data in the \textit{uv}-plane. Our galaxies are at an average redshift of $\sim 0.045$, making their \hi\ angular extent compact ($\lesssim4'$) and alleviating the short-spacing issue as compared to very nearby galaxies. Nevertheless, we want to verify whether the VLA measurements are missing a fraction of the \hi\ mass in the form of extended emission. To do so, we compare our \hi\ masses with the 21cm LARS measurement from \citet{Pardy2014} using the Robert C. Byrd Green Bank Telescope (GBT). Our measurements either agree with, or are higher than the GBT values. We also measure higher \hi\ fluxes and masses compared to the values derived for galaxies that had available VLA observations in \citet{Pardy2014}. The difference in estimated mass values is likely due to a combination of several factors, including difference in emission detection (SoFIA 2 using S+C algorithm with large spatial and spectral kernels VS visual detection favoring only high surface brightness emission), \hi\ flux estimation methods (summation in the detection mask VS profile fitting), and zero-flux offset. In the Appendix, we present a comparison of our mass estimations for 3 galaxies to those obtained from available imaging in the APERTIF survey \citep{Adams2022}, a wide-field \hi\ survey targeting the Northern-hemisphere sky. Our estimations are in agreement with the APERTIF values within 2.5$\sigma$ for all objects, we thus consider our \hi\ estimates robust. One LARS galaxy, namely LARS08, has available \hi\ measurements in the ALFALFA-SDSS sample. The \hi\ mass measured in ALFALFA for LARS08 is $M_{HI,ALFALFA}=2.04\pm0.25\times 10^{10}\, M_{\odot}$, in agreement within error bars with the value $M_{HI,t}=1.62\pm0.20\times 10^{10}\, M_{\odot}$ we measure. We conclude that the fraction of extended flux missed by the interferometer is likely negligible. 

\subsection{Comparing 21cm and \texorpdfstring{\lya}{Lya} emission}
\begin{figure*}[!t]
    \centering
     \includegraphics[width=0.9\textwidth]{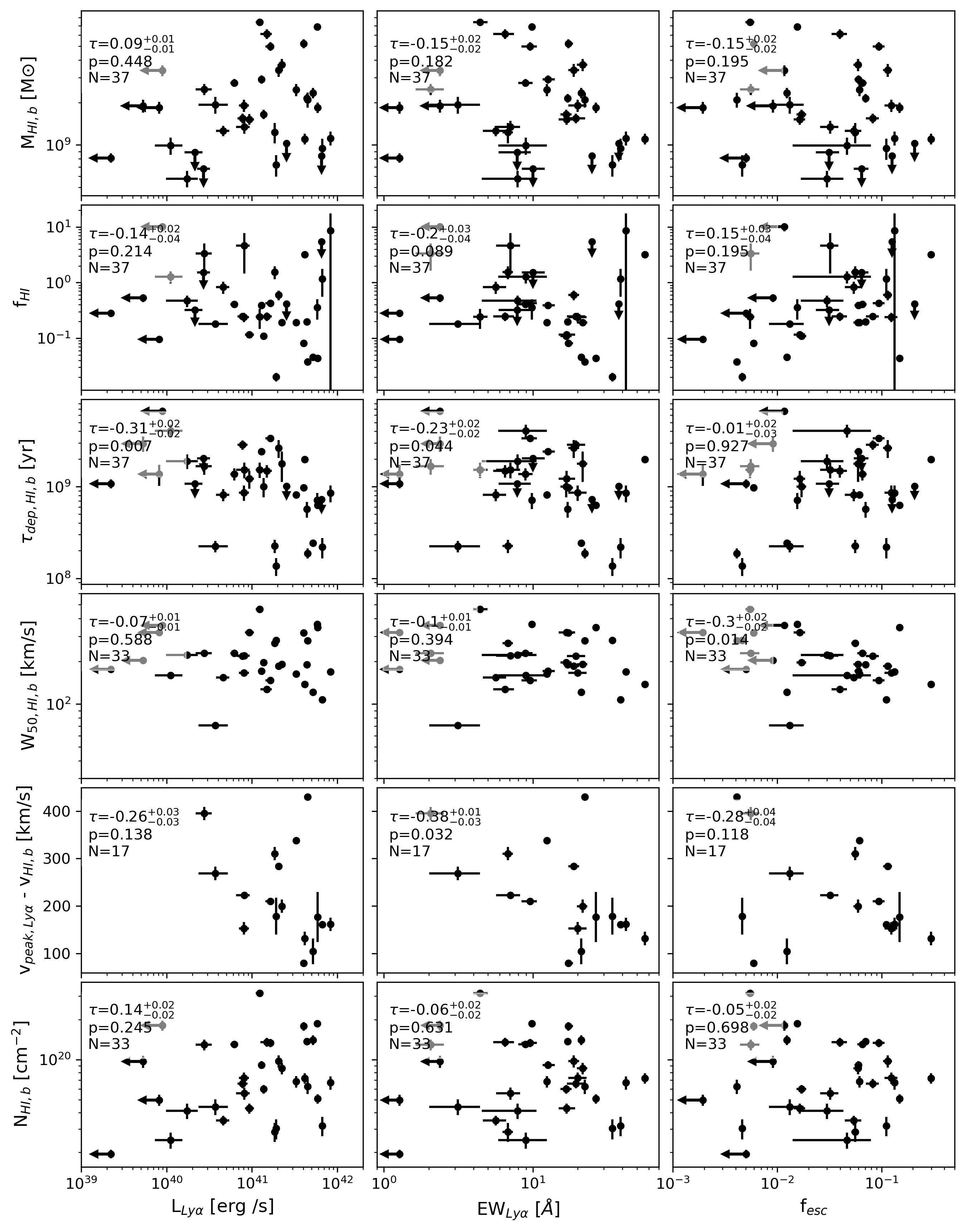}
    \caption{\hi\ properties extracted in the synthesized beam-shaped aperture as a function of \lya\ observables. From left to right, the x-axis shows the \lya\ luminosity, \lya\ equivalent width and \lya\ escape fraction. From top to bottom, the y-axis shows the \hi\ mass, the \hi\ gas to stellar mass fraction, the \hi\ mass per unit star formation rate, the \hi\ linewidth and the offset between \lya\ red peak and \hi\ centroid velocity. Generalized Kendall $\tau$, p-values and sample sizes associated with each scatter plot are shown on the top left corner of each panel.}
    \label{fig:Lya_hi_props}
\end{figure*}

We have established that galaxies in the (e)LARS samples have neutral gas properties similar to those of optically-selected galaxies in the local Universe, and that \lya-emitters cover a parameter space similar to that of weak and non-emitters. We now look at the impact of 21cm properties extracted in the beam on \lya\ observables, and in particular on the \lya\ luminosity, EW and escape fraction. In Figure \ref{fig:Lya_hi_props}, we present scatter plots of various \hi\ properties characterizing the content and kinematics of neutral gas in the (e)LARS galaxies as a function of \lya\ observables. In particular, we compare \lya\ properties to the \hi\ mass, \hi\ fraction, \hi\ depletion time $\tau_{dep,HI}=M_{HI}/SFR$, the \hi\ line width W$_{50}$, the offset between \lya\ red peak velocity and \hi\ velocity centroid, and the average \hi\ column density. For each scatter plot, we show on the top left corner the Kendall $\tau$ parameter, p-value and sample size that characterize the degree of correlation between \lya\ and \hi\ variables. 
The Kendall $\tau$ and p-value presented here are calculated following the methodology developed in \citet{Isobe1986}, that takes into account variable sets with upper limits. Additionally, errors on the Kendall $\tau$ parameter due to the uncertainties on variables are calculated using a Monte-Carlo framework similar to that in \citet{Curran2014}. The code\footnote{Publicly available on \url{https://github.com/Knusper/kendall}.} used for the calculation of the generalised Kendall $\tau$ parameter was first described and applied in \citet{Herenz2024} to assess the degree of correlation between ionized gas and \lya\ observables for LARS and eLARS galaxies. To estimate the errors associated with the $\tau$ parameter values, we run $10^4$ Monte-Carlo simulations for each sets of variables presented in Figure \ref{fig:Lya_hi_props}. 
As in \citet{Melinder2023}, we consider a correlation/anti-correlation is significant if $p\leq0.003$. Given $p\leq0.003$ and the sample sizes $N=\{37,33,17\}$ considered here, following the approach in \citet{Herenz2024} yields a requirement on $|\tau|>\{0.34,0.36,0.53\}$ respectively for the different samples considered. Specifically, if $p>0.003$, we cannot confidently reject the null hypothesis, which states that the two variables being tested are not associated. If $p>0.003$ and $|\tau|$ is above any of the specific thresholds for a given sample size, there is a non-negligible probability that the strong correlation between two variables is due to chance. Instead if $p\leq0.003$ and $|\tau|$ is below a given threshold, we can confidently state that the correlation between the two variables tested is extremely weak.
As can be seen from the p-values shown in Figure \ref{fig:Lya_hi_props}, with $p>0.007$, none of the \hi\ properties we measure is robustly correlated with any of the \lya\ observables presented here.

 We have further tried adding a third variable by color-coding the plots, such as the dust extinction E(B-V), which is anti-correlated with the \lya\ escape fraction \citep{Melinder2023} or the gas-phase metallicity, but no clear trends emerge from such scatter plots either. The global \lya\ emission observables thus appear mostly independent of the properties of the \hi\ gas reservoir in a galaxy.

\subsection{The impact of mergers on \hi\ and \lya\ emission }
 A large fraction of (e)LARS galaxies show disturbed \hi\ envelopes, sometimes encompassing one or several companion objects. Several 21cm ancillary detections were made in the LARS and eLARS fields, some of them in close spatial and spectral proximity to galaxies in our sample (see e.g. the object South East of eLARS05 and 10 on Fig. \ref{fig:opt_nhic}). A few galaxies are interacting with massive, HI-rich spirals in their vicinity. This is the case for  LARS06 in particular, which is found a few kpc away from UGC 10028 and is embedded within the neutral gas envelope of this galaxy (see Fig.~\ref{fig:opt_hi_l06}). While the 21cm measurements of objects with HI-rich companions do not characterize the neutral gas content of the targets only, they still probe the neutral gas environment that is relevant to \lya\ escape. Therefore, we have chosen to keep these objects in the study. Note however that since LARS06 is a \lya\ absorber, it is not included in correlation analysis on the impact of \hi\ properties on \lya\ emission. 

 In Table \ref{tab:mergers} we have identified galaxies considered as gravitationally interacting via spectroscopic close-pair search and visual identification using their optical and \hi\ maps.
 We find that 74\% of galaxies in the (e)LARS sample are gravitationally interacting with companion galaxies. This fraction slightly higher for \lya-emitters (83\%) than for  weak or non-emitters (70\%). Notably, a larger fraction of \lya-emitters show disturbed morphologies due to the interaction in the optical (50\% for \lya-emitters, 33\% for weak or non-emitters). The fraction of galaxies identified as interacting through their \hi\ morphology is much larger  for \lya-emitters (63\%) as compared to weak/non-emitters (32\%). This indicates that the merger-driven disruption of a galaxy's neutral gas envelope could increase the  probability of a galaxy emitting strongly in \lya.

\begin{figure}
    \centering
    \includegraphics[width=\linewidth]{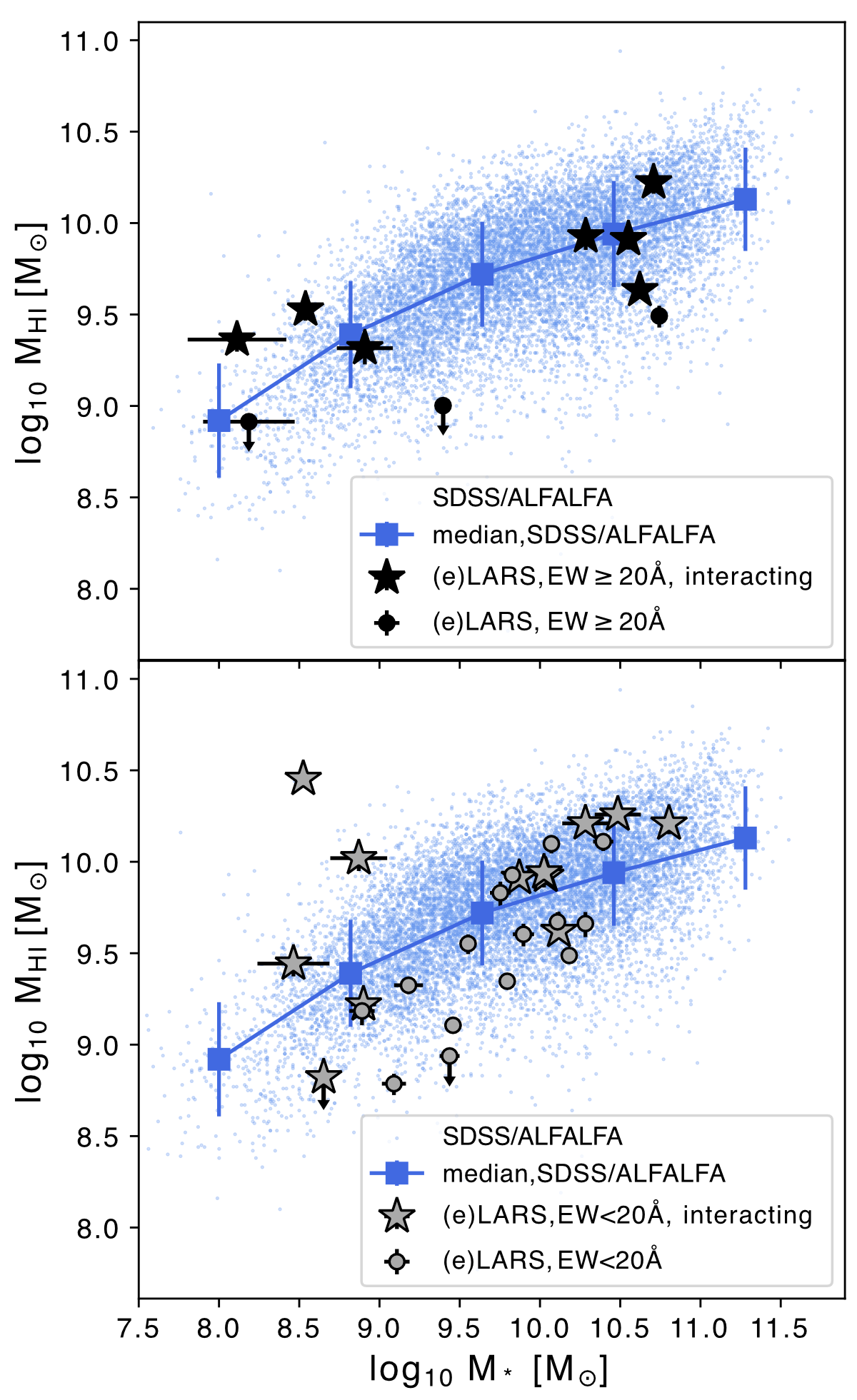}
    \caption{\hi\ mass as a function of the stellar mass for SDSS/ALFALFA galaxies (in blue) and (e)LARS galaxies for objects classified as interacting and non-interacting. We show the position of \lya-emitters with $EW_{Ly\alpha}>20\AA$ on the top panel and weak or non-emitters on the bottom panel. The square blue markers show the median in bins of stellar mass for SDSS/ALFALFA galaxies. Circles indicate galaxies not identified as undergoing gravitational interaction, while stars show the objects classified as interacting. Only galaxies observed in 21cm are shown here.}
    \label{fig:mstarmhi_int}
\end{figure}

To further explore the impact of mergers on the \hi\ and \lya\ properties of (e)LARS galaxies, we show in Figure \ref{fig:mstarmhi_int} the galaxies identified as interacting via their morphology for \lya-emitters and weak or non-emitters on the $M_{HI}-M_{*}$ diagram. This Figure is similar to that presented in the top right panel of Figure \ref{fig:hi_prop}, but highlights the position of \lya-emitting mergers on the parameter space. The top panel of Figure \ref{fig:mstarmhi_int} shows that interacting \lya-emitters are found across the \hi\ mass parameter space of the $z=0$ populations, with several interacting \lya-emitters having a larger \hi\ mass than most of the $z=0$ at a similar stellar mass. Interestingly, non-interacting \lya-emitters have either significantly lower \hi\ masses for their stellar mass as compared to the bulk of $z=0$ galaxies, or are non-detections in \hi. This could suggest two possible modes for \lya-emission: one where \lya\ emission occurs in high \hi\ mass galaxies, with mergers facilitating the emission, and another in the low \hi-mass regime, where \lya\ can escape without the need for merger interaction. Looking at the bottom panel of Figure \ref{fig:mstarmhi_int}, one can observe that mergers and low \hi-mass galaxies are not always strong \lya-emitters. We conduct K-S tests to determine if galaxy interaction significantly impacts the \lya\ properties of galaxies in (e)LARS samples. The p-values for the different properties tested in the interacting and non-interacting (e)LARS samples are presented in Table \ref{tab:kstest}. We find that galaxies in (e)LARS identified as interacting have \lya, galaxy, and \hi\ properties that are statistically consistent with those of non-interacting galaxies ($p\geq0.03$, when $p<0.003$ is required for the two samples to be considered as statistically different). This also applies to galaxies identified only through \hi\ morphologies characteristic of merger-interaction, that have properties consistent with those of galaxies not identified as mergers through this method ($p\geq0.06$ for all properties tested).
From these results, we hypothesize that certain processes or properties, such as an ongoing merger interaction or a low-HI mass, might facilitate the emission of \lya\ from galaxies. Nevertheless, the geometry of the \hi\ reservoir and the small-scale \hi\ distribution around a star-forming region are crucial to the emission of \lya\ photons. The observation of \lya-emission from a system thus likely depends on the line-of-sight observed, and additionally for mergers, on the mass ratio of the two galaxies and timescale of the merger interaction.

\section{Discussion}\label{sec:discussion}
\subsection{The impact of the global neutral gas content on \texorpdfstring{\lya}{Lya} emission}
We have presented the \hi\ properties of local \lya-emitting galaxies and assessed whether \hi\ properties impact the \lya\ output of galaxies. We have found no significant correlation between global \hi\ properties and \lya\ luminosity, equivalent width, and escape fraction. Additionally, the global \hi\ properties of \lya-emitting galaxies ($EW\geq20\AA$) are similar to those of optically-selected $z=0$ galaxies, and \lya-emitters occupy a very similar parameter space in terms of \hi\ content as weak and non \lya-emitters. From these results, we conclude that \lya-emitting galaxies do not constitute a separate galaxy population as per their neutral gas properties.
We note that the \hi\ properties presented here are calculated in large apertures ($\sim$40kpc) as compared to \lya\ properties measured in HST imaging. This suggests that if there is any dependence between \lya\ and \hi\ properties, it is found at the local level, on similar physical scales as those characterizing \lya\ emission. The resolved neutral gas maps presented in \citet{LeReste2022} for LARS08 and eLARS01 also support this argument: eLARS01 for example has the second largest \hi\ mass of the (e)LARS sample ($M_{HI}=1.67\pm0.22\,10^{10}\,\rm{M_{\odot}}$), yet it is a strong \lya\ emitter, with $EW=21.2\AA$. This can be explained by the geometry of the \hi\ around the \lya-emitting regions, with most of the \hi\ mass being offset from the regions where \lya\ emission has the highest luminosity, equivalent width, and escape fraction. Statistical measures of the \hi\ content and geometry of \lya-emitters on scales below 30kpc will be important to firmly establishing the role neutral gas properties play into the emission of \lya\ from galaxies.

\subsection{Large scale environment/Mergers} 
    Numerous galaxies in the original LARS sample were known to be mergers due to clear signs of interaction in the optical \citep{Guaita2015, Micheva2018}. However, the eLARS sample has been selected with lower specific star formation rate requirements, and with the aim to cover a wider range in FUV luminosities and morphologies. This resulted in an increase of apparently normal star-forming disk galaxies compared to the original sample. Despite the difference in sample selection, a large fraction of the eLARS galaxies also have neutral gas distributions characteristic of interacting galaxies. This includes galaxies showing very regular morphologies in the optical, such as eLARS05 and eLARS10 which appear to share a common \hi\ envelope. Out of all (e)LARS galaxies observed, 74\% display clear evidence of ongoing galaxy interaction in their optical or \hi\ morphology, or via the presence of a close companion confirmed spectroscopically. This is not completely surprising, as galaxies above the main sequence in the local universe are known to show signs of dynamical disturbances linked to interaction, with systems ranging from major mergers to galaxies with a close companion. Galaxy interactions typically cause an increase in availability of gas and in the efficiency at which neutral gas is converted to molecular gas, and ultimately, stars \citep{Solomon1988}.

    Interestingly, the subsample of galaxies formally considered as \lya-emitters through their high equivalent widths have a higher interacting fraction (83\%) than weak or non-emitters (70\%).
    When looking more closely at the method used to characterize interacting galaxies, we find that a much larger fraction of \lya-emitters present \hi\ morphological features characteristic of interactions (63\%, as compared to  32\% for weak and non-emitters), and to a lesser extent, optical merger morphology (50\% for \lya-emitters, and 33\% for weak and non-emitters). Similarly, a larger fraction of \lya-emitters show signs of interaction through disturbed \ha\ kinematics (33\%) than the weak and non-emitters (23\%).
    Additionally, \lya-emitters identified as interacting were found to be spread across the \hi\ parameter space, and some are found above the median $M_{HI}-M_{*}$ relation at $z=0$, while non-interacting \lya-emitters have significantly lower \hi\ mass than $z=0$ galaxies.  Together, these results suggest that merger interactions could play a role in the emission of \lya\ photons from star-forming galaxies through disruption of the neutral gas envelope. This process could be more important for \lya\ emission in objects with high \hi\ gas fractions. In low \hi\ mass galaxies however, while interaction might help \lya\ emission production and escape, other mechanisms such as stellar feedback could be sufficient to produce the conditions required for the emission and escape of \lya\ radiation. However, a merger interaction or low \hi\ mass does not necessarily guarantee the observability of \lya\ with large EW, as a large fraction of interacting and low \hi\ mass galaxies are found amongst the weak and non \lya-emitters. This might be due to the angle at which we are seeing a galaxy, which is especially relevant for mergers as interactions typically result in anisotropic and asymmetric neutral gas geometry.  Therefore, while galaxy interactions might increase the production of \lya\ in galaxies through enhanced star formation, the escape of this radiation does not necessarily take place on our line of sight. 
    Alternatively, \lya\ emission might depend on the timescale of the interaction. Peak starburst activity typically occurs at nuclear coalescence, in the later stages of mergers \citep{Georgakakis2000}. Therefore, late-stage mergers would likely have turned a larger fraction of their gas into new stars, increasing both intrinsic production of \lya\ and  \lya\ escape by lowering the neutral gas fraction. The fact that a larger fraction of \lya-emitters is identified as interacting via optical and \hi\ morphological features, as opposed to close pairs for weak and non \lya-emitters, supports that argument.

    Our study is not the first to point out the role of galaxy interaction on \lya\ emission. The role of mergers as a general mechanism for \lya\ emission from galaxies has been explicitly suggested both in the low and high-redshift universe \citep[e.g.][]{Hayes2005,Cooke2010}. In \citet{Cooke2010}, a relation was found between \lya\ emission and galaxy pair separation in galaxies selected using the Lyman Break technique at $z>3$. However, a blind spectroscopic study of 28 Lyman Alpha Emitters in the HETDEX pilot survey could not replicate this result \citep{Hagen2016}. This was interpreted as a result from selection effects in \citet{Cooke2010} when considering Lyman Break Galaxies, which are biased towards massive systems, and could require interaction for \lya\ emission. Observations of \lya\ in ultra Luminous Infrared Galaxies (uLIRGs) has also shown a large detection rate of \lya\ emission originating, suprisingly, from extremely massive dusty mergers \citep{Martin2015}. Several 21cm observations of other \lya-emitting galaxies have shown the neutral gas of these objects to be strongly disturbed by merger interactions \citep{Cannon2004,Purkayastha2022,LeReste2023,Dutta2024}. Mergers often lead to higher star formation rate density and are thus linked to \lya\ emission through increase in intrinsic UV production, and stronger feedback. However, the gas could be playing another role than simply feeding star formation in an efficient manner. The offset or perturbed neutral gas geometries caused by mergers have been suggested as key in the escape of \lya\ and even Lyman Continuum emission by some 21cm studies \citep{LeReste2022,Purkayastha2022,LeReste2023}. Finally, recent James Webb Space Telescope observations combined with hydrodynamical simulations have suggested galaxy mergers could also facilitate \lya\ emission from galaxies into the Epoch of Reionization \citep{Witten2024}.
    Whether mergers play a role through their impact on the interstellar medium geometry, kinematics, or simply by increasing the star formation rate, our study indicates that they could be an important process for \lya\ production and escape in the local universe. Detailed analysis of the impact of mergers on neutral gas geometry and \lya\ will be presented in subsequent work using higher angular resolution data (VLA C-array).

\subsection{Caveats}
    One of the major caveats to our study is the mismatch between the scales on which \hi\ and \lya\ are observed, despite the use of a synthesized beam-shaped aperture to reduce the scale discrepancy. Indeed, 20 of the 45 galaxies studied in \citet{Melinder2023} show \lya\ emission at the edge of the detector, thus several of the \lya\ quantities we use here are actually lower limits or estimates to the global \lya\ properties. The larger scales probed by our 21cm observations are likely to overestimate the \hi\ properties corresponding to a given set of \lya\ properties estimated from photometry. Since the difference in angular scales probed by the VLA varies from galaxy to galaxy and the \hi\ distribution is not uniform, interpolating the \hi\ quantities observed to the scales on which \lya\ properties as measured is not possible. However, this aperture difference certainly increases the scatter when comparing \lya\ and HI. Comparative studies of \lya\ and 21cm  emission on smaller physical scales would alleviate the issue.

\section{Conclusion}\label{sec:conclusion}
We have presented the 21cm \hi\ properties of \lya-emitting galaxies in the Lyman Alpha Reference Samples. We detect 21cm emission from 33 of the 37 galaxies observed with the VLA in D-configuration ($\sim$55" synthesized beam size). A majority of (e)LARS galaxies show disturbed neutral gas and optical morphologies due to galaxy interaction, or have spectroscopically-confirmed companions in close proximity (74\%). This is likely due to the sample selection scheme, and in particular the \ha\ EW and FUV luminosity requirements that select galaxies above the $z=0$ star-forming main-sequence, more likely to be involved in minor and major mergers. 

We compare the 21cm neutral gas properties of \lya-emitters ($\rm{EW}\geq20 \AA$) to those of weak and non \lya-emitters in the (e)LARS samples, and to those of optically-selected $z=0$ galaxies. We find that (e)LARS galaxies have similar \hi\ gas masses and gas fraction for a given stellar mass as the bulk of optically selected galaxies at $z=0$, regardless of their \lya\ EW. \lya-emitters with high EW tend to have a larger offset to the star-forming main sequence at a given \hi\ mass than weak and non-emitters, but the samples overlap and cannot be distinguished statistically.

We compare the fraction of interacting galaxies to their \lya\ emission class as defined by their equivalent width. A majority of (e)LARS galaxies are gravitationally interacting with a companion (74\%), and this interacting fraction is slightly higher for objects with strong \lya\ emission (83\% for galaxies with $\rm{EW}\geq20 \AA$) as compared to those with weak or no \lya\ emission (70\% for objects with $\rm{EW}<20 \AA$). A larger fraction of \lya-emitters show \hi\ morphologies (63\% vs 32\%), optical morphologies (50\% vs 33\%), and disturbed \ha\ kinematics (33\% vs 23\%) characteristic of galaxy interaction than weak and non \lya-emitters. Additionally, interacting \lya-emitters present a large range of \hi\ gas fractions, while non-interacting \lya-emitters all have \hi\ gas fraction significantly lower than the bulk of $z=0$ galaxies. While interacting and non-interacting subsamples significantly overlap in \lya\ and \hi\ properties, and cannot be differentiated statistically, our results indicate two possible modes of emission for \lya\ depending on the \hi\ mass of the host galaxy. One of the modes would require mergers to sufficiently disturb neutral gas in the ISM and CGM to enable \lya\ emission, while in the other, emission could happen in normal star-forming galaxies provided their \hi\ mass is low. However, the viewing angle of a system, and additionally for mergers, the mass ratio or stage of interaction at which the galaxies are observed ultimately likely plays a determining role on \lya\ observables.

Finally, global properties derived with the 21cm line show no statistically robust correlation with \lya\ observables (L$_{Ly\alpha}$, EW or $f_{esc}$), whether they characterize \hi\ content or kinematics. 
We note that the aperture used to extract \lya\ information is significantly smaller than the minimum resolution element in 21cm to which they are compared. Nevertheless, this indicates that if neutral gas plays a role in \lya\ transfer, it likely does on smaller scales than those probed here (<30 kpc). Additional VLA data obtained with the C-array configuration will be used to establish the relation between \lya\  and \hi\ observables on scales of $\sim$10kpc in a future publication. These data will allow investigations of the impact of neutral gas content and geometry on \lya\ emission on matched scales that will help determine if neutral \hi\ gas significantly impacts \lya\ emission from galaxies.

\bibliographystyle{aa}
\bibliography{bibli.bib}
\section*{Acknowledgements}
ALR thanks Stephen Pardy, Brian Eisner, Kathleen Fitzgibbon and Bridget Reilly for sharing their VLA data reduction manuscripts for the LARS and eLARS projects, part of which were used to reduce the data presented here; Hayley Roberts and Kameswara Bharadwaj Mantha for helpful discussions on galaxy merger identification; and Kelley Hess for providing preliminary APERTIF data of (e)LARS galaxies for verification of VLA measurements.
\\
The National Radio Astronomy Observatory is a facility of the National Science Foundation operated under cooperative agreement by Associated Universities, Inc.\\
This research is based on observations made with the
NASA/ESA Hubble Space Telescope obtained from the
Space Telescope Science Institute, which is operated by
the Association of Universities for Research in Astron-
omy, Inc., under NASA contract NAS 5–26555. These
observations are associated with programs 12310, 13483,
13027, 14923, and 13656.\\
The Legacy Surveys consist of three individual and complementary projects: the Dark Energy Camera Legacy Survey (DECaLS; Proposal ID \#2014B-0404; PIs: David Schlegel and Arjun Dey), the Beijing-Arizona Sky Survey (BASS; NOAO Prop. ID \#2015A-0801; PIs: Zhou Xu and Xiaohui Fan), and the Mayall z-band Legacy Survey (MzLS; Prop. ID \#2016A-0453; PI: Arjun Dey). DECaLS, BASS and MzLS together include data obtained, respectively, at the Blanco telescope, Cerro Tololo Inter-American Observatory, NSF’s NOIRLab; the Bok telescope, Steward Observatory, University of Arizona; and the Mayall telescope, Kitt Peak National Observatory, NOIRLab. Pipeline processing and analyses of the data were supported by NOIRLab and the Lawrence Berkeley National Laboratory (LBNL). The Legacy Surveys project is honored to be permitted to conduct astronomical research on Iolkam Du’ag (Kitt Peak), a mountain with particular significance to the Tohono O’odham Nation. The full acknowledgment for the legacy surveys can be found at \url{https://www.legacysurvey.org/acknowledgment/}. 
A.L.R. was supported by Stockholm University.
M.J.H. is supported by the Swedish Research Council (Vetenskapsr\aa{}det) and is Fellow of the Knut \& Alice Wallenberg Foundation. J.M. is funded by the Swedish National Space Agency (SNSA, grant 2021-00083). G.O. acknowledges support from Vetenskapsrådet (VR) and the Swedish National Space Agency (SNSA).
D.K. is supported by the Centre National d’Etudes Spatiales (CNES)/Centre National de la Recherche Scientifique (CNRS); convention no 230400. 
\section*{Appendix}
\subsection*{Appendix A: VLA observations properties}
\begin{table*}[h]
    \centering
    \scriptsize
        \caption{VLA observation and clean image properties of the LARS and eLARS galaxies. We show the average and standard deviation for the sample on the last row, where applicable.}
    \label{tab:vla_obs}
    \begin{tabular}{ccccccccccc}
         \hline
          ID & Int. time & flagged & BP cal & Phase calibrator & b$_{\text{min}}$ & b$_{\text{maj}}$& p.a. & $\overline{\rm{beam\,size}}$ &rms\\
             & h                & \%      &                     &                  & "   & " & $^{\circ}$ &kpc & mJy\\
         \hline
  LARS01 &          1.65 &    18.70 &  3C286 &  J1219+4829 &  44.892 &  56.930 &  77.514 & 28.6 &  1.33 \\
  LARS02 &          8.33 &    37.05 &  3C147 &  J0834+5534 &  41.180 &  48.815 &  51.386 & 26.9 &  0.65 \\
  LARS03 &          3.69 &    65.75 &  3C286 &  J1400+6210 &  58.154 &  73.539 & -48.043 & 40.5 &  1.05 \\
  LARS04 &          2.19 &    32.09 &  3C286 &  J1252+5634 &  43.598 &  58.138 &   6.693 & 33.1 &  1.07 \\
  LARS05 &          3.25 &     7.55 &  3C286 &  J1400+6210 &  47.128 &  56.144 &  35.683 & 34.8 &  0.82 \\
  LARS06 &          1.63 &    27.14 &  3C286 &  J1545+4751 &  41.902 &  47.456 &  43.306 & 30.4 &  1.31 \\
  LARS07 &          1.63 &     7.74 &  3C286 &  J1330+2509 &  44.773 &  48.332 &   7.700 & 34.9 &  1.09 \\
  LARS08 &          2.18 &    16.36 &  3C286 &  J1254+1141 &  46.384 &  56.508 &   8.266 & 39.2 &  1.15 \\
  LARS09 &         11.78 &    41.01 &  3C147 &  J0741+3112 &  44.899 &  48.069 & -76.232 & 43.0 &  0.53 \\
 eLARS01 &          1.63 &    17.39 &  3C286 &  J1035+5628 &  42.381 &  54.819 &  66.113 & 28.7 &  1.34 \\
 eLARS02 &          1.62 &     6.01 &  3C286 &  J1400+6210 &  44.464 &  57.527 &  24.479 & 43.1 &  1.13 \\
 eLARS03 &          3.60 &    12.95 &  3C286 &  J1206+6413 &  46.507 &  53.423 &  -6.180 & 35.0 &  0.72 \\
 eLARS04 &          3.25 &    17.40 &  3C286 &  J1634+6245 &  42.989 &  56.131 & -77.940 & 28.4 &  0.87 \\
 eLARS05 &          3.58 &    53.38 &  3C286 &  J1035+5628 &  52.839 &  78.109 &  71.092 & 44.0 &  1.03 \\
 eLARS06 &          3.67 &    30.13 &  3C286 &  J1206+6413 &  43.883 &  82.728 &  66.386 & 42.6 &  1.01 \\
 eLARS07 &          3.32 &     9.47 &  3C286 &  J1035+5628 &  42.955 &  59.887 & -73.329 & 35.6 &  0.94 \\
 eLARS08 &          3.32 &    16.25 &  3C286 &  J1035+5628 &  53.469 &  73.315 &  70.906 & 38.9 &  0.78 \\
 eLARS09 &          3.25 &    27.17 &  3C286 &  J1313+6735 &  54.580 &  92.753 &  71.463 & 44.7 &  0.83 \\
 eLARS10 &          3.58 &    53.38 &  3C286 &  J1035+5628 &  52.839 &  78.109 &  71.092 & 43.4 &  1.03 \\
 eLARS11 &          3.24 &    35.07 &  3C147 &  J0834+5534 &  49.539 &  61.650 &  47.309 & 33.6 &  1.10 \\
 eLARS12 &          3.66 &    30.56 &  3C286 &  J1400+6210 &  46.088 &  96.921 &  73.567 & 45.7 &  0.95 \\
 eLARS13 &          3.31 &    32.18 &  3C286 &  J0949+6614 &  46.378 &  59.024 &  -8.645 & 34.2 &  1.09 \\
 eLARS14 &          3.32 &    20.29 &  3C286 &  J1206+6413 &  87.077 &  56.325 &  61.739 & 46.7 &  0.53 \\
 eLARS15 &          3.25 &    31.25 &  3C286 &  J1400+6210 &  41.474 &  62.341 &  66.801 & 36.5 &  1.13 \\
 eLARS16 &          1.62 &    14.23 &  3C286 &  J1400+6210 &  41.386 &  49.637 &  10.827 & 31.7 &  1.21 \\
 eLARS17 &          3.32 &    32.34 &  3C286 &  J1035+5628 &  53.087 &  61.413 & -69.553 & 35.6 &  0.89 \\
 eLARS18 &          1.63 &    11.73 &  3C286 &  J1438+6211 &  41.924 &  56.297 &  63.041 & 29.0 &  1.35 \\
 eLARS19 &          3.31 &    16.77 &  3C286 &  J1219+4829 &  55.689 &  64.156 & -83.041 & 37.0 &  0.80 \\
 eLARS20 &          3.25 &    35.92 &  3C286 &  J1400+6210 &  48.935 &  80.160 &  75.257 & 40.3 &  1.02 \\
 eLARS21 &          3.25 &    37.37 &  3C286 &  J1400+6210 &  48.352 &  91.108 & -83.540 & 45.7 &  1.14 \\
 eLARS22 &          3.25 &    26.82 &  3C286 &  J1634+6245 &  45.628 &  52.726 & -69.898 & 45.4 &  0.88 \\
 eLARS23 &          1.63 &    15.53 &  3C286 &  J1438+6211 &  42.295 &  52.896 &  11.273 & 47.5 &  1.11 \\
 eLARS24 &          3.26 &    42.28 &  3C286 &  J1438+6211 &  43.212 &  48.371 & -42.271 & 43.0 &  1.07 \\
 eLARS25 &          3.32 &    20.67 &  3C286 &  J1035+5628 &  44.021 &  50.474 &  80.999 & 41.8 &  0.87 \\
 eLARS26 &          3.32 &    21.21 &  3C286 &  J1206+6413 &  45.908 &  49.563 &  -5.614 & 43.1 &  0.99 \\
 eLARS27 &          1.63 &    21.12 &  3C286 &  J1438+6211 &  40.740 &  52.801 &  24.316 & 41.0 &  1.15 \\
 eLARS28 &          3.32 &    26.82 &  3C286 &  J1206+6413 &  47.418 &  48.480 & -53.413 & 43.5 &  0.95 \\
    \hline  
Average & 3.24$\pm$1.84 &26.19$\pm$13.49  & - & - & 50.1$\pm$8.0 & 61.5$\pm$13.5 & 13.23$\pm$56.0 & 38.3$\pm5.9$  & 1.00$\pm$0.20\\
    \hline  
    \end{tabular}
\end{table*}
 We present parameters characterizing the VLA D-array observations of the (e)LARS samples in Table \ref{tab:vla_obs}. These include the integration time, percentage of flagged visibilities, the sources used as bandpass and phase calibrators, the synthesized beam properties and the cube rms.

\subsection*{Appendix B: 21cm Maps and Spectra of LARS and eLARS galaxies}
\begin{figure*}[ht]
    \centering
    \includegraphics[width=\textwidth]{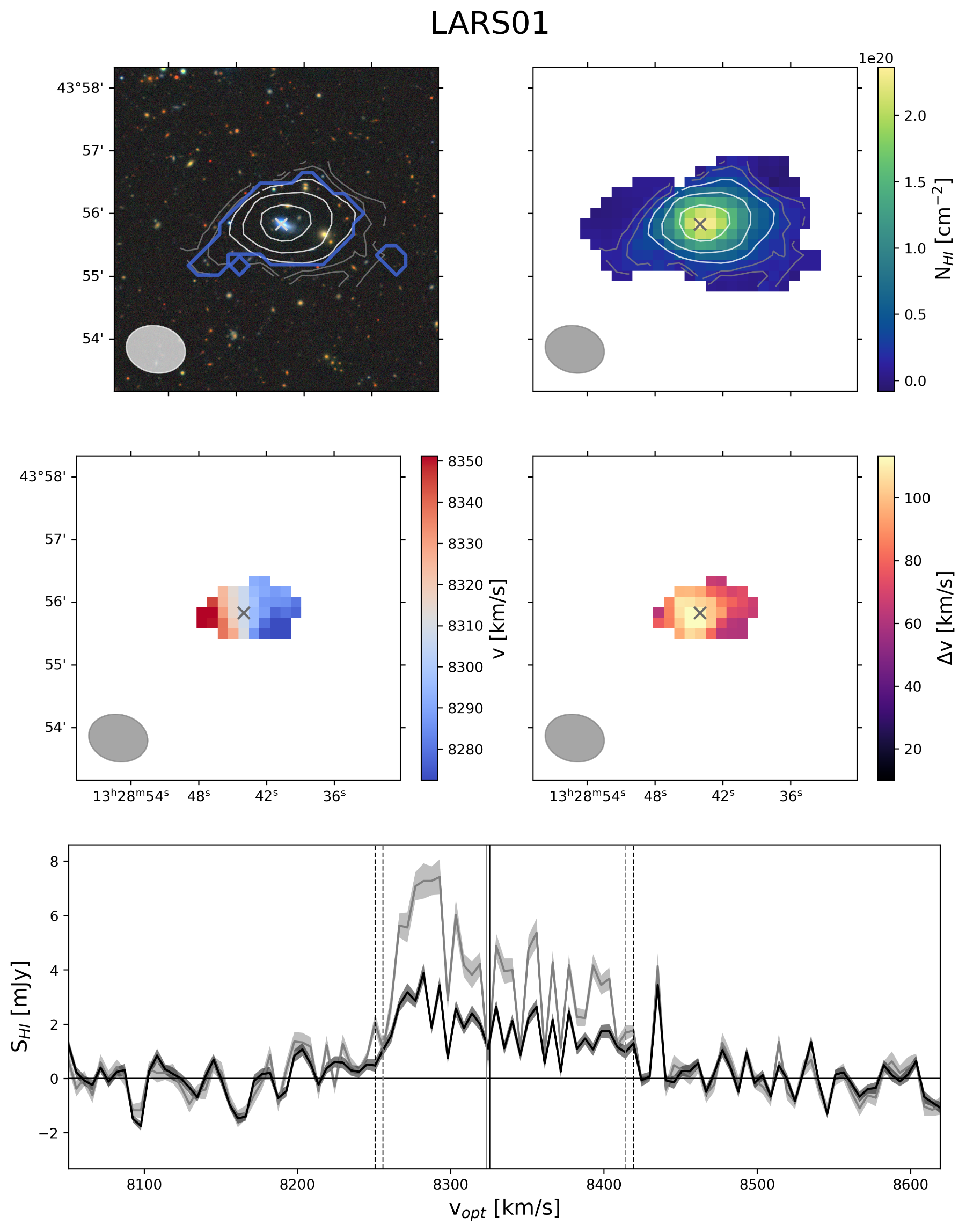}
    \caption{Same caption as Figure \ref{fig:opt_hi_el04}, but for LARS01.}
    \label{fig:opt_hi_l01}
\end{figure*}

\begin{figure*}[ht]
    \centering
    \includegraphics[width=\textwidth]{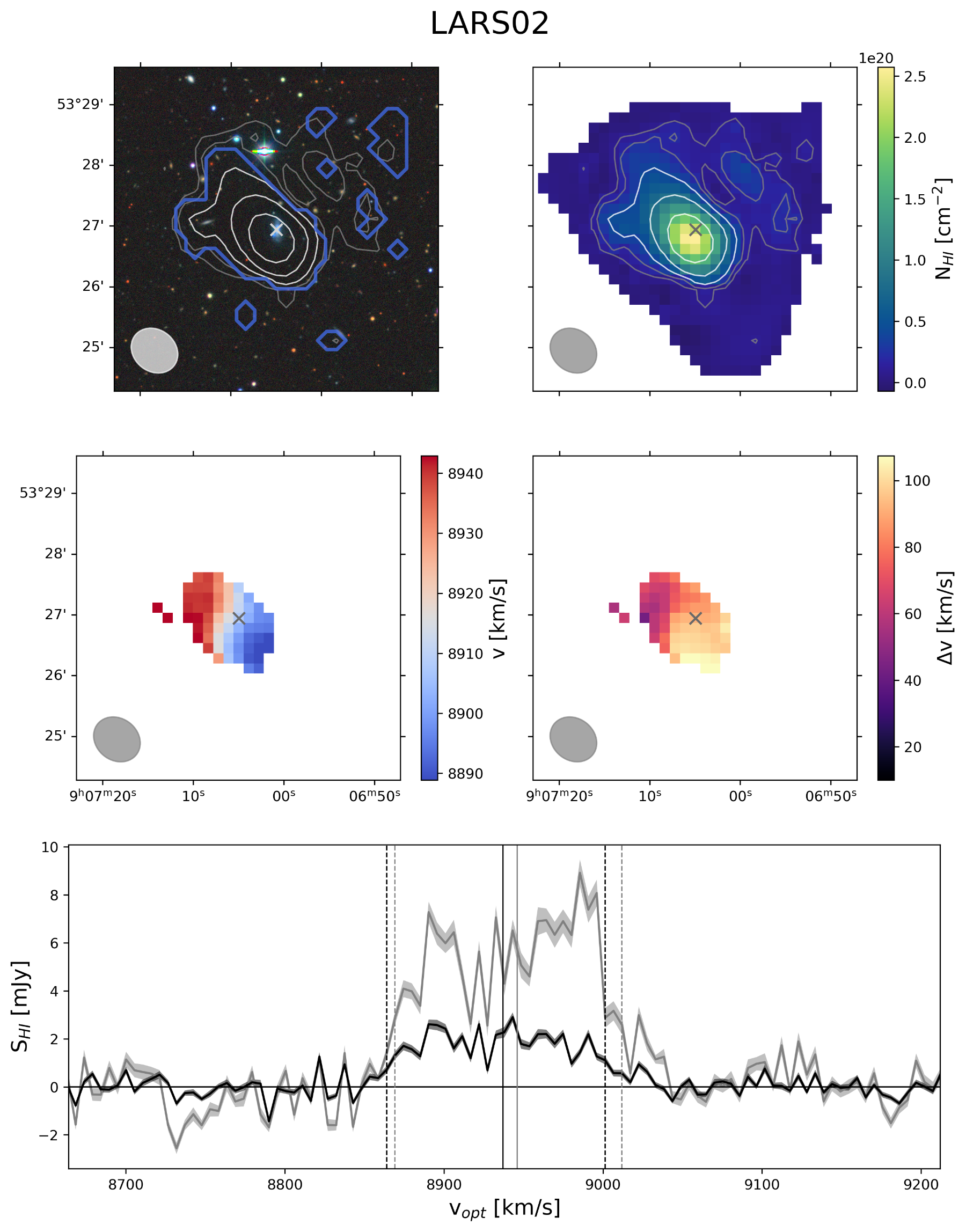}
    \caption{Same caption as Figure \ref{fig:opt_hi_el04}, but for LARS02.}
    \label{fig:opt_hi_l02}
\end{figure*}

\begin{figure*}[ht]
    \centering
    \includegraphics[width=\textwidth]{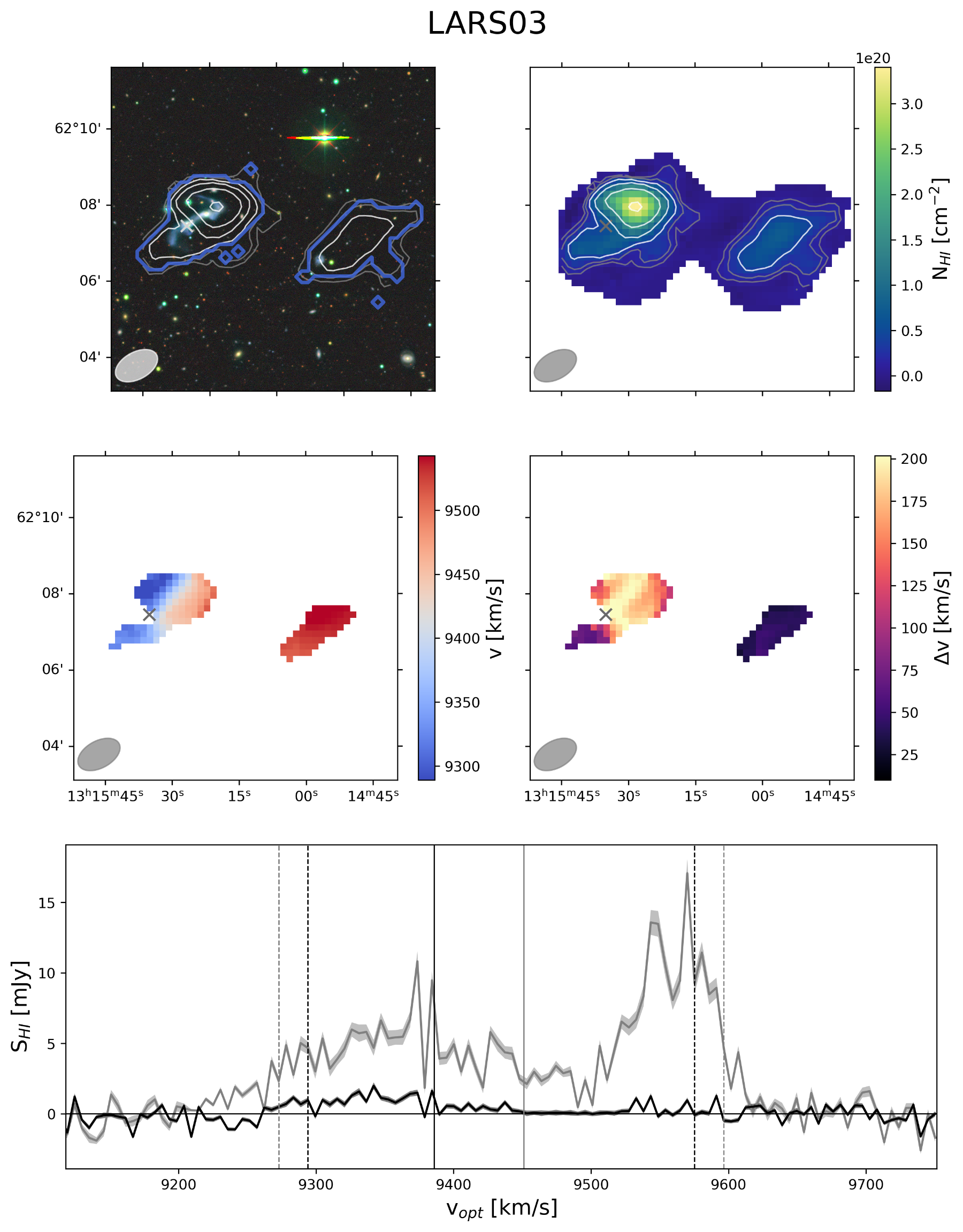}
    \caption{Same caption as Figure \ref{fig:opt_hi_el04}, but for LARS03.}
    \label{fig:opt_hi_l03}
\end{figure*}

\begin{figure*}[ht]
    \centering
    \includegraphics[width=\textwidth]{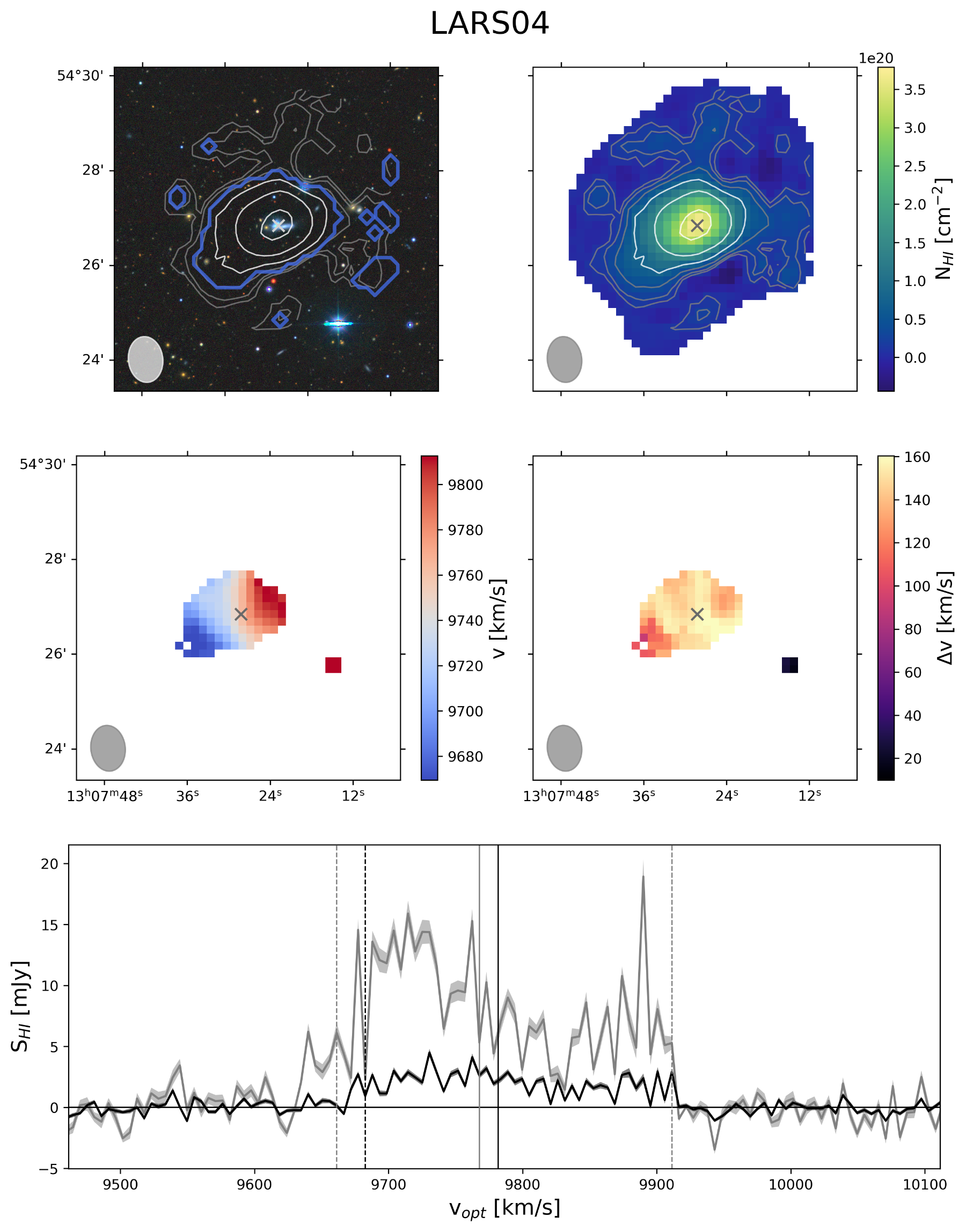}
    \caption{Same caption as Figure \ref{fig:opt_hi_el04}, but for LARS04.}
    \label{fig:opt_hi_l04}
\end{figure*}

\begin{figure*}[ht]
    \centering
    \includegraphics[width=\textwidth]{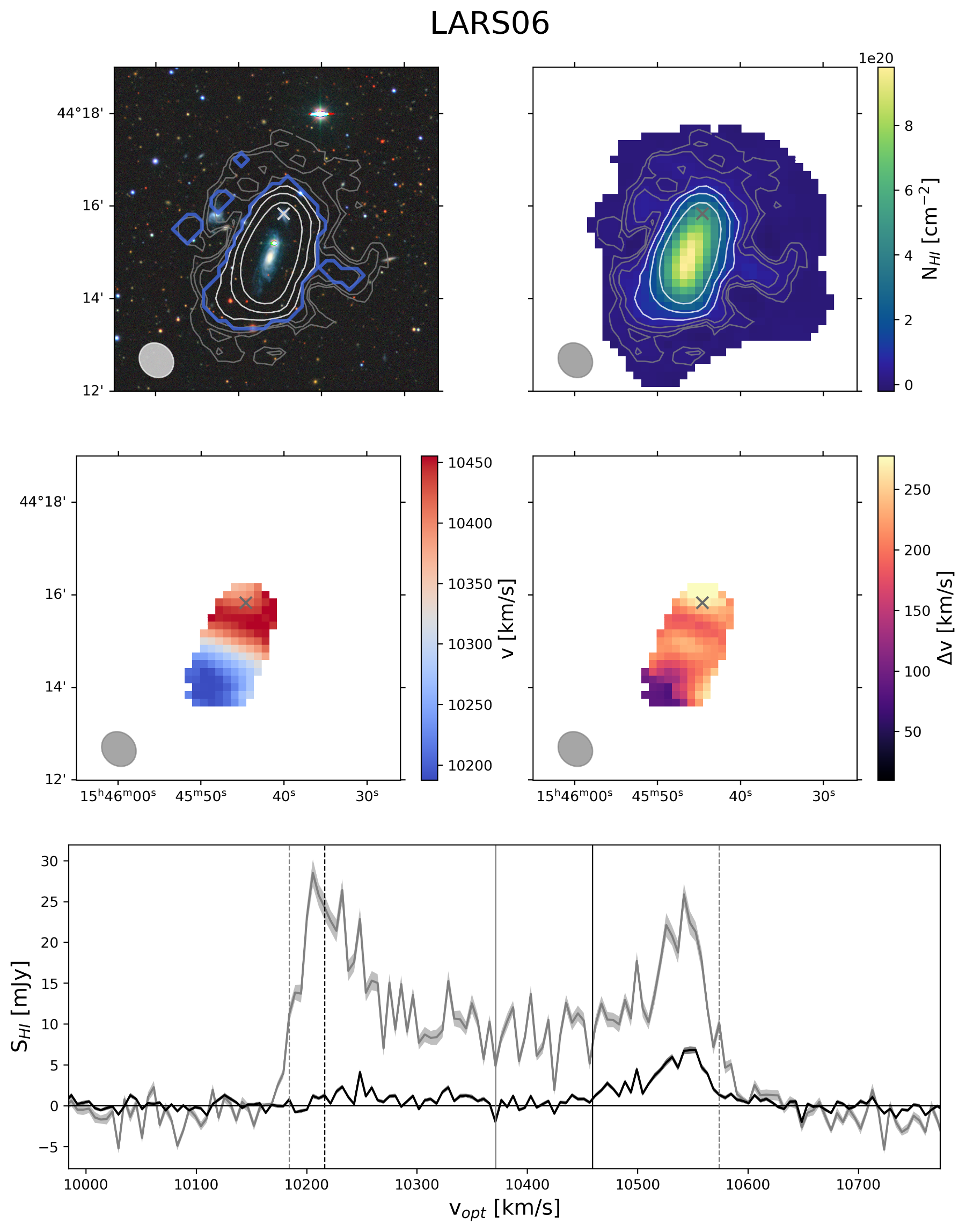}
    \caption{Same caption as Figure \ref{fig:opt_hi_el04}, but for LARS06.}
    \label{fig:opt_hi_l06}
\end{figure*}

\begin{figure*}[ht]
    \centering
    \includegraphics[width=\textwidth]{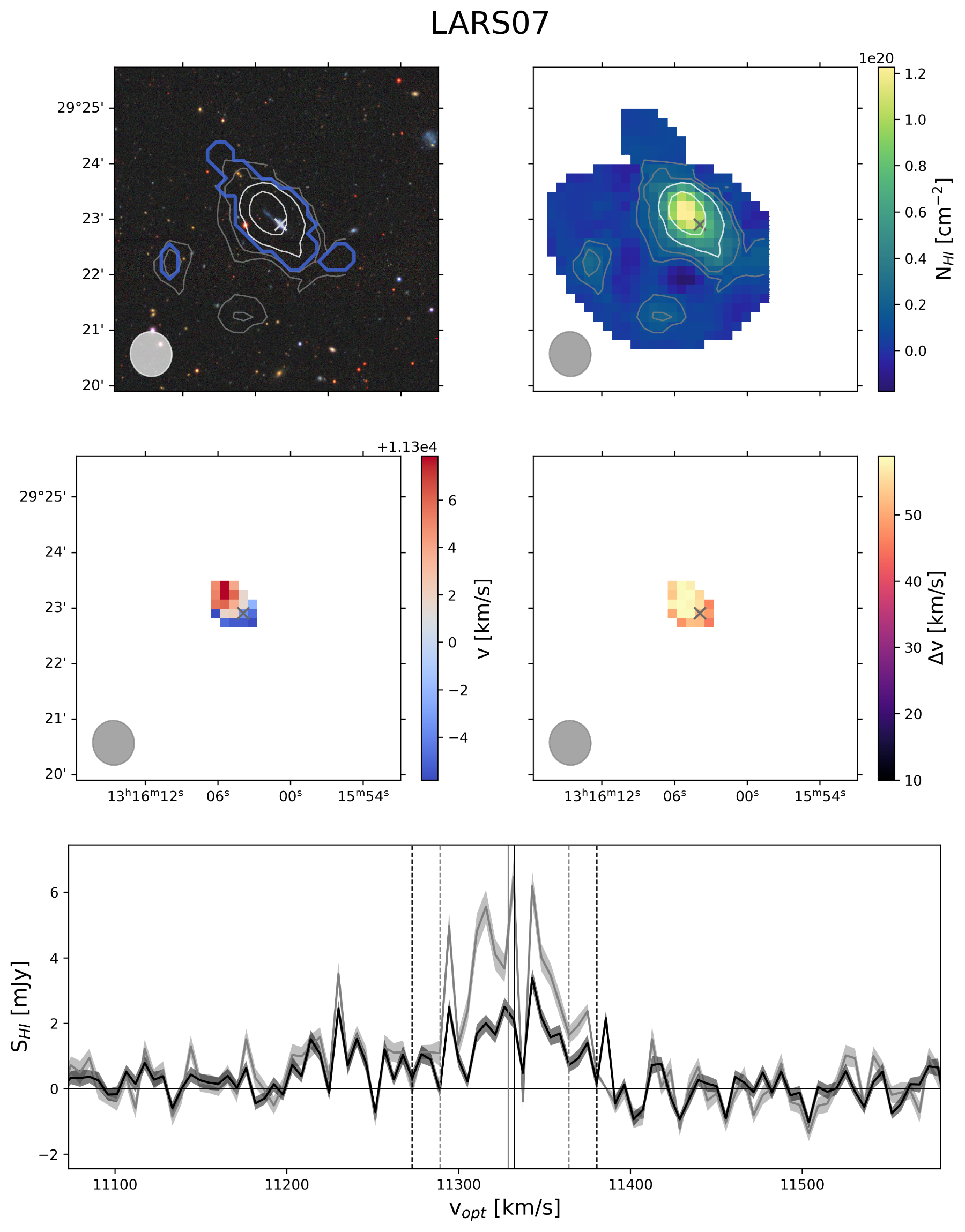}
    \caption{Same caption as Figure \ref{fig:opt_hi_el04}, but for LARS07.}
    \label{fig:opt_hi_l07}
\end{figure*}

\begin{figure*}[ht]
    \centering
    \includegraphics[width=\textwidth]{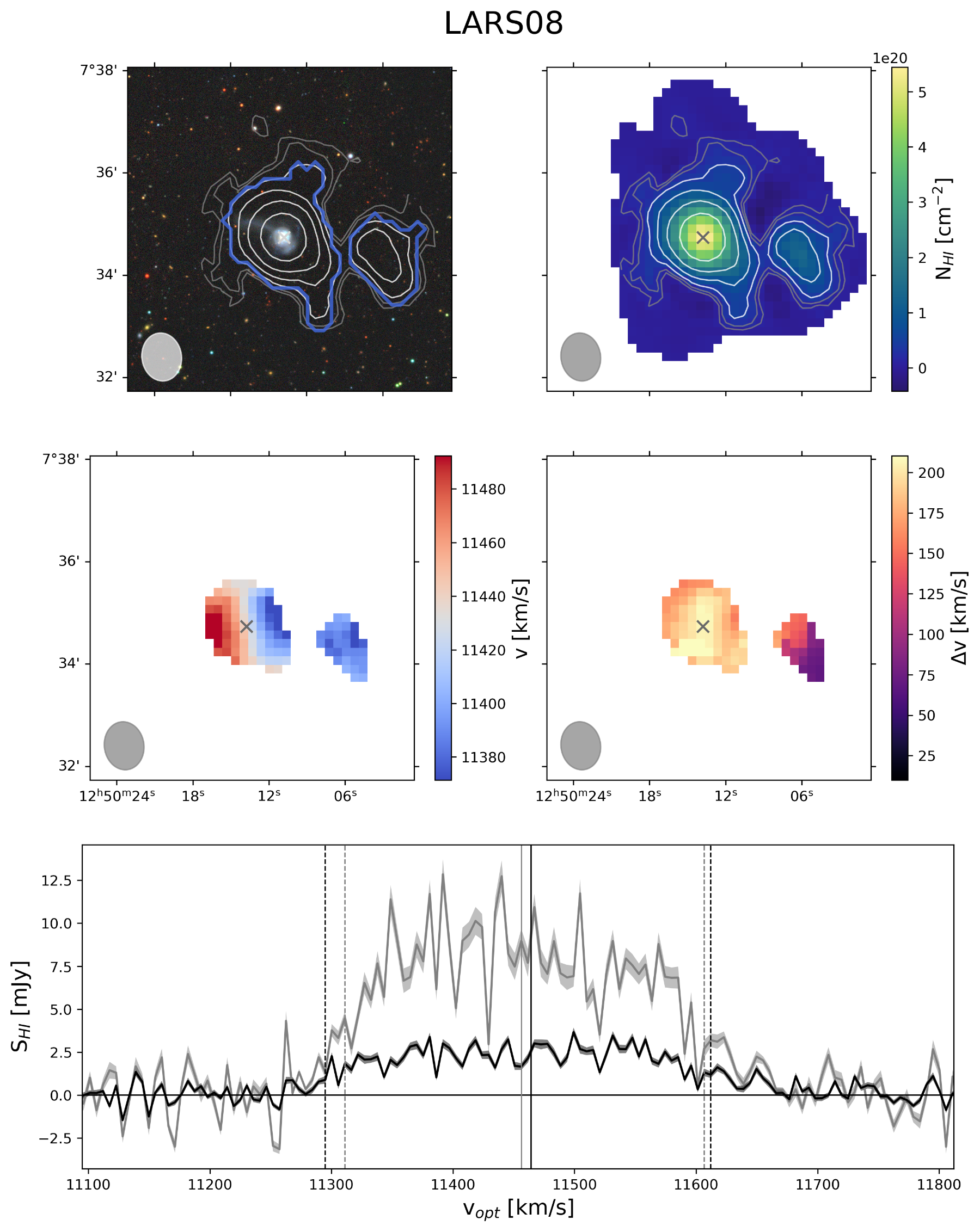}
    \caption{Same caption as Figure \ref{fig:opt_hi_el04}, but for LARS08.}
    \label{fig:opt_hi_l08}
\end{figure*}

\begin{figure*}[ht]
    \centering
    \includegraphics[width=\textwidth]{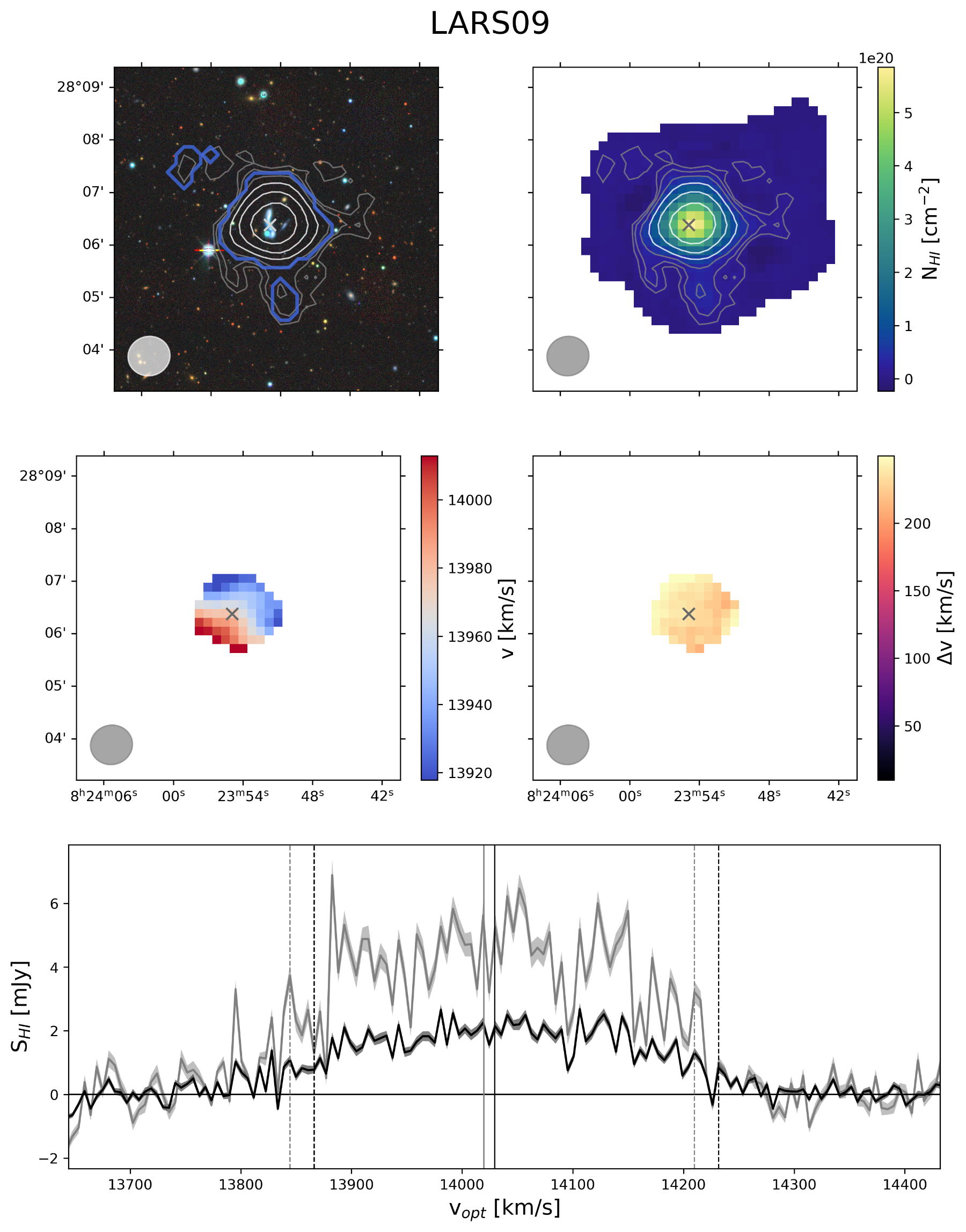}
    \caption{Same caption as Figure \ref{fig:opt_hi_el04}, but for LARS09.}
    \label{fig:opt_hi_l09}
\end{figure*}

\begin{figure*}[ht]
    \centering
    \includegraphics[width=\textwidth]{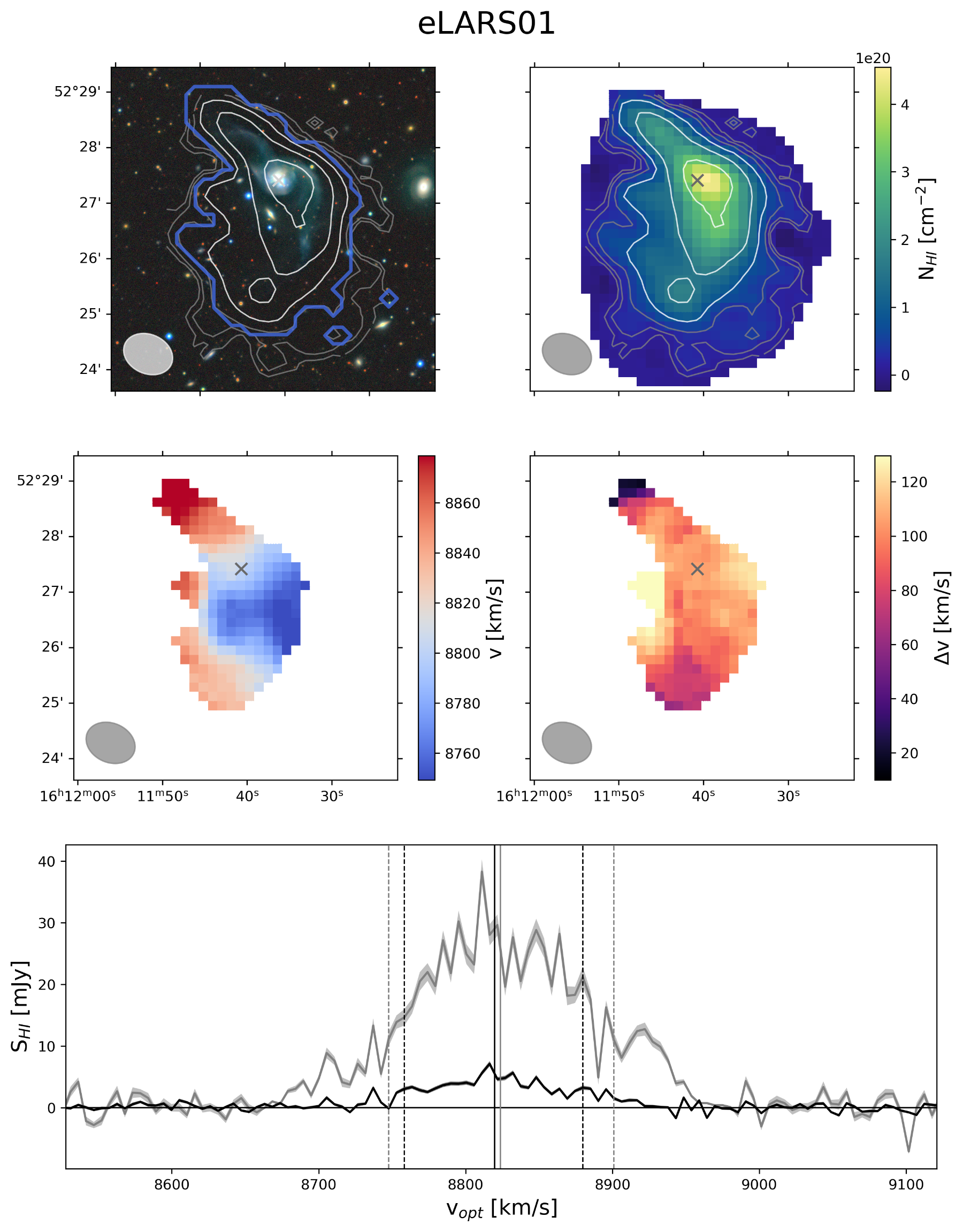}
    \caption{Same caption as Figure \ref{fig:opt_hi_el04}, but for eLARS01.}
    \label{fig:opt_hi_el01}
\end{figure*}

\begin{figure*}[ht]
    \centering
    \includegraphics[width=\textwidth]{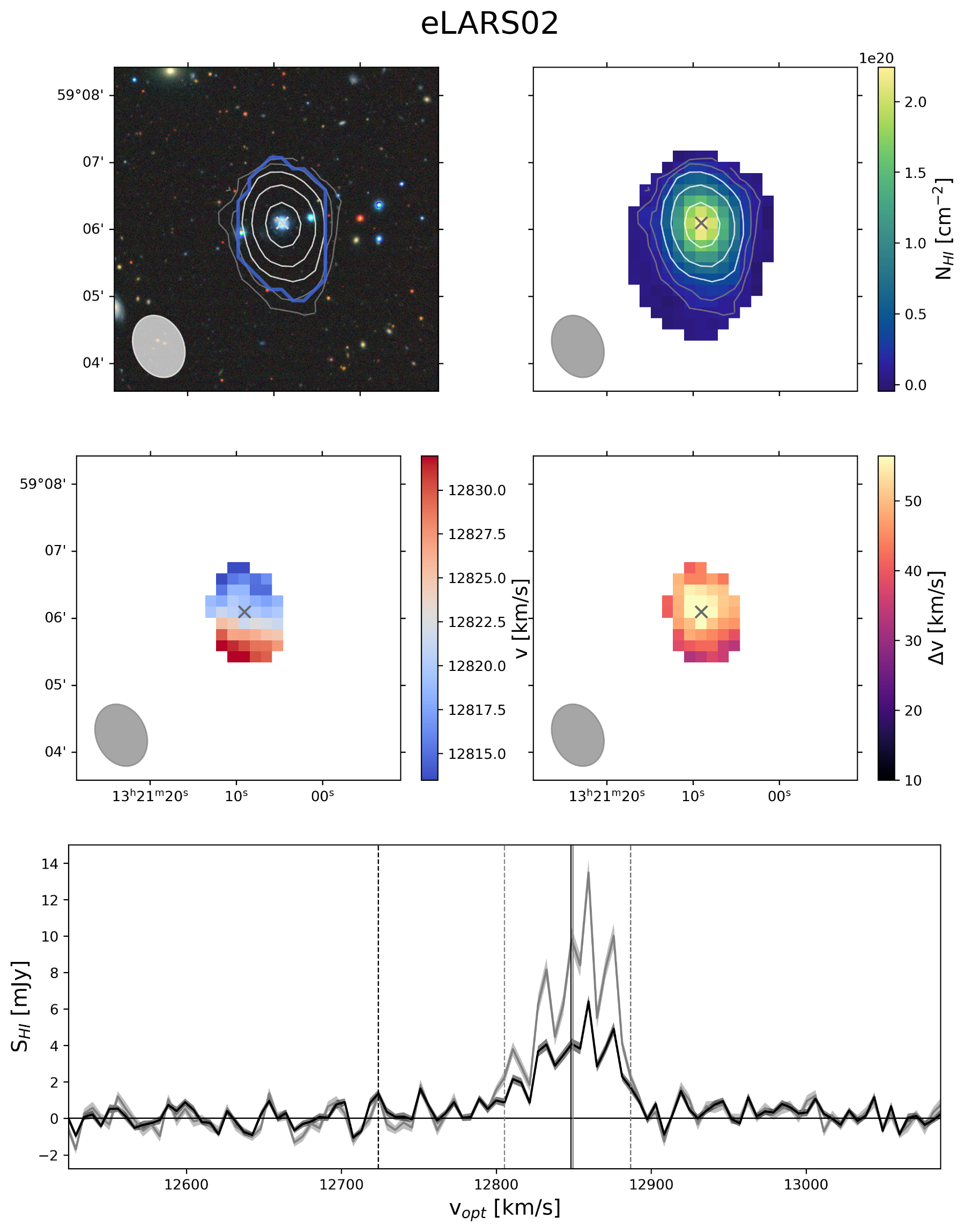}
    \caption{Same caption as Figure \ref{fig:opt_hi_el04}, but for eLARS02.}
    \label{fig:opt_hi_el02}
\end{figure*}

\begin{figure*}[ht]
    \centering
    \includegraphics[width=\textwidth]{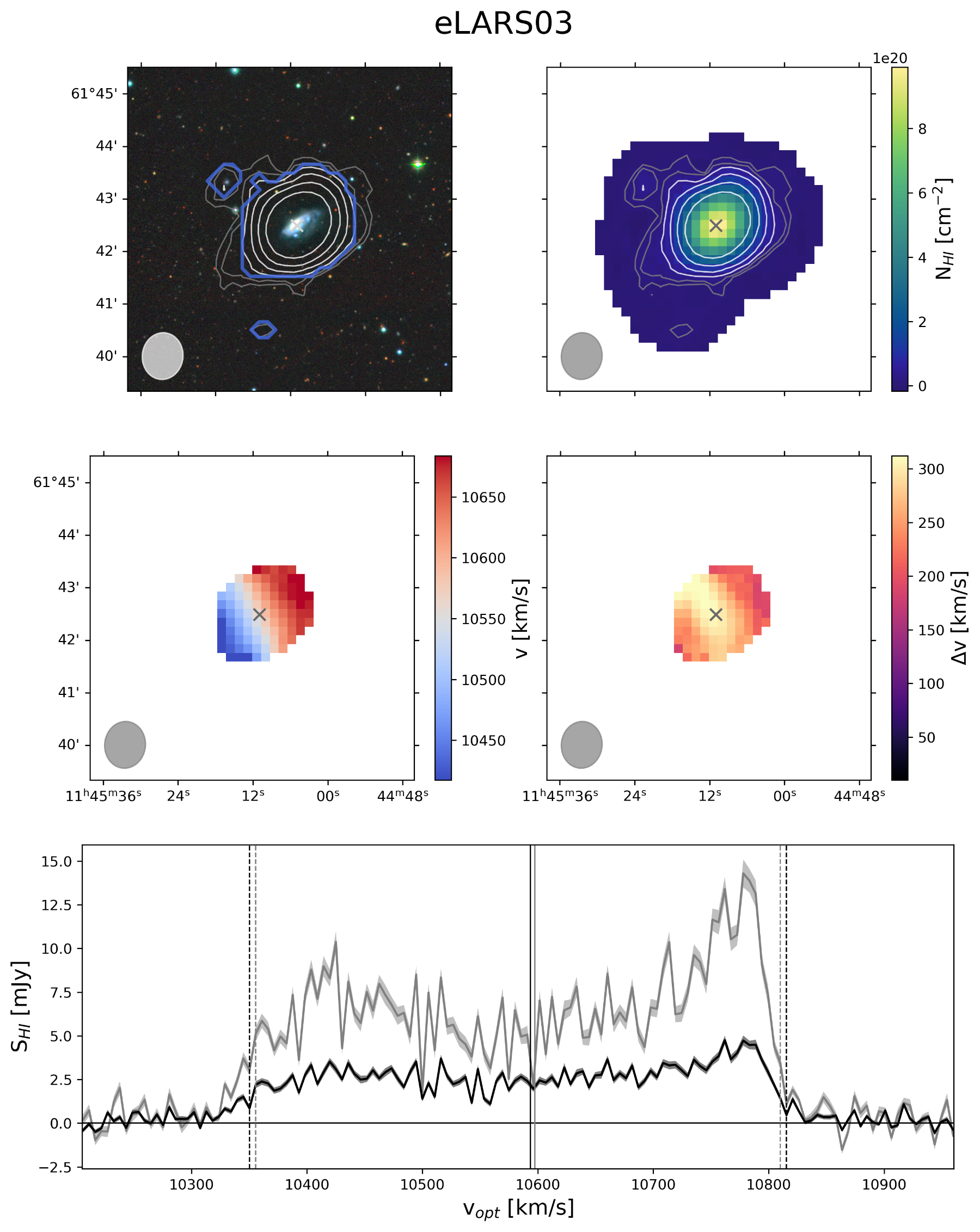}
    \caption{Same caption as Figure \ref{fig:opt_hi_el04}, but for eLARS03.}
    \label{fig:opt_hi_el03}
\end{figure*}

\begin{figure*}[ht]
    \centering
    \includegraphics[width=\textwidth]{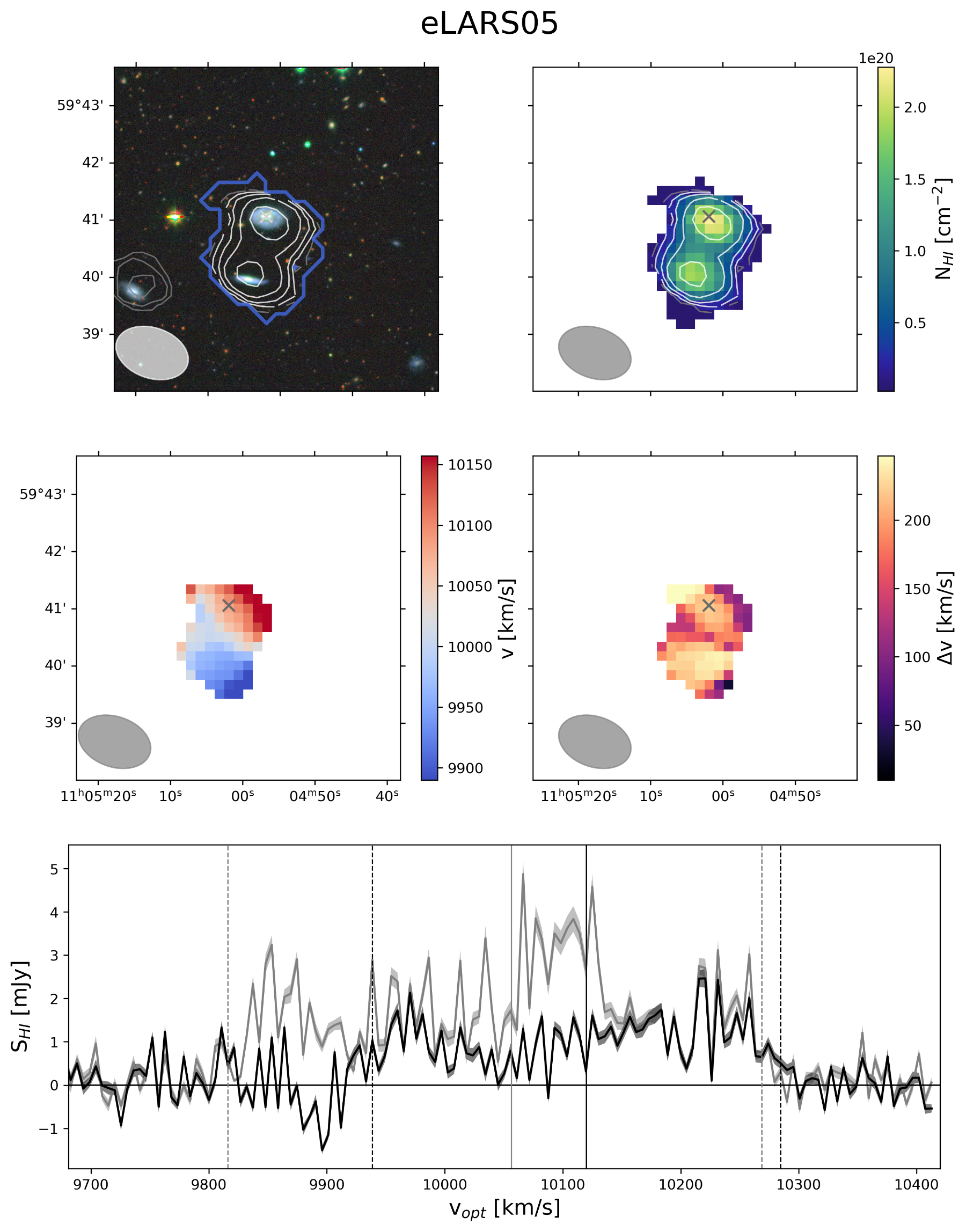}
    \caption{Same caption as Figure \ref{fig:opt_hi_el04}, but for eLARS05. The object on the bottom left of the top left panel is a separate detection by SoFIA-2.}
    \label{fig:opt_hi_el05}
\end{figure*}

\begin{figure*}[ht]
    \centering
    \includegraphics[width=\textwidth]{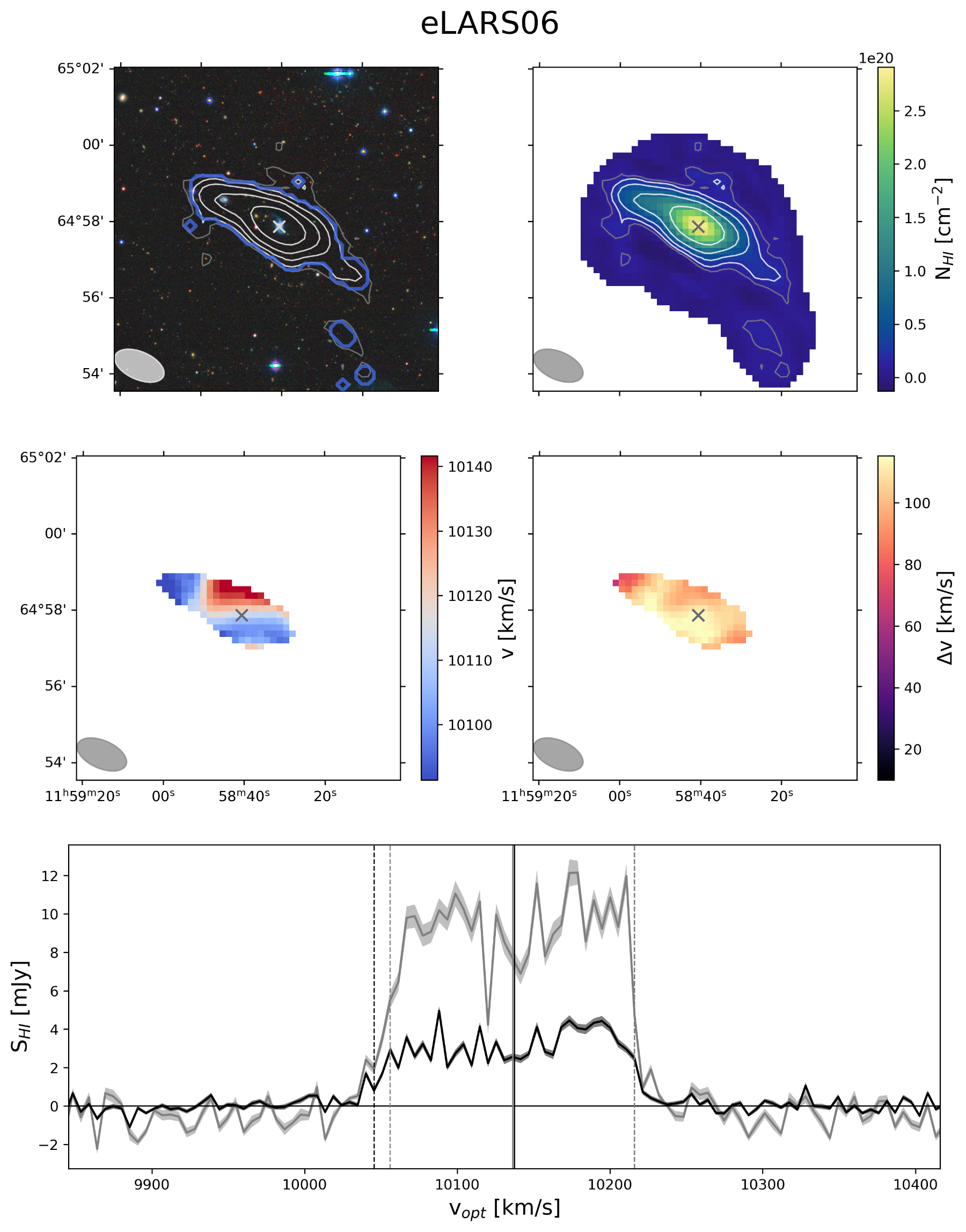}
    \caption{Same caption as Figure \ref{fig:opt_hi_el04}, but for eLARS06.}
    \label{fig:opt_hi_el06}
\end{figure*}

\begin{figure*}[ht]
    \centering
    \includegraphics[width=\textwidth]{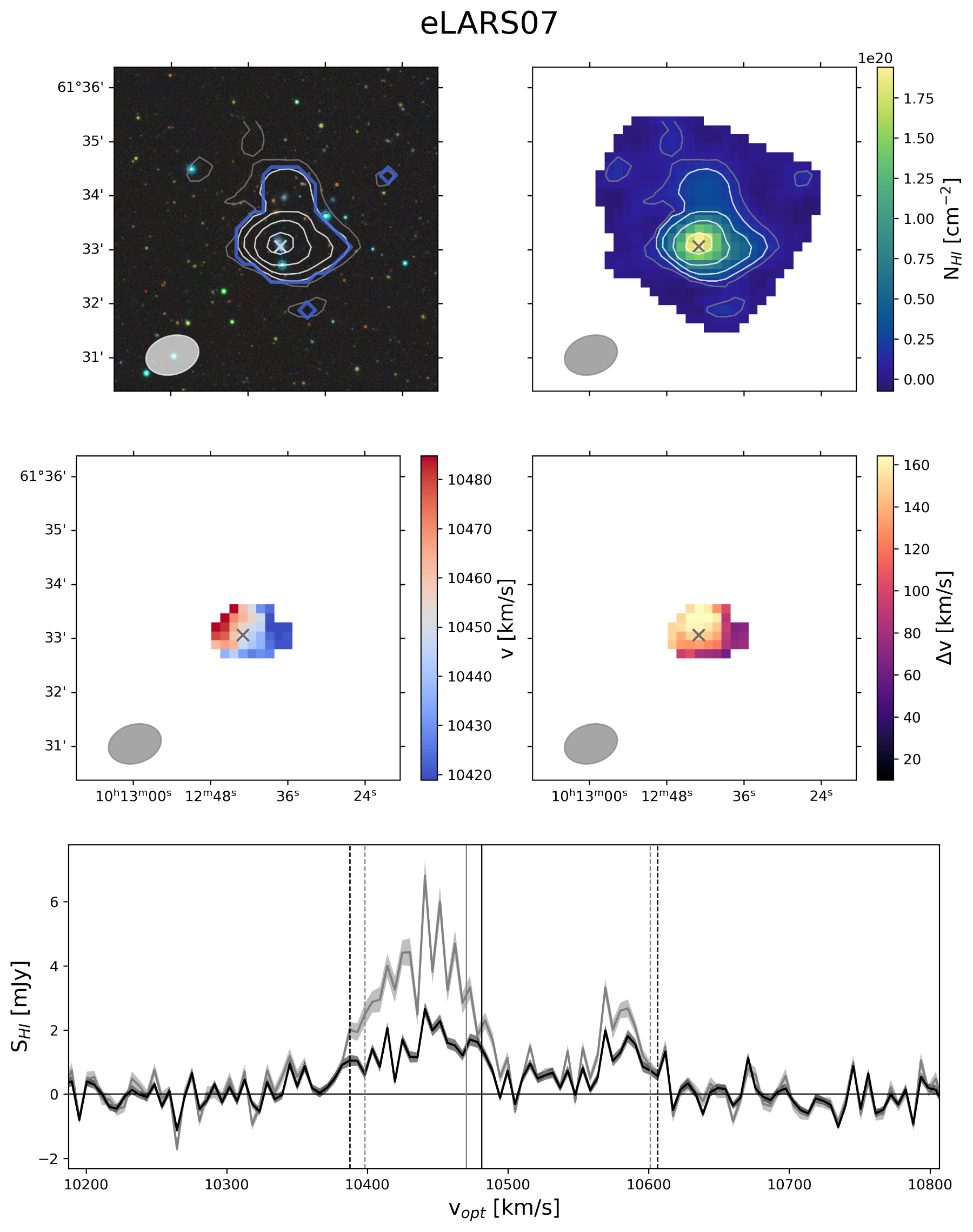}
    \caption{Same caption as Figure \ref{fig:opt_hi_el04}, but for eLARS07.}
    \label{fig:opt_hi_el07}
\end{figure*}

\begin{figure*}[ht]
    \centering
    \includegraphics[width=\textwidth]{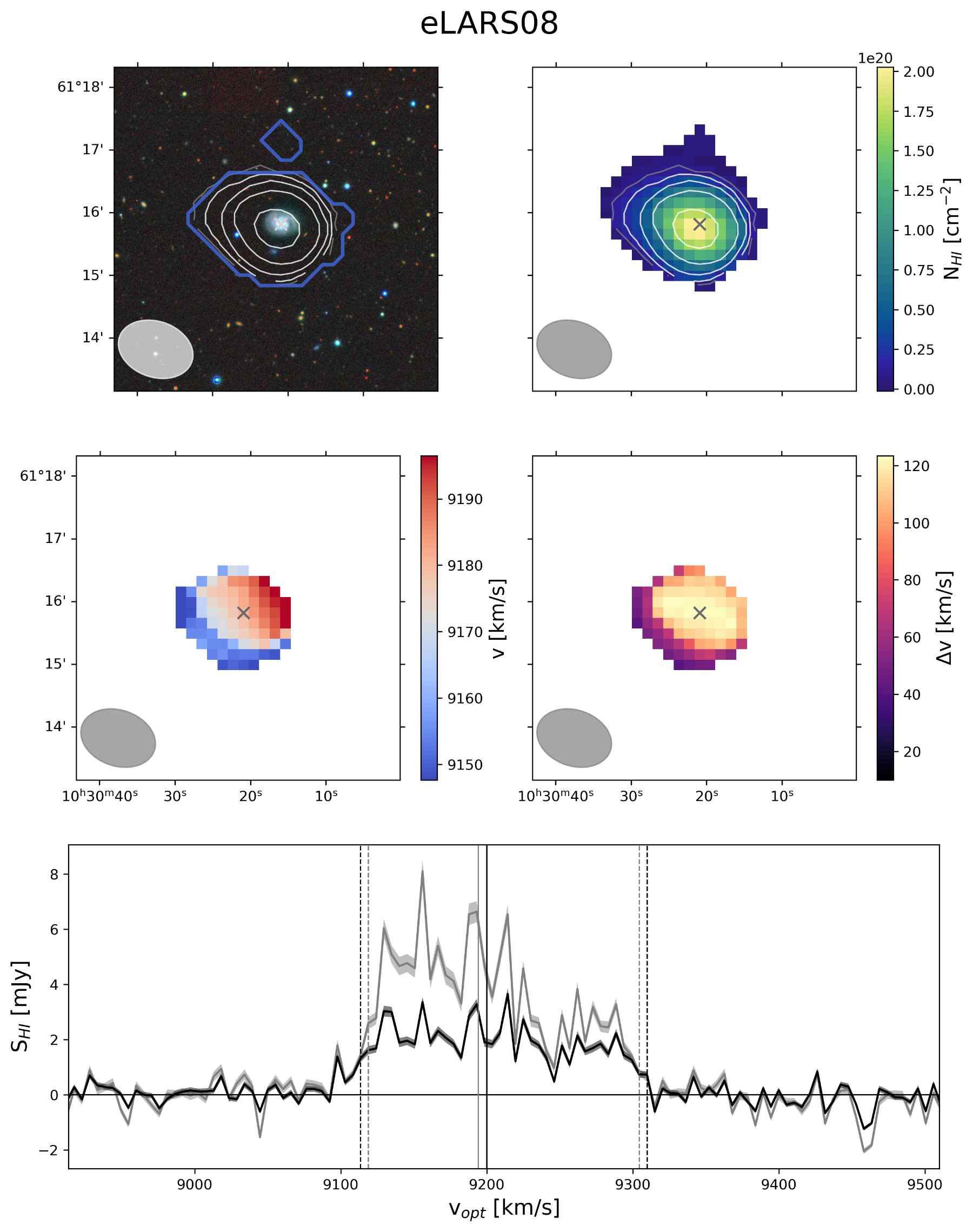}
    \caption{Same caption as Figure \ref{fig:opt_hi_el04}, but for eLARS08.}
    \label{fig:opt_hi_el08}
\end{figure*}

\begin{figure*}[ht]
    \centering
    \includegraphics[width=\textwidth]{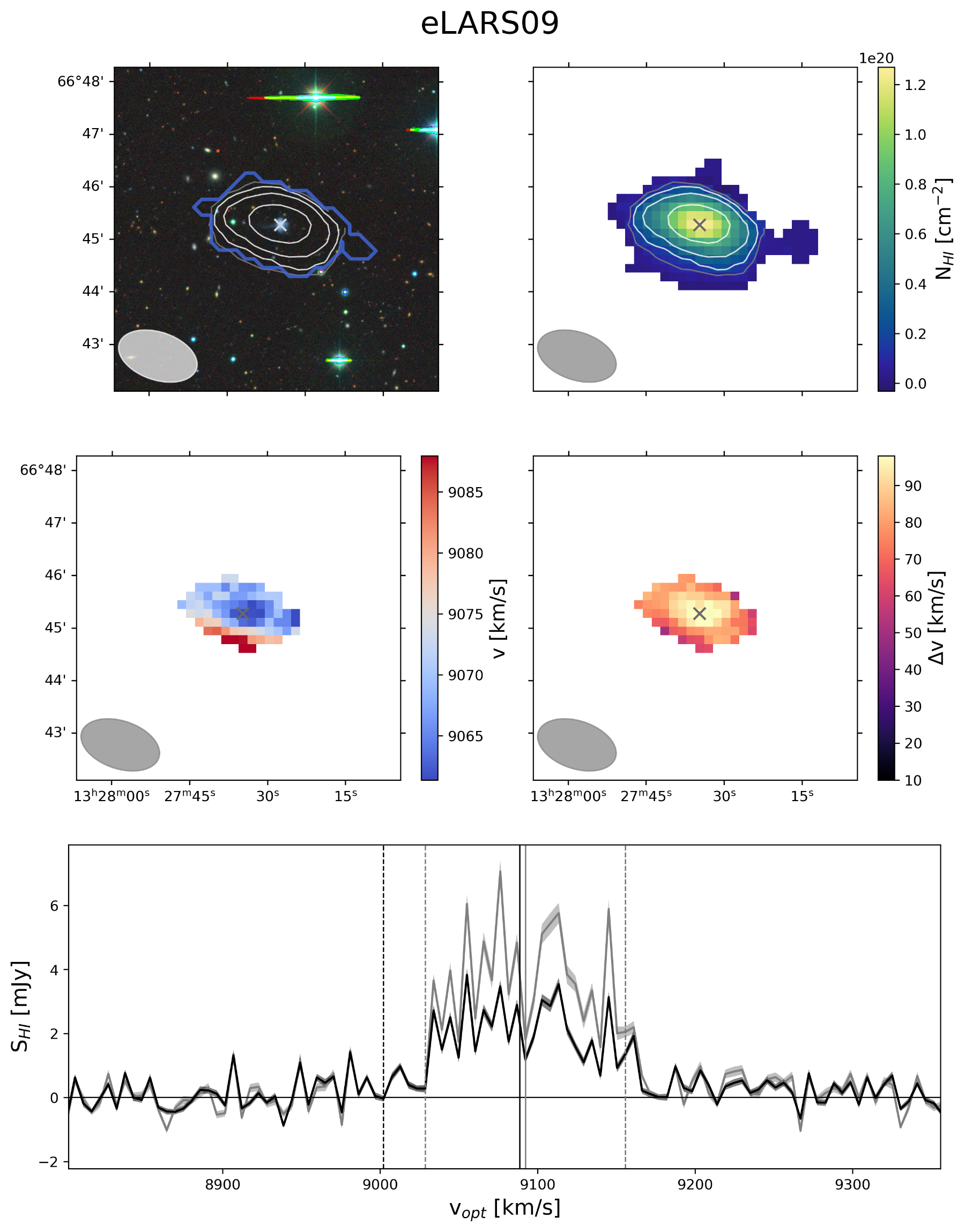}
    \caption{Same caption as Figure \ref{fig:opt_hi_el04}, but for eLARS09.}
    \label{fig:opt_hi_el09}
\end{figure*}

\begin{figure*}[ht]
    \centering
    \includegraphics[width=\textwidth]{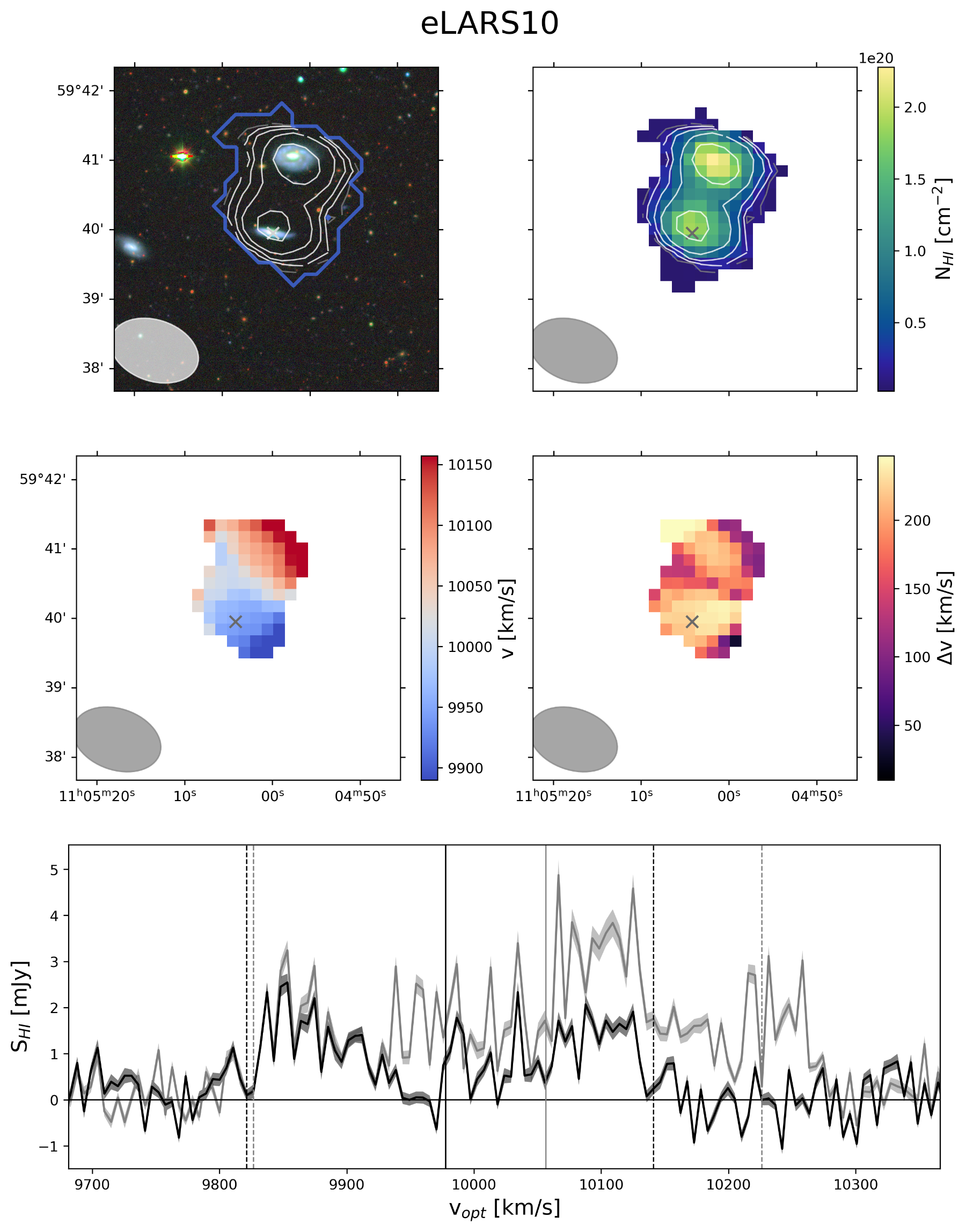}
    \caption{Same caption as Figure \ref{fig:opt_hi_el04}, but for eLARS10.}
    \label{fig:opt_hi_el10}
\end{figure*}

\begin{figure*}[ht]
    \centering
    \includegraphics[width=\textwidth]{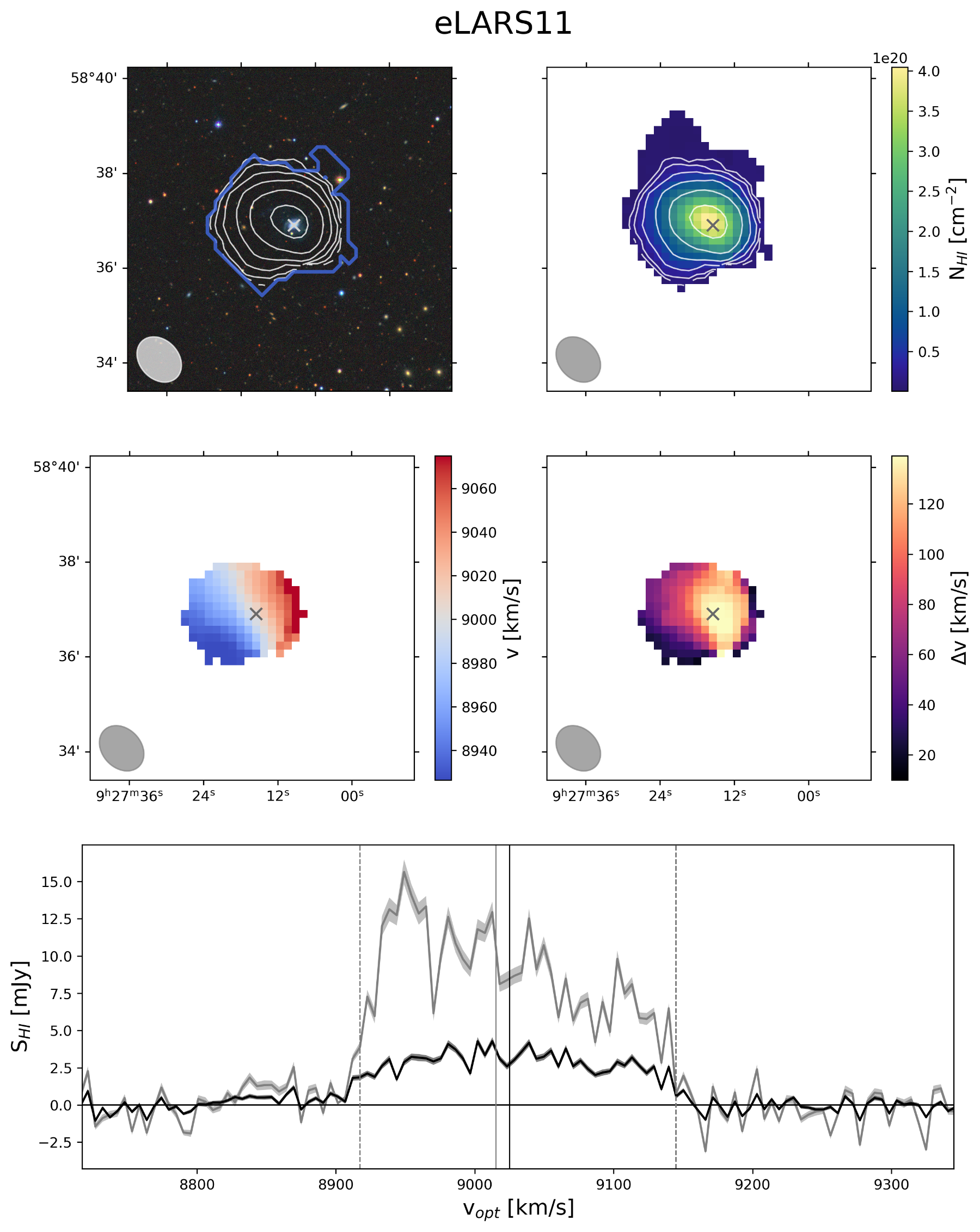}
    \caption{Same caption as Figure \ref{fig:opt_hi_el04}, but for eLARS11.}
    \label{fig:opt_hi_el11}
\end{figure*}

\begin{figure*}[ht]
    \centering
    \includegraphics[width=\textwidth]{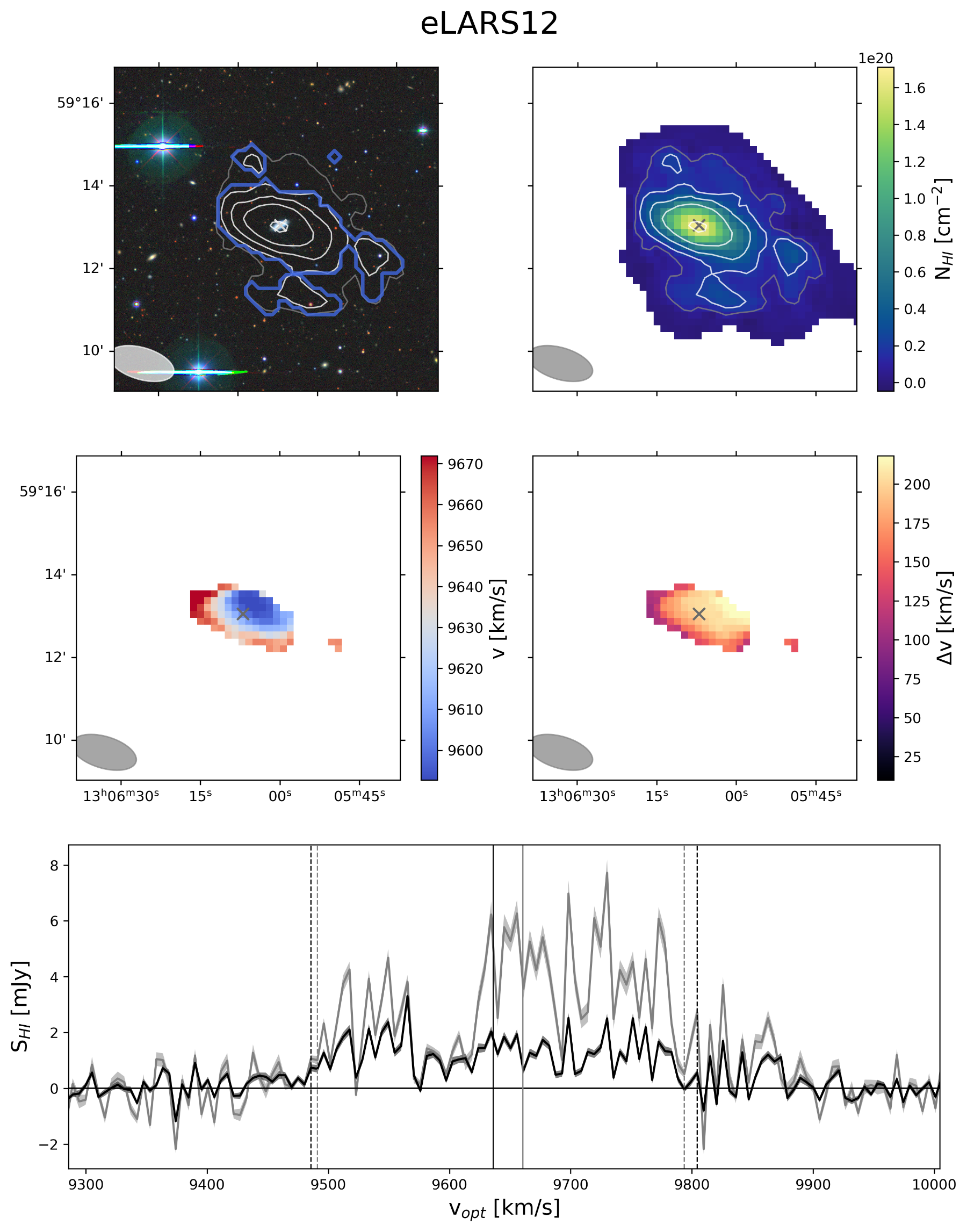}
    \caption{Same caption as Figure \ref{fig:opt_hi_el04}, but for eLARS12.}
    \label{fig:opt_hi_el12}
\end{figure*}

\begin{figure*}[ht]
    \centering
    \includegraphics[width=\textwidth]{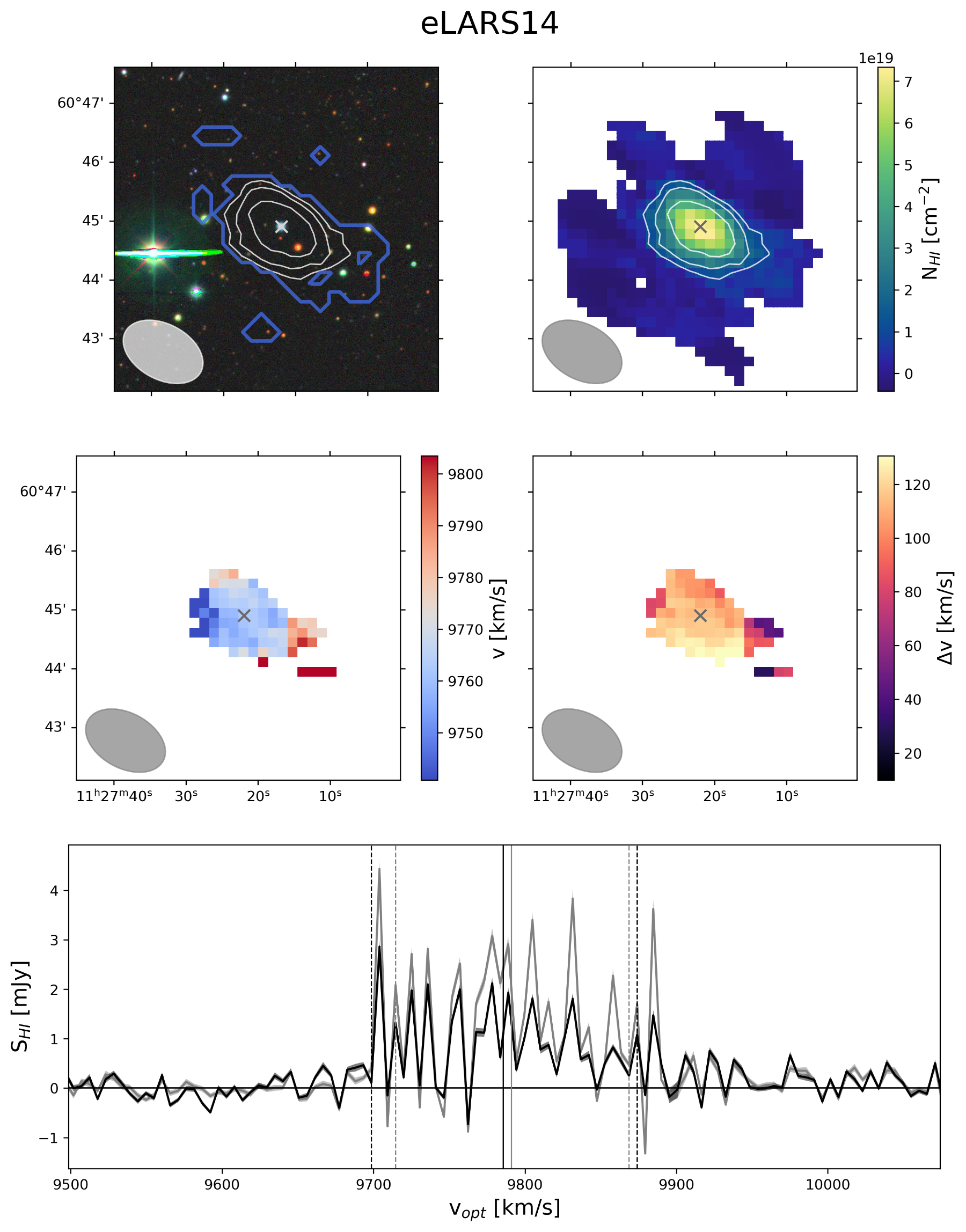}
    \caption{Same caption as Figure \ref{fig:opt_hi_el04}, but for eLARS14.}
    \label{fig:opt_hi_el14}
\end{figure*}

\begin{figure*}[ht]
    \centering
    \includegraphics[width=\textwidth]{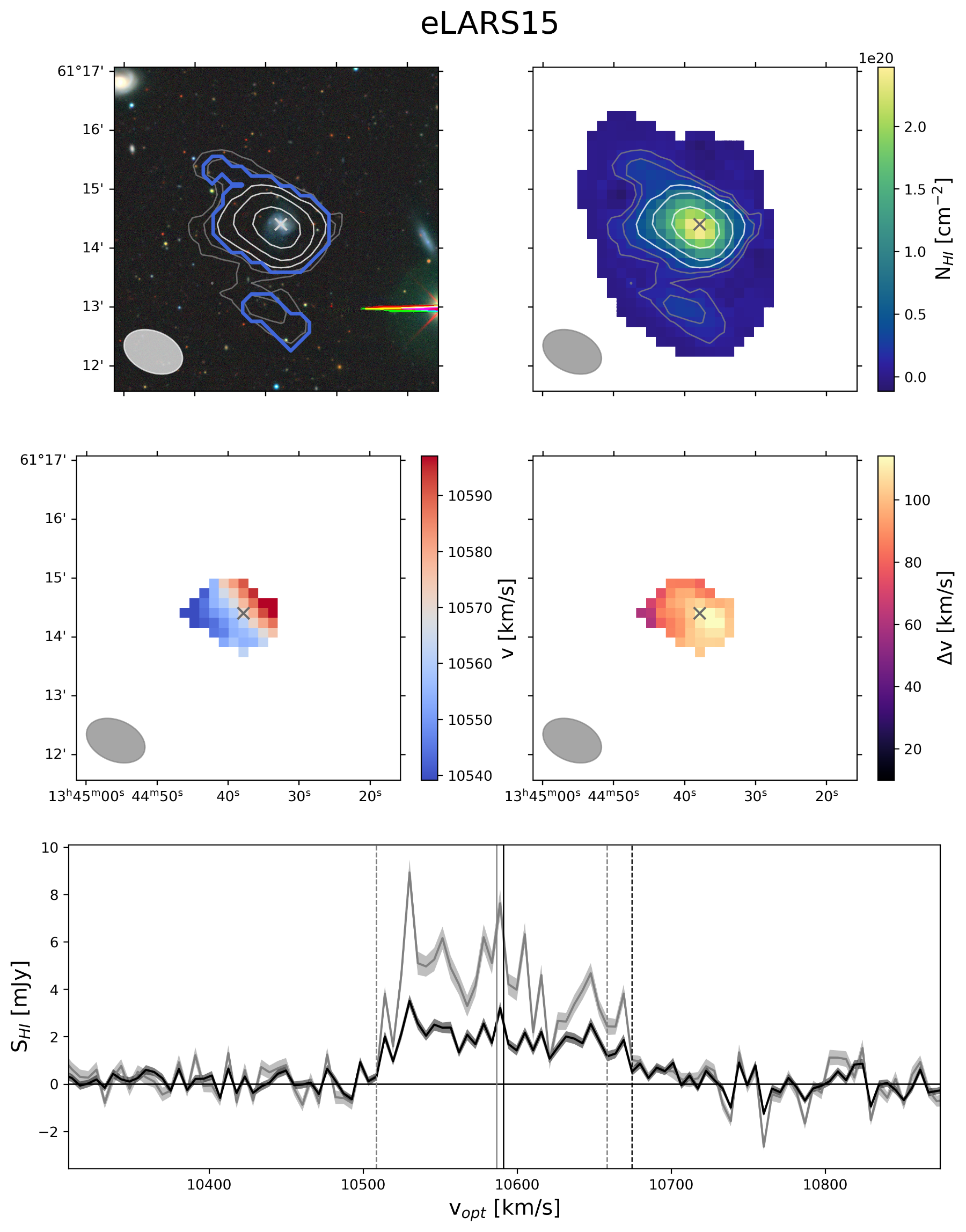}
    \caption{Same caption as Figure \ref{fig:opt_hi_el04}, but for eLARS15.}
    \label{fig:opt_hi_el15}
\end{figure*}

\begin{figure*}[ht]
    \centering
    \includegraphics[width=\textwidth]{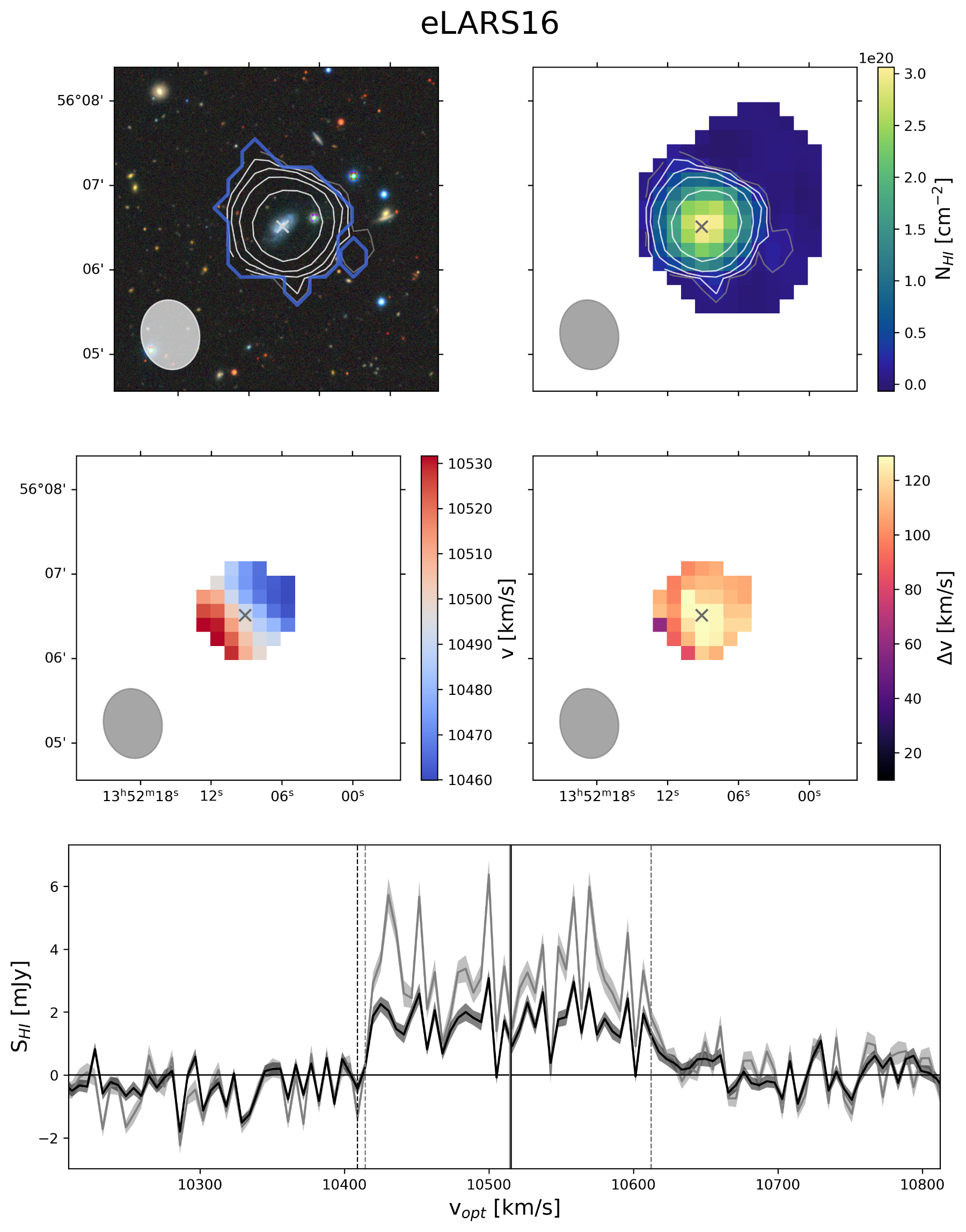}
    \caption{Same caption as Figure \ref{fig:opt_hi_el04}, but for eLARS16.}
    \label{fig:opt_hi_el16}
\end{figure*}

\begin{figure*}[ht]
    \centering
    \includegraphics[width=\textwidth]{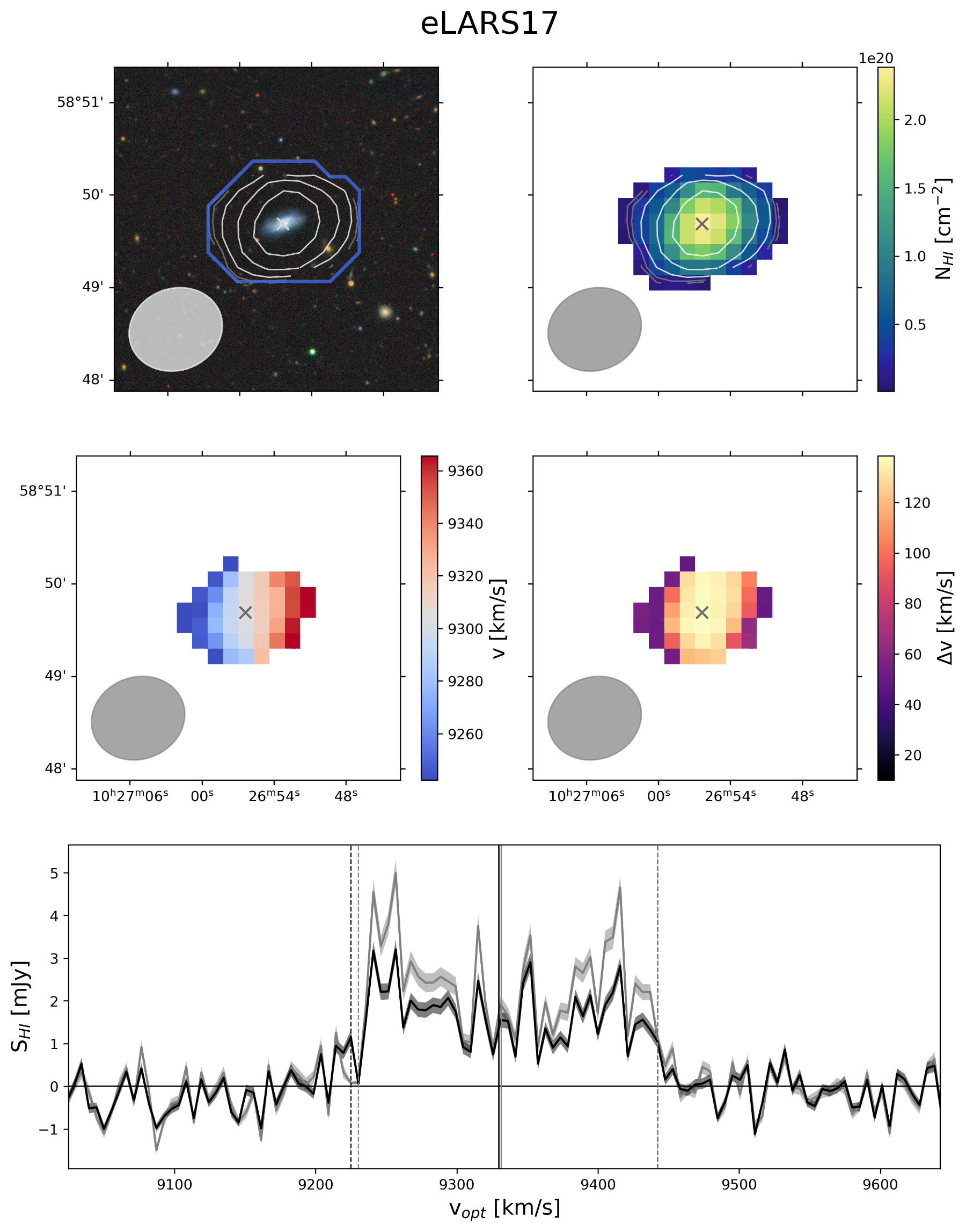}
    \caption{Same caption as Figure \ref{fig:opt_hi_el04}, but for eLARS17.}
    \label{fig:opt_hi_el17}
\end{figure*}

\begin{figure*}[ht]
    \centering
    \includegraphics[width=\textwidth]{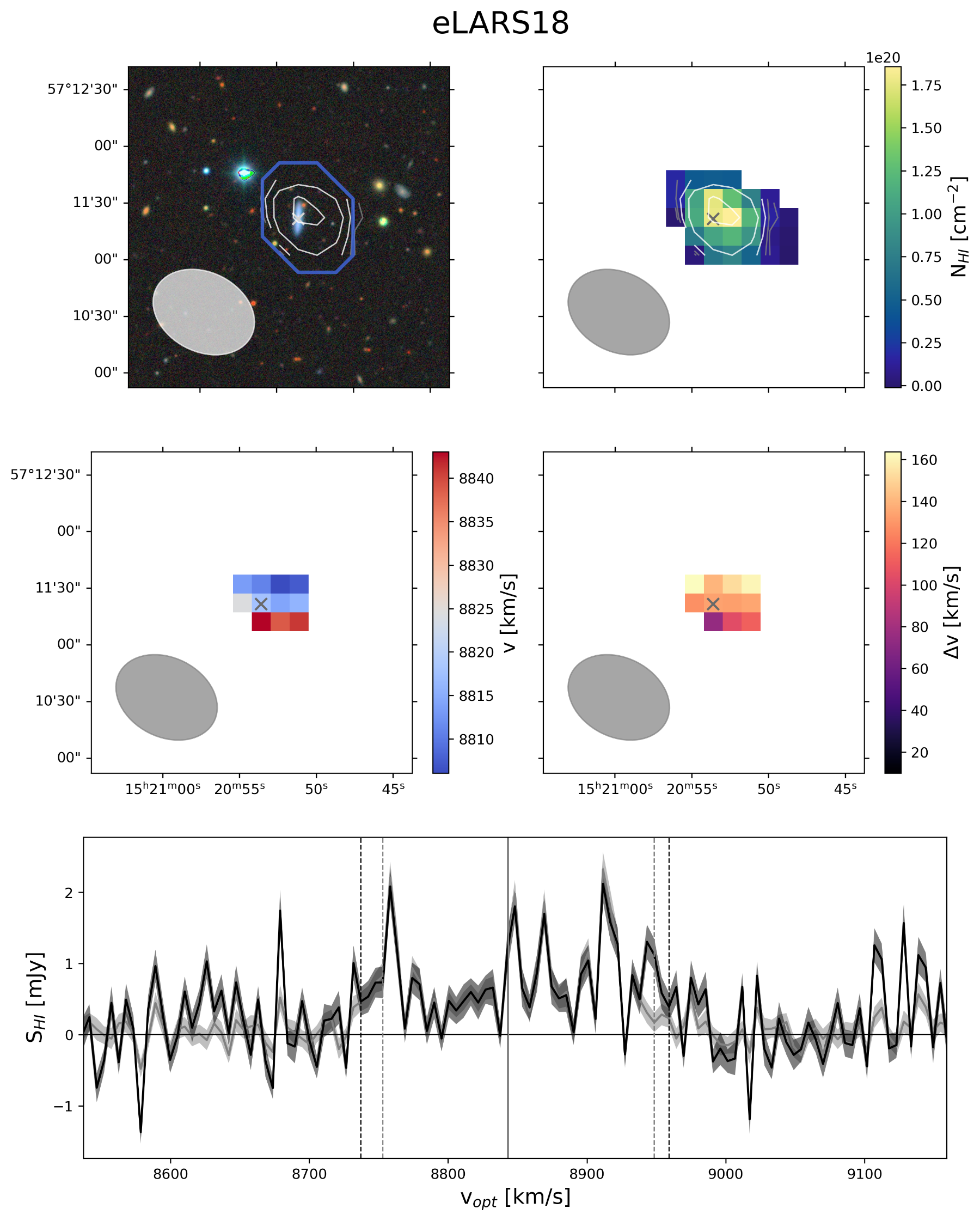}
    \caption{Same caption as Figure \ref{fig:opt_hi_el04}, but for eLARS18.}
    \label{fig:opt_hi_el18}
\end{figure*}

\begin{figure*}[ht]
    \centering
    \includegraphics[width=\textwidth]{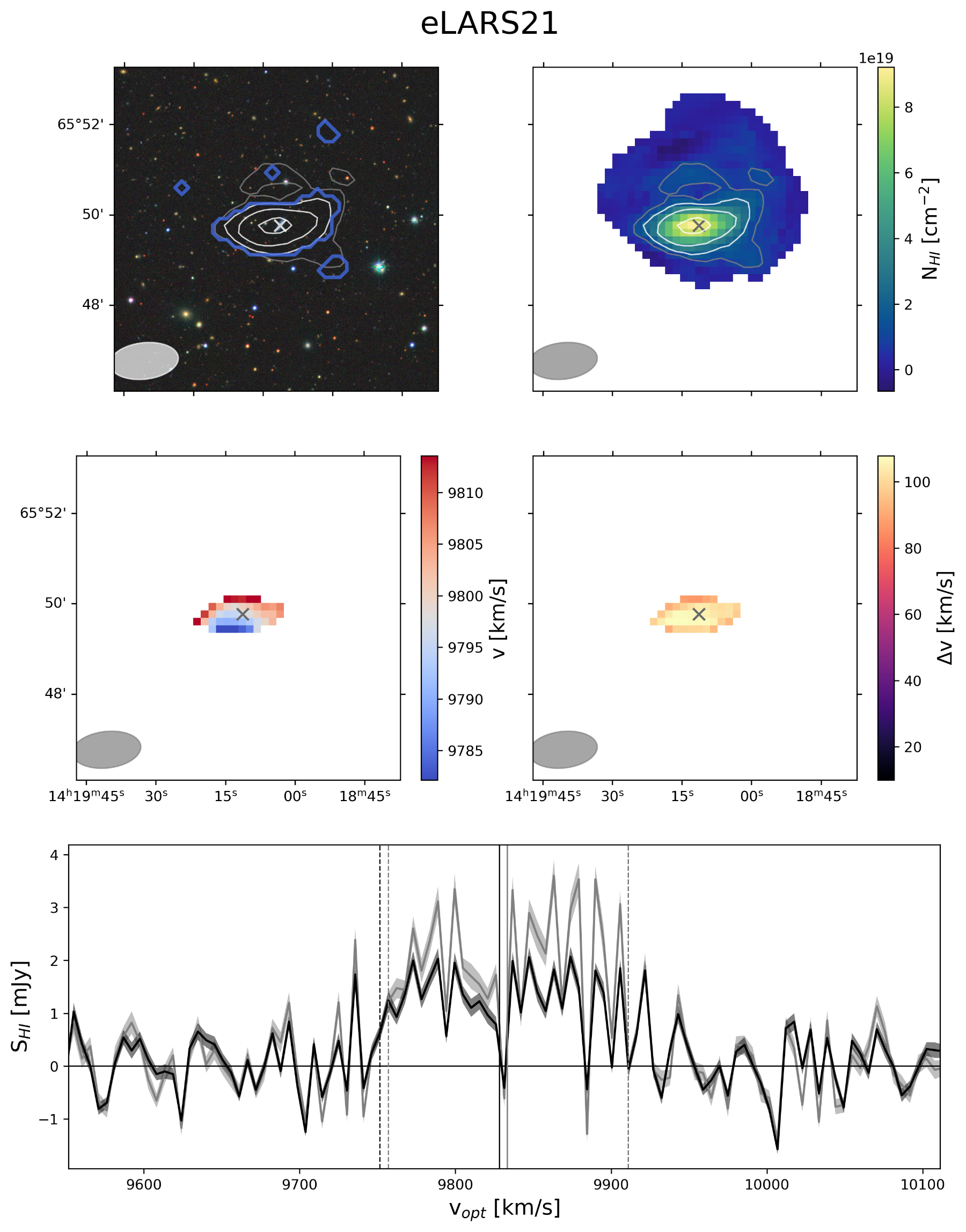}
    \caption{Same caption as Figure \ref{fig:opt_hi_el04}, but for eLARS21.}
    \label{fig:opt_hi_el21}
\end{figure*}

\begin{figure*}[ht]
    \centering
    \includegraphics[width=\textwidth]{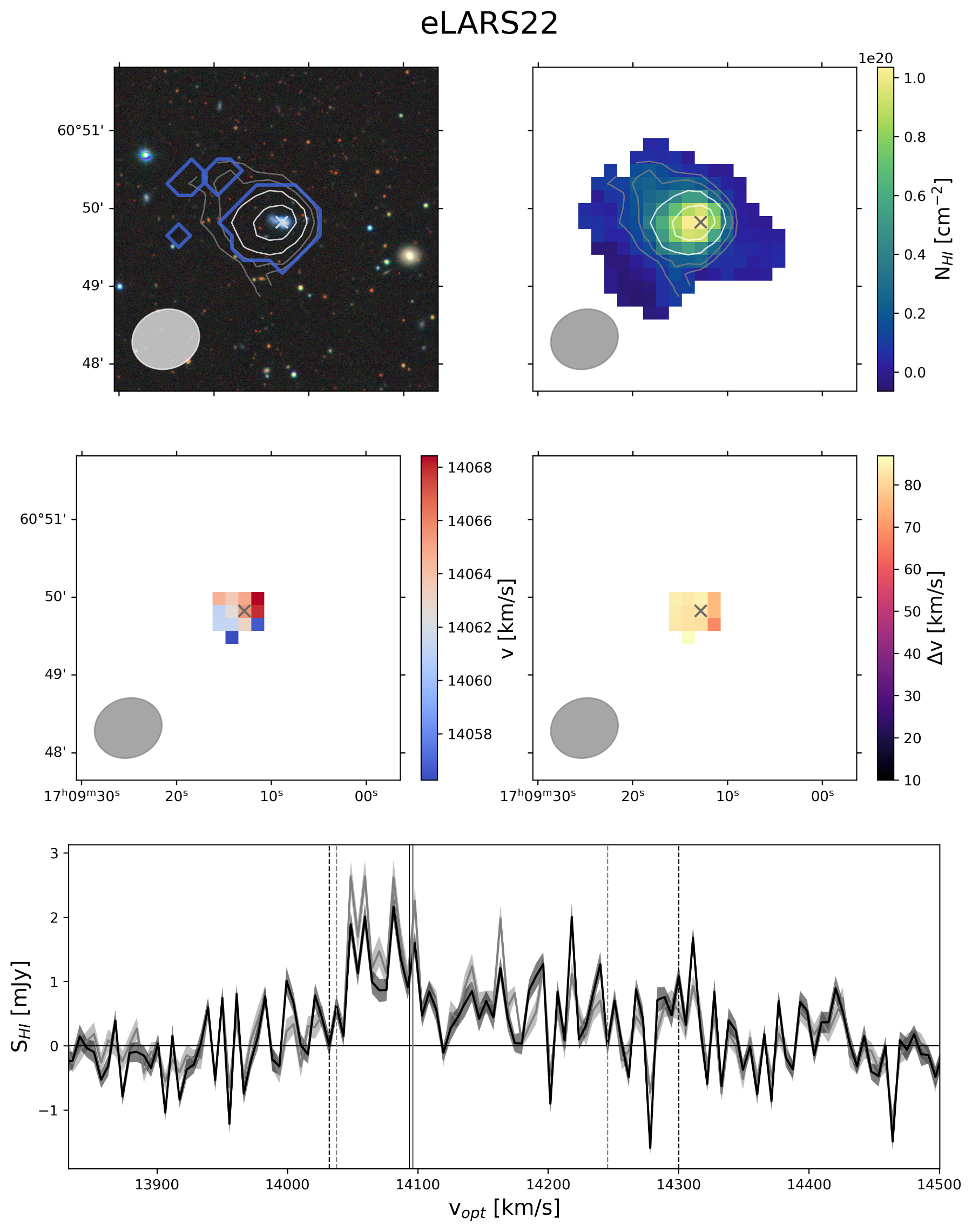}
    \caption{Same caption as Figure \ref{fig:opt_hi_el04}, but for eLARS22.}
    \label{fig:opt_hi_el22}
\end{figure*}

\begin{figure*}[ht]
    \centering
    \includegraphics[width=\textwidth]{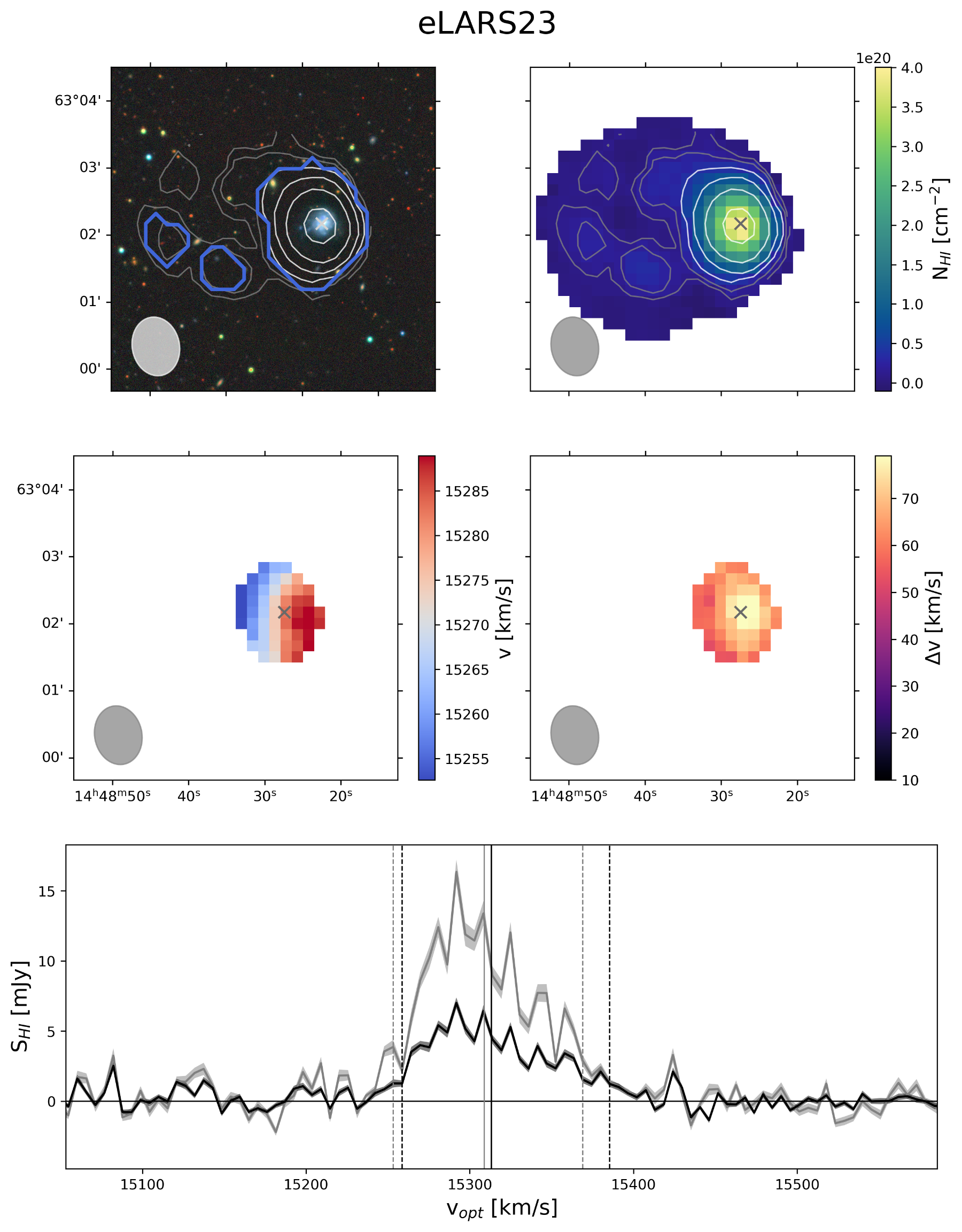}
    \caption{Same caption as Figure \ref{fig:opt_hi_el04}, but for eLARS23.}
    \label{fig:opt_hi_el23}
\end{figure*}

\begin{figure*}[ht]
    \centering
    \includegraphics[width=\textwidth]{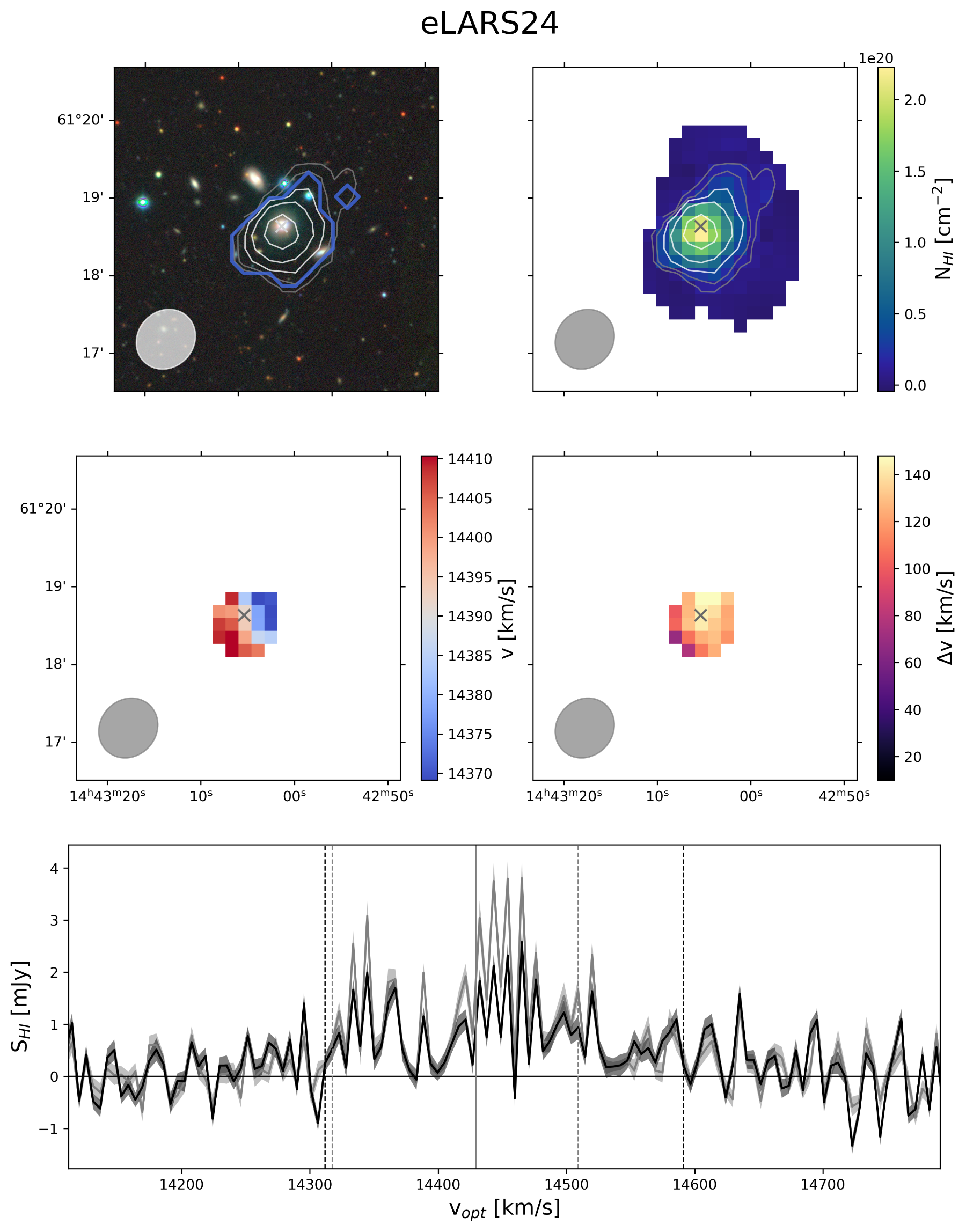}
    \caption{Same caption as Figure \ref{fig:opt_hi_el04}, but for eLARS24.}
    \label{fig:opt_hi_el24}
\end{figure*}

\begin{figure*}[ht]
    \centering
    \includegraphics[width=\textwidth]{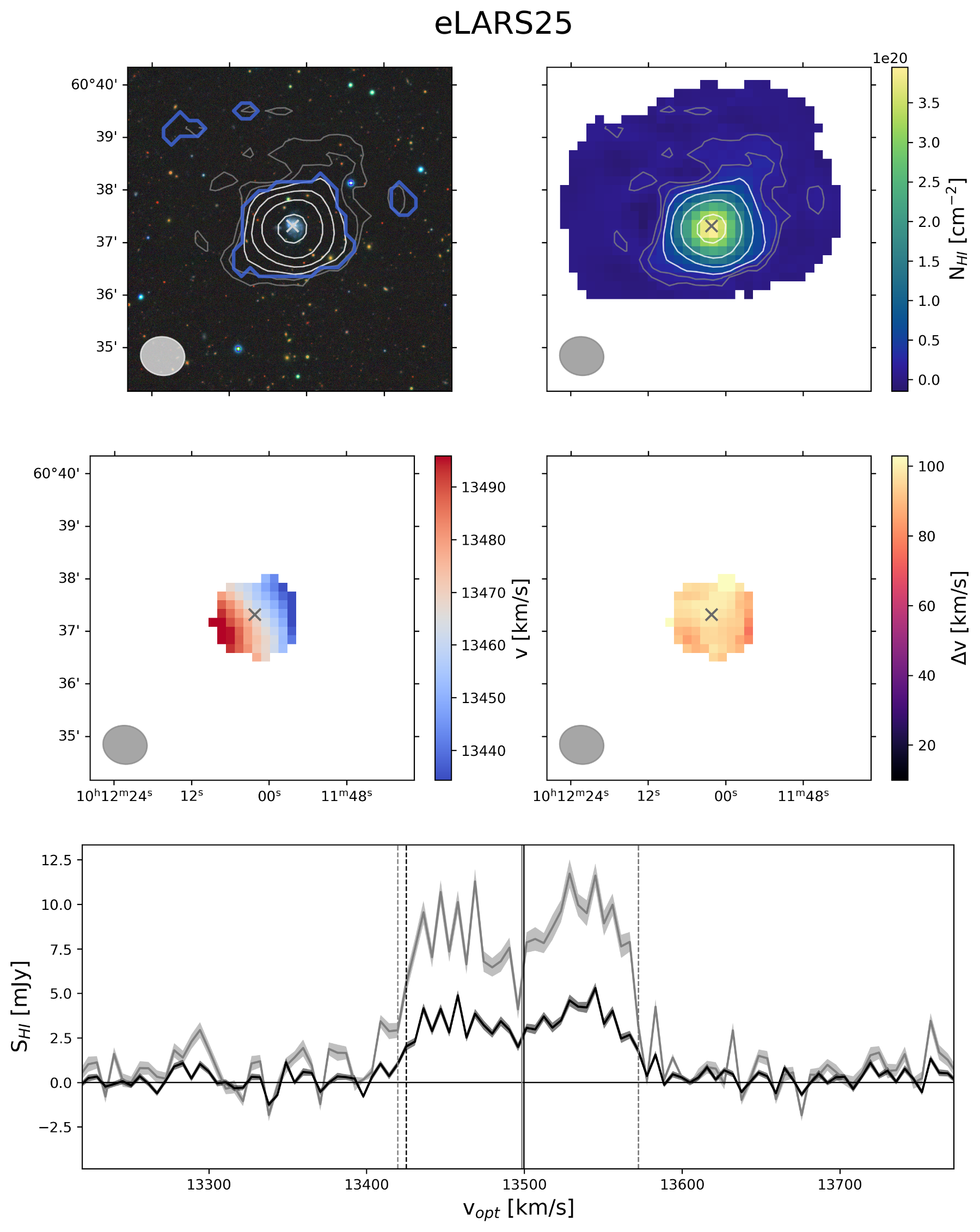}
    \caption{Same caption as Figure \ref{fig:opt_hi_el04}, but for eLARS25.}
    \label{fig:opt_hi_el25}
\end{figure*}

\begin{figure*}[ht]
    \centering
    \includegraphics[width=\textwidth]{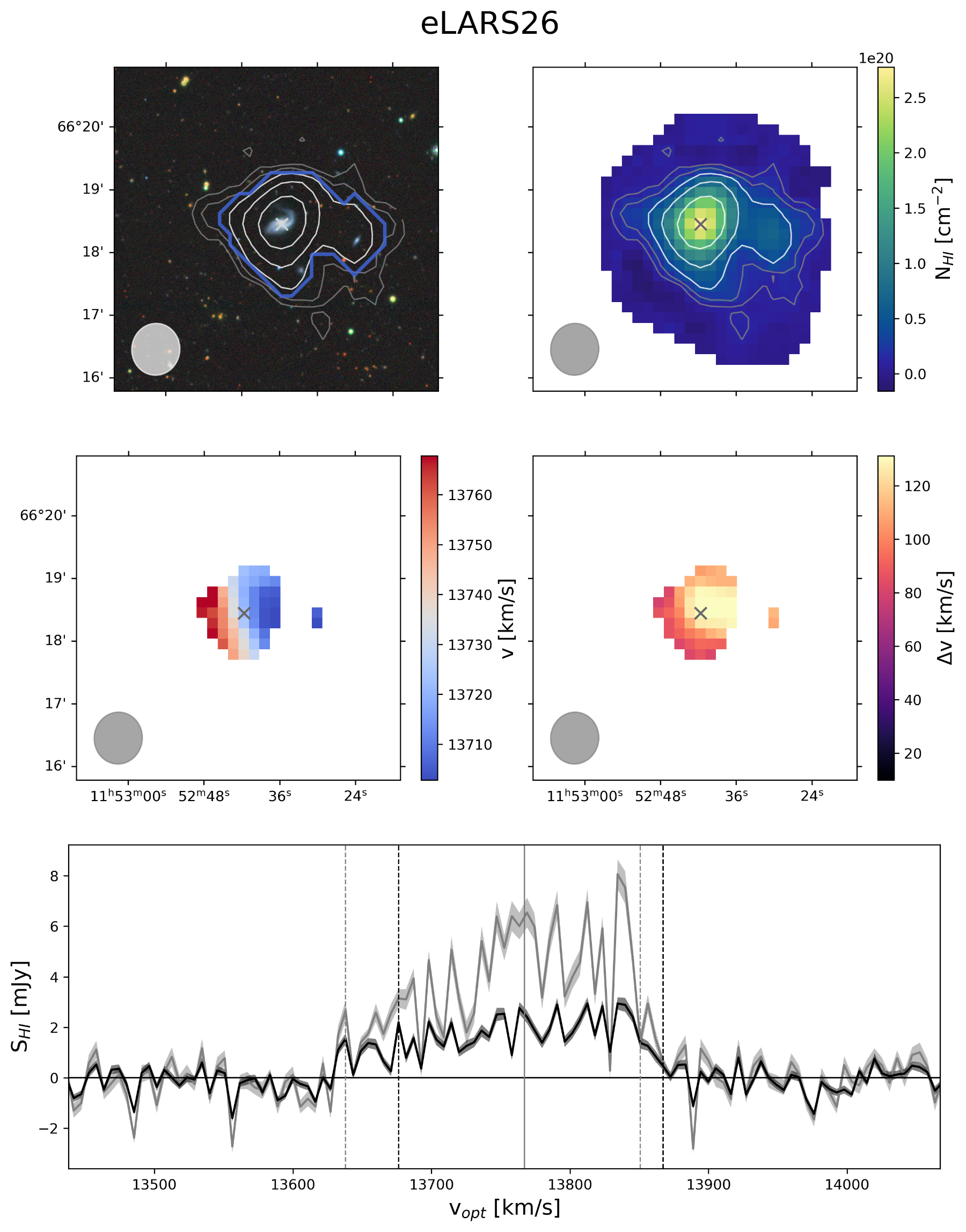}
    \caption{Same caption as Figure \ref{fig:opt_hi_el04}, but for eLARS26.}
    \label{fig:opt_hi_el26}
\end{figure*}

\begin{figure*}[ht]
    \centering
    \includegraphics[width=\textwidth]{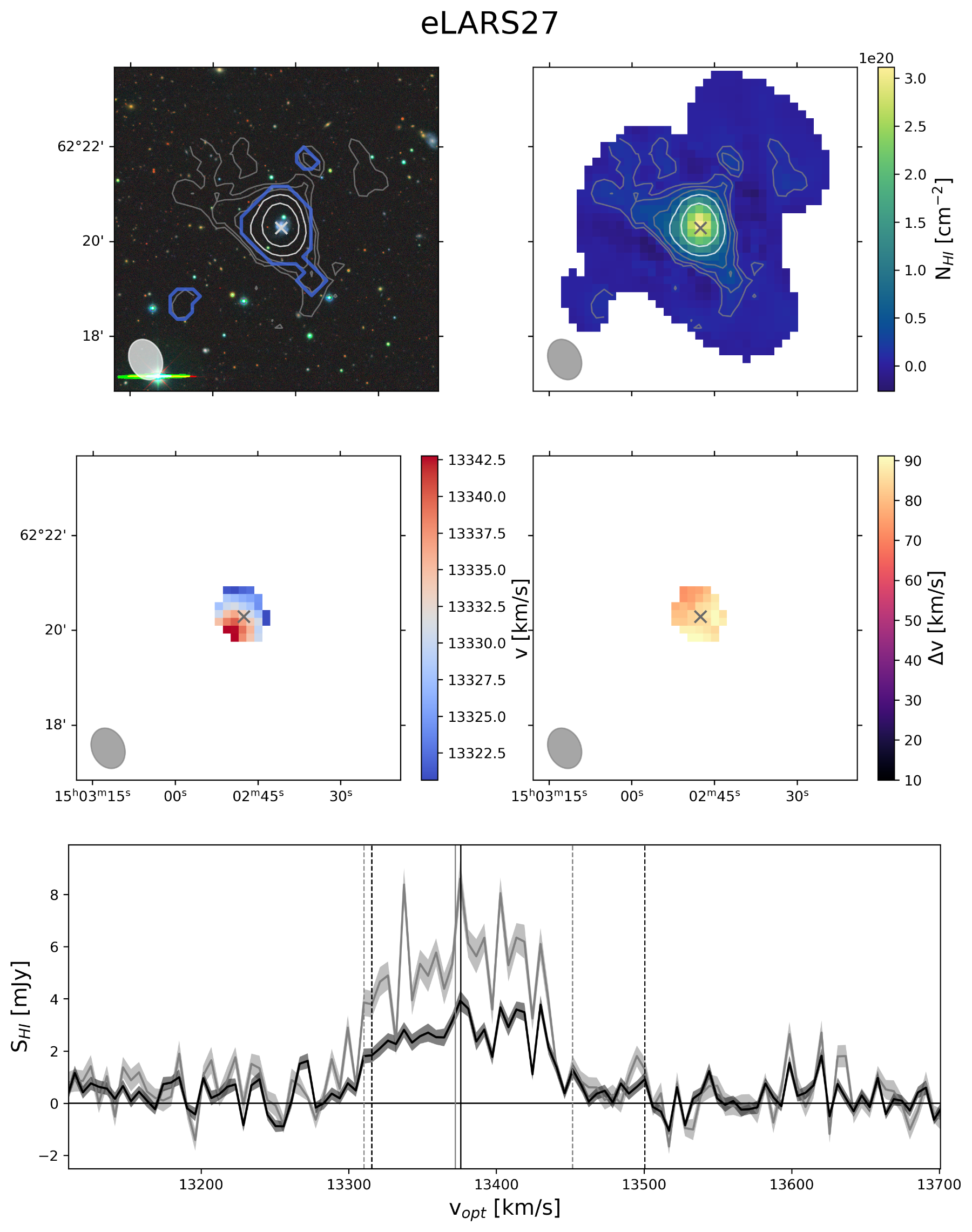}
    \caption{Same caption as Figure \ref{fig:opt_hi_el04}, but for eLARS27.}
    \label{fig:opt_hi_el27}
\end{figure*}

\begin{figure*}[ht]
    \centering
    \includegraphics[width=\textwidth]{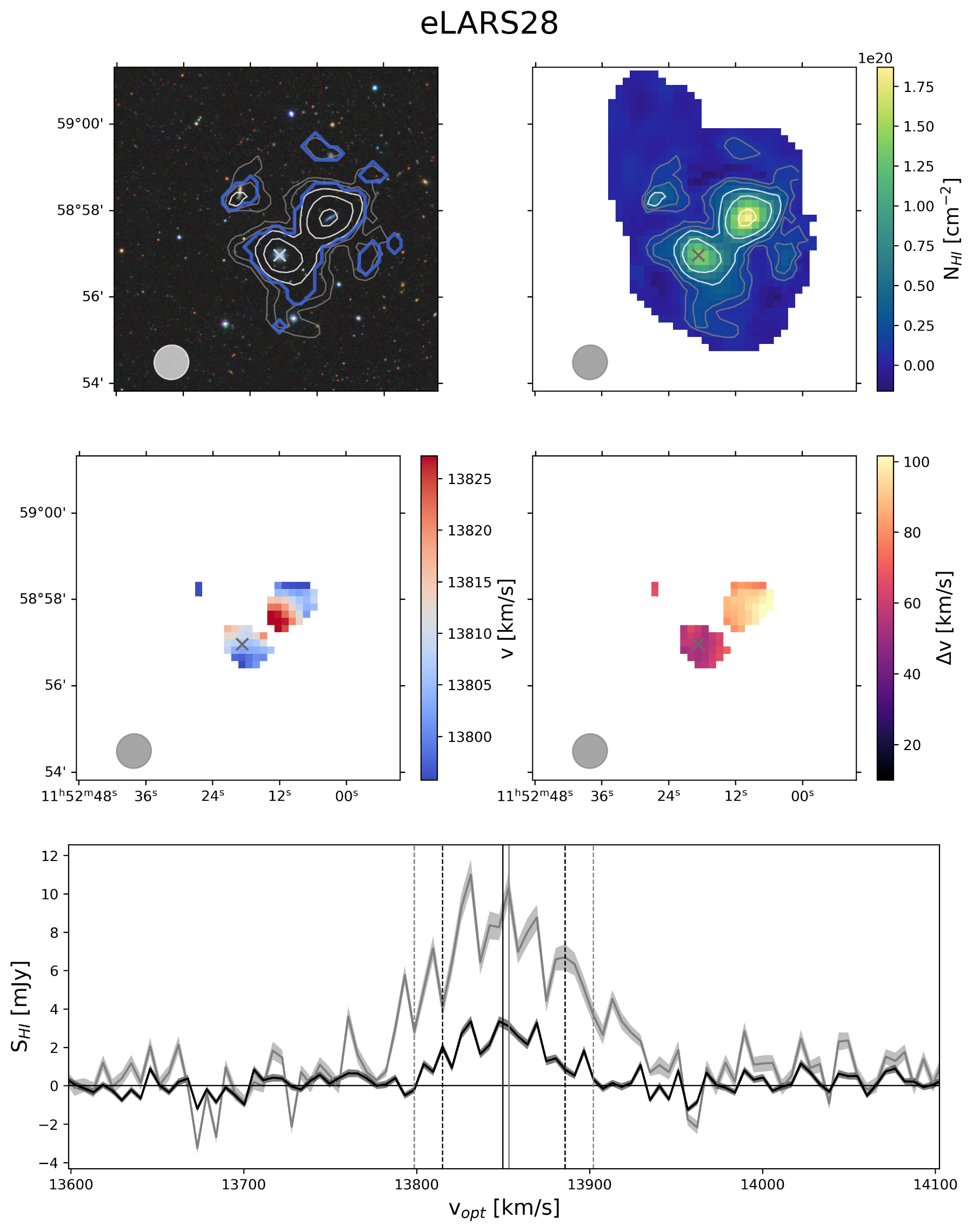}
    \caption{Same caption as Figure \ref{fig:opt_hi_el04}, but for eLARS28.}
    \label{fig:opt_hi_el28}
\end{figure*}

We present 21cm moment maps and spectra for all LARS and eLARS galaxies on Figure \ref{fig:opt_hi_l01} to \ref{fig:opt_hi_el28}. The contours are shown for the main SoFIA 2 detection including the galaxy of interest. A blue contour shows the regions with $SNR>3$.

\subsection*{Appendix C: Comparison with APERTIF and GBT data}
\label{sec:appendix-Apertif}
Here, we compare the \hi\ mass values measured in the manuscript to those obtained with the GBT and VLA, presented in \cite{Pardy2014}, and to preliminary data from the APERTIF survey \citep[][data for (e)LARS obtained through private communication with K. Hess, $\sim$20" beam]{Adams2022} to evaluate the quality of our measurements and data reduction. 
At the time when we conducted this test, three (e)LARS galaxies had available APERTIF imaging: LARS04, eLARS08 and eLARS17.
If we assume 10\% flux calibrator uncertainties, the APERTIF and VLA measurements made in this manuscript agree within 1$\sigma$ for LARS04 and eLARS08, and the difference between measurements for eLARS17 is within $2.4\sigma$. In comparison, the GBT mass for LARS04 is lower than the APERTIF and present VLA measurements by $2.5\sigma$, while the VLA mass reported in \citet{Pardy2014} is lower by more than 3$\sigma$. The GBT has a beam of $8'$, and should recover flux from the diffuse gas that is missed by the interferometric measurements, thus the difference in measurements is surprising. Inspecting the GBT profiles shown in \citet{Pardy2014}, it is apparent that the spectra are strongly impacted by noise. We thus attribute the significant offset to a misplacement of the baseline in the GBT spectra, leading to underestimating of the 21cm flux as measured with the GBT. The VLA measurements made in \cite{Pardy2014} are also lower, with a value different by over 3$\sigma$ for LARS04 than our own estimates and the APERTIF value. Since the APERTIF data was independently reduced by an experienced team using a different telescope, and since they agree with our estimates, we believe the measurements presented here should supersede the values in \cite{Pardy2014}.

\subsection*{Appendix D: GSWLC and SDSS stellar masses and SFRs}
\label{sec:appendix-GSWLC }

\begin{figure}
    \centering
    \includegraphics[width=\linewidth]{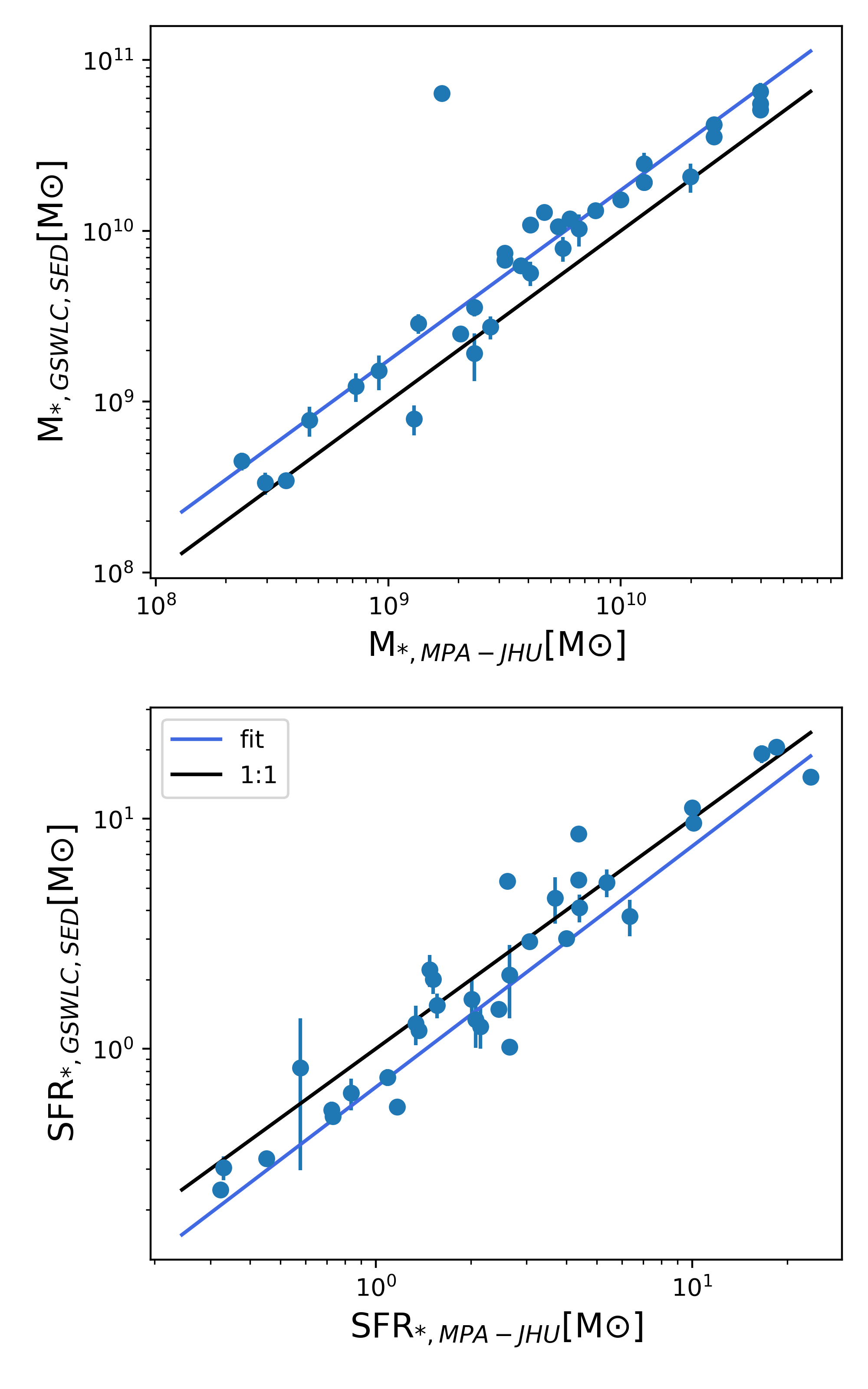}
    \caption{Galaxy properties for (e)LARS galaxies in the GSWLC and SDSS DR8 MPA-JHU catalog. The top panel shows the stellar masses, the bottom panel presents the star formation rates in the two different catalogs. The black line shows the 1:1 relation, the blue line shows the fit to the data.}
    \label{fig:GSWLC}
\end{figure}
Here, we compare the stellar masses and SFRs obtained through SED fitting in GSWLC to the values retrieved from the SDSS DR8 MPA-JHU catalog. The masses and star formation rates for the two catalogs are presented in Figure \ref{fig:GSWLC}, they are in relatively good agreement. We performed a linear least square fit to the data (with single iteration 10$\sigma$ outliers removal) to infer the GSWLC masses and SFRs from the SDSS values for the seven galaxies with missing GSWLC data. The fit to the stellar mass and SFR is shown on Figure \ref{fig:GSWLC}, inferred values are presented in Table \ref{tab:gal_params}.
\end{document}